\documentclass[sn-mathphys]{sn-jnl}

\usepackage{aas_macros}
\jyear{2022}%

\newcommand{\be}{\begin{equation}}
\newcommand{\ee}{\end{equation}}
\def\bear#1\ear{\begin{align}#1\end{align}}
\newcommand{\nline}{\notag \\}
\renewcommand{\f}{\frac}
\newcommand{\de}{\mathrm{d}}
\newcommand{\del}{\partial}

\renewcommand{\e}{\mathrm{e}}

\newcommand{\la}{\left\langle}
\newcommand{\ra}{\right\rangle}

\newcommand{\Msun}{\mbox{M}_{\odot}}

\renewcommand{\mathbf}[1]{\mbox{\boldmath $#1$}}

\newcommand{\vect}{\mathbf}

\newcommand{\eqn}[1]{equation~(\ref{#1})}
\newcommand{\eqns}[2]{equations~(\ref{#1}) and (\ref{#2})}
\newcommand{\secn}[1]{Section~\ref{#1}}

\newcommand{\fig}[1]{Figure~\ref{#1}}

\begin{document}

\title[Introduction to Reionization]{A Short Introduction to Reionization Physics}

\author{\fnm{Tirthankar Roy} \sur{Choudhury}}\email{tirth@ncra.tifr.res.in}

\affil{\orgdiv{National Centre for Radio Astrophysics}, \orgname{Tata Institute of Fundamental Research}, \orgaddress{\street{Pune University Campus}, \city{Pune}, \postcode{411007}, \country{India}}}

\abstract{The epoch of reionization probes the state of our universe when the very first stars formed and ionized the hydrogen atoms in the surrounding medium. Since the epoch has not yet been probed observationally, it is often called the ``final frontier'' of observational cosmology. This final frontier is attracting a lot of attention because of the availability of a large number of telescopes in a wide variety of wavebands. This review article summarizes some of the concepts required to understand the interesting physics of reionization and how to analyze the high-redshift universe using related observations.}




\maketitle

\section{Introduction}\label{sec1}

This review is written in memory of T.~Padmanabhan (or Paddy, as he was known to his colleagues and students), who passed away in September 2021. He was not only my Ph.D. supervisor but also someone who I looked up to throughout my research career. Paddy's excellence in research was ably matched by his interests in pedagogy, in particular, his efforts in making cutting edge research areas accessible to young researchers through textbooks and review articles. As a tribute to his efforts in this direction, this is my humble attempt to introduce a contemporary and relevant research topic, namely, the epoch of reionization, to students beginning to work in this field. Along the way, my aim would be to also highlight some of Paddy's contribution in related areas. In addition, I would take this opportunity to recapitulate some interesting results obtained by me and my collaborators.

The epoch of reionization is the cosmic era when hydrogen atoms in the Universe got ionized because of the radiation produced by the very first stars. The study of reionization thus concerns the ionization and thermal history of hydrogen, the element which forms the bulk of the baryonic matter in our universe. Within the framework of hot Big Bang model, we know that the hydrogen atoms formed for the first time during the \emph{recombination epoch} which is well-probed by the Cosmic Microwave Background (CMB). Right after the recombination epoch, the Universe entered a phase called the ``dark ages'' where no radiation sources (stars or active galaxies) existed. The hydrogen thus remained largely neutral at this phase. The small inhomogeneities in the dark matter density field which were present during the recombination epoch (and well-probed by the CMB) started growing via gravitational instability, eventually forming the first stars inside galaxies. Once these stars form, the dark ages end and the ``cosmic dawn'' begins. The first population of luminous stars, and possibly some early population of accreting black holes (quasars), will generate ultraviolet (UV) radiation which can ionize the hydrogen atoms in the surrounding intergalactic medium (IGM). This process is known as ``reionization''. This is the second major change in the ionization state of hydrogen in the universe (the first being the recombination).

\begin{figure}
\begin{center}
\includegraphics[width=\textwidth]{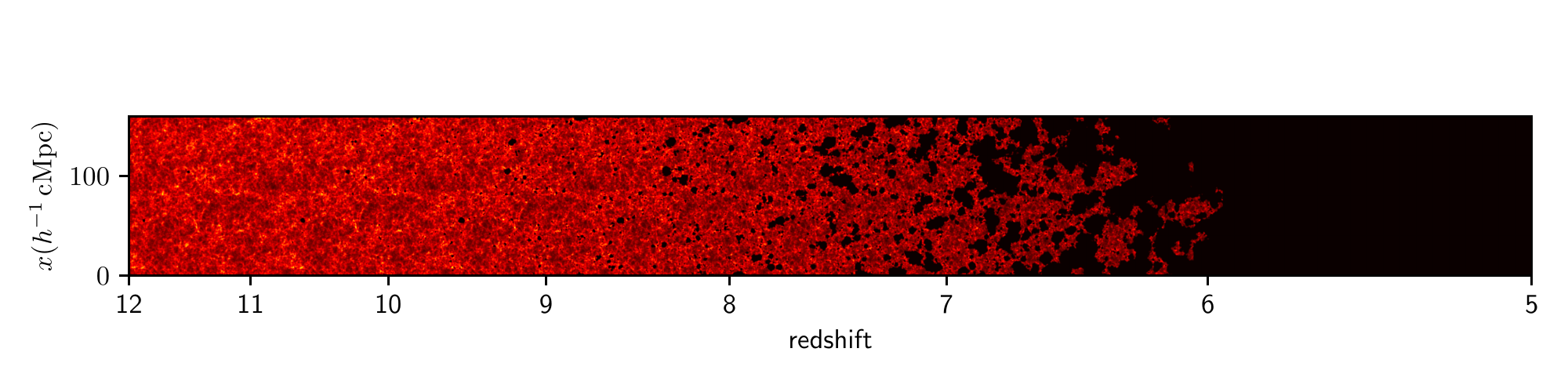}
\end{center}
\caption{A lightcone map, obtained using a semi-numerical simulation of reionization \cite{2018MNRAS.481.3821C}, showing the evolution of the neutral and ionized hydrogen density field. The red points correspond to neutral regions while the black areas represent the ionized ones. The universe undergoes a phase transition from being highly neutral at high redshifts to a fully ionized one at lower redshifts.}
\label{fig:lightcone}
\end{figure}

Reionization started around the time when first structures formed. The exact timing of this start is still unknown but is probably in the redshift range $z \sim 20-30$. In the simplest picture, each source produced an ionized region around it; these regions then overlapped and percolated into the IGM. The process of overlapping completed around $z \sim 5-6$ at which point the bulk of the hydrogen returned back to being ionized. This process can be visualized through a lightcone of the hydrogen distribution, shown in \fig{fig:lightcone}. The map shown was made using a semi-numerical simulation of reionization \cite{2018MNRAS.481.3821C}, to be discussed later in the article. For the moment, we want to highlight that the hydrogen at high redshifts is largely neutral (red regions). At some redshift, ionized regions (black points) start to appear and these keep growing and overlapping with decreasing redshift. The universe finally transitions to a completely ionized state marking the completion of reionization.

The process of reionization is of immense importance in the study of structure formation since, on the one hand, it is a direct consequence of the formation of first structures and luminous sources while, on the other, it affects subsequent structure formation. Observationally, the reionization era represents a phase of the universe which is yet to be probed; the earlier phases ($z \sim 1000$) are probed by the CMB while the post-reionization phase ($z < 6$) is probed by various observations based on UV galaxies, quasars and other sources. In addition to the importance outlined above, the study of cosmic reionization has acquired increasing significance over the last few years because of the availability of good quality data from different telescopes. In fact, almost any experiment which probes the Universe at $z \gtrsim 5$ has some relevance to the epoch of reionization and thus reionization can be thought of an ideal probe for the high-redshift universe.

It is clear that the study of reionization has a rather broad scope and it is not possible to cover each and every aspect in a short review like this. We will hence concentrate on discussing some basic concepts relevant to the physics of reionization and on deriving some of the most fundamental equations in the subject area. We will also indicate how to use the solutions of these equations to compare the theory with related observations. The limited scope of this review will not allow us to discuss many of the more recent developments in great detail, we refer the readers to other more comprehensive review articles \cite{2001PhR...349..125B,2005SSRv..116..625C,2018PhR...780....1D,2022arXiv220802260G}. We will also not cover the related interesting topics of the dark ages and cosmic dawn, and limit ourselves to the epoch of reionization.

The review article is written keeping those in mind who already have some exposure to cosmology and large-scale structure formation \cite{2002thas.book.....P,1999coph.book.....P,2010gfe..book.....M} and are looking for some quick introductory material to understand how theoretical models of reionization are constructed. For the most part, we will be concentrating on a formalism to compute the evolution of the globally averaged ionized fraction; the evolution of this fraction is essentially what we call the reionization history and what researchers in the field are trying to understand.

The plan of the article is as follows:  We begin by discussing the properties of the sources of reionization in \secn{sec:sources}. This is followed by a discussion on the radiative transfer of photons through the IGM in \secn{sec:RT}. These two sections are the most essential parts of the article and lead to some of the most useful equations needed for constructing reionization models. In \secn{sec:observables}, We connect these calculations to the main observables related to the reionization history. Beyond the globally averaged reionization history, one can also study reionization using the fluctuations in the ionized field. We briefly summarize some of the methods used for calculating these fluctuations in \secn{sec:fluctuations}. We end by presenting the future prospects in \secn{sec:future}.

\section{Sources of reionization}
\label{sec:sources}

This section is devoted to modelling the properties of the reionization sources. Among the various possible sources, we will concentrate on the stars inside galaxies which are believed to be the primary contributors to the hydrogen ionizing photon budget.

\subsection{Dark matter haloes}

The stars, which constitute the galaxies and are believed the main sources of ionizing photons, form inside dark matter haloes.\footnote{The formation of the collapses objects is an important problem of study in large-scale structure formation. Paddy was one of the first people to write a textbook addressing many of these issues \cite{1993sfu..book.....P}.} These haloes are nothing but gravitationally bound dense structures and they can attract the baryons (i.e., gas particles) to form dense clouds. The first step in any reionization model is to calculate the number of haloes per unit \emph{comoving} volume per unit mass range, known as the \emph{halo mass function} and denoted by $n(M, z)$.

In analytical calculations, the mass function can be computed using the excursion set formalism, which makes use of linear perturbation theory of matter fluctuations and spherical collapse \cite{1991ApJ...379..440B}. The result is a simple form known as the \emph{Press-Schechter mass function} \cite{1974ApJ...187..425P}
\be
n(M, z) = \f{\bar{\rho}_m}{M^2} f(\nu) \left\vert\f{\de \ln \nu}{\de \ln M}\right\vert,
\label{eq:dndM_fnu}
\ee
where $\bar{\rho}_m$ is the mean comoving matter density and
\be
\nu = \f{\delta_c(z)}{\sigma(M)}.
\ee
In the above relation, $\delta_c(z) = 1.686 / D(z)$ is the critical density of collapse at redshift $z$, $D(z)$ being the linear growth factor of matter fluctuations and $\sigma^2(M)$ is the variance of the density field smoothed in a spherical region of mass $M$. The function $f(\nu)$ has the simple form
\be
f(\nu) = \sqrt{\f{2}{\pi}} \nu~\e^{-\nu^2 / 2}.
\label{eq:fnu_PS}
\ee

\begin{figure*}
\begin{center}
\includegraphics[width=0.6\textwidth]{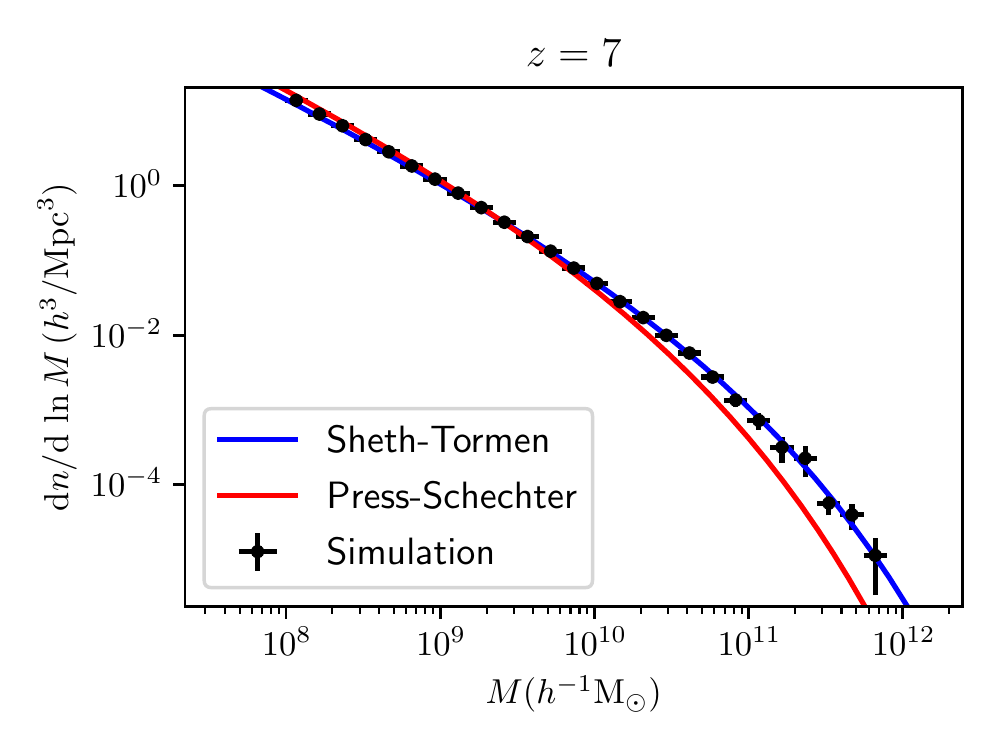}
\end{center}
\caption{The comparison of the halo mass function obtained from $N$-body simulations (points with error-bars) and two different analytical forms: Press-Schechter (red curve) and Sheth-Tormen (blue curve). The results are shown for $z = 7$. The $N$-body simulation has been run using the publicly available code \texttt{GADGET-2} \cite{2005MNRAS.364.1105S} which can be downloaded from \url{https://wwwmpa.mpa-garching.mpg.de/gadget/}. The haloes in the simulation have been identified using a Friends-of-friends group-finding algorithm.}
\label{fig:mf}
\end{figure*}

While the Press-Schechter mass function provides quite decent qualitative match with more accurate results obtained from $N$-body simulations, the quantitative match is not satisfactory, particularly at high masses. As an example, let us look at \fig{fig:mf} where we have plotted the mass function at $z = 7$ (appropriate for studying reionization) obtained from $N$-body simulations (points with error-bars) and compared with the Press-Schechter mass function (red curve). It should be obvious that the analytical form given above is not an acceptable match to the more accurate simulation results for halo masses $M \gtrsim 10^{10} \Msun$.

A more accurate formalism, based on the ellipsoidal collapse, also provides an analytical form of the mass function. In fact, the basic form as in \eqn{eq:dndM_fnu} remains the same. The change is in the form of $f(\nu)$ which is given by \cite{2001MNRAS.323....1S,2002MNRAS.329...61S}
\be
f(\nu) = A \sqrt{\f{2 a}{\pi}} \left[1 + (a \nu^2)^{-p}\right] \nu~\e^{-a \nu^2 / 2},
\label{eq:fnu_ST}
\ee
where $A$, $a$ and $p$ are constants. This is known as the \emph{Sheth-Tormen mass function}. It is possible to tune the values of the parameters $A$, $a$ and $p$ to match the simulations and a good match is found to be for the values $A = 0.353, a = 0.73, p = 0.175$ \cite{2001MNRAS.321..372J}. The match can be seen in \fig{fig:mf} where the blue curve represents the Sheth-Tormen mass function.

Another option available for analytical modelling is to use numerical fits to the mass functions obtained from large $N$-body simulations, particularly the ones which have been validated at high redshifts and for small halo masses, appropriate for reionization studies \cite{2013MNRAS.433.1230W}.

A related quantity that plays an important role in reionization studies is the \emph{collapsed fraction}
\be
f_{\mathrm{coll}}(M_{\mathrm{min}}, z) = \f{1}{\bar{\rho}_m} \int_{M_{\mathrm{min}}}^{\infty} \de M~M~n(M, z),
\ee
which determines the mass fraction in collapsed haloes of mass $M \geq M_{\mathrm{min}}$. For the Press-Schechter mass function, this fraction has a remarkably simple form
\be
f_{\mathrm{coll}}(M_{\mathrm{min}}, z) = \text{erfc} \left[\f{\delta_c(z)}{\sqrt{2}~\sigma(M_{\mathrm{min}})}\right]. 
\ee
For the other cases, $f_{\mathrm{coll}}$ needs to calculated by numerically integrating the halo mass function.

\subsection{Cooling of gas in dark matter haloes}

If one assumes that the halo is in virial equilibrium, several other quantities related to its physical properties can be calculated. The first obvious quantity is the radius of the halo, also known as the \emph{virial radius}, which is given by
\be
R_{\mathrm{vir}} = 1.48~\text{kpc} \left(\f{M}{10^8 \Msun}\right)^{1/3} \left(\f{\Omega_m h^2}{0.15}\right)^{-1/3} \left(\f{\Delta_{\mathrm{vir}}}{18 \pi^2}\right)^{-1/3} \left(\f{1+z}{10}\right)^{-1},
\ee
where $\Delta_{\mathrm{vir}}$ is the overdensity of the halo. Using the theory of spherical collapse, one can show that this overdensity for an Einstein-deSitter universe (which is a very good assumption for the high-redshift universe) is given by $\Delta_{\mathrm{vir}} = 18 \pi^2$.

The gas particles will be gravitationally attracted to the dark matter haloes because of their deep potential well. They will subsequently relax to a virial equilibrium leading to a definite relation between their kinetic energy $K$ and the potential energy $U$
\be
2 K + U = 0.
\ee
Let us define a \emph{virial temperature} $T_{\mathrm{vir}}$ through the relation
\be
K = \f{k_B T_{\mathrm{vir}}}{\mu m_p}, ~~\mu \equiv \f{\rho_{\mathrm{tot}}}{n_{\mathrm{tot}} m_p},
\ee
where $m_p$ is the proton (or hydrogen) mass, $\mu$ is the mean molecular weight, defined in terms of the total baryonic number density $n_{\mathrm{tot}}$ (including the free electrons, if any) and mass density $\rho_{\mathrm{tot}}$. The virial relation then reduces to a relation between $T_{\mathrm{vir}}$ and the \emph{circular velocity} $v_c$ of the halo
\be
T_{\mathrm{vir}} = \f{\mu m_p v_c^2}{2 k_B},~~v_c = \sqrt{\f{G M}{R_{\mathrm{vir}}}}.
\ee
For haloes typical of hosting the first stars, the virial temperature has a value
\be
T_{\mathrm{vir}} = 1.06 \times 10^4~\mathrm{K} \left(\f{\mu}{0.6}\right) \left(\f{M}{10^8 \Msun}\right)^{2/3} \left(\f{\Omega_m h^2}{0.15}\right)^{1/3} \left(\f{\Delta_{\mathrm{vir}}}{18 \pi^2}\right)^{1/3} \left(\f{1+z}{10}\right).
\ee

The gas particles inside the halo will acquire speeds $\sim v_c$ which will heat them up to temperatures $\sim T_{\mathrm{vir}}$, a process known as \emph{shock heating}. At these temperatures, the gas will reach an equilibrium where the inward gravitational pull is balanced by the outward thermal pressure. The density of the gas in virial equilibrium inside a halo forming as $z \sim 10$ is (assuming a virial overdensity $\sim 200$) $\rho_b \sim 10^{-25} \mathrm{gm~cm}^{-3}$, with the corresponding number densities being (assuming $\mu \sim 1$) $n_b \sim 6 \times 10^{-2} \mbox{cm}^{-3}$. This is too small a density for nuclear reactions to occur, hence in the absence of any further interactions, the gas would simply remain in this low-density hot state. Formation of stars would require the gas to condense, which would need processes that can cool the gas.\footnote{The relations presented here may be more accurately computed assuming a density profile of the halo, composition etc. These would lead to some corrections which are only of order unity. To the order of magnitude, such details can be ignored.}

For the gas to be able to compress, it needs to cool so that the average kinetic energy of the gas particles decrease and the gravitational forces can make the gas collapse. There are many possible processes which can achieve the cooling. The dominant process at high redshifts, appropriate for a gas consisting of primarily hydrogen and helium atoms, is the collisional excitation of atoms followed by radiative cooling. Essentially, the atoms are excited to a higher energy level because of a collision with another atom. This is followed by a de-excitation to the lower state resulting in an emission of a photon. If this photon can escape the halo (i.e., optically thin condition), the energy of the system will decrease and hence it will cool. Other processes for cooling (e.g., collisional ionization and recombination) are relatively less prominent at redshifts we are interested in.

It is possible to compute the cooling rate from the basics of atomic physics. Without going into the details of such calculations and results, let us summarize the main points which are important for understanding the formation of early stars, for details the reader can look up, e.g., Chapter 7 of Paddy's book on Theoretical Astrophysics \cite{2002thas.book.....P}. For a gas composed of primordial elements (i.e., hydrogen and helium), the cooling rate is dominated by the collisional excitation of neutral hydrogen and singly ionized helium and shows two peaks around $2 \times 10^4$ K and $9 \times 10^4$ K and, more importantly, drops drastically for $T \lesssim 10^4$ K. This is because the atoms are mostly in the ground state and the gas particles do not have sufficient energy to excite the atoms (recall that the temperature corresponding to the energy difference between $n=1$ and $n=2$, Lyman-$\alpha$, is $\sim 10^5$ K). This implies that haloes which have $T_{\mathrm{vir}} \lesssim 10^4$~K will not be able to cool (unless something other than hydrogen and helium atoms is present), which leads to a minimum mass of haloes that can form stars to be $\sim 10^8 \Msun$.

At lower redshifts, the presence of metals (e.g., oxygen, carbon, nitrogen) can help in increasing the cooling rate, though the rise is marginal for $T \lesssim 10^4$~K. However, the presence of metals is not common at high redshifts before the star formation becomes significant. The cooling rate at low temperatures can be boosted by presence of molecules like H$_2$ and CO. Their rotational and vibrational levels can be excited by electrons of smaller energies. At high redshifts, the ones which are relevant for reionization, the only molecules which are possible to form in substantial quantities are the H$_2$, although their formation is not straightforward and requires presence of a catalyst (e.g., dust grains or free electrons). The presence of H$_2$ allows cooling even at smaller virial temperatures $T_{\mathrm{vir}} \sim 500$ K, which corresponds to halo masses $M \sim 10^6 \Msun$. These haloes are called \emph{minihaloes} and can be important sources of stars at very early redshifts. The difficulty with molecular cooling inside minihaloes is that the formation of the first stars can lead to suppression in molecular cooling, which can then quench the star formation (a process known as feedback, which we will discuss in \secn{sec:feedback}). The photons responsible for this quenching have energies in the range 11.26 -- 13.6 eV (Lyman-Werner radiation) which can dissociate the H$_2$ molecules. 

So, to summarize the discussion related to the cooling of gas inside haloes, there exists a minimum halo mass $M_{\mathrm{min}}$ below which it is not possible to cool the gas and form stars and its value depends on the composition of the gas.

\subsection{Formation of stars}

The cooling of gas leads to fall in the pressure, hence the gravitational force can be balanced by the outward force only for regions of smaller mass than before. This leads to \emph{fragmentation} of the gas in the halo, where the gas in separated into clumps of smaller sizes. A detailed discussion of fragmentation is beyond the scope of this review, there are reviews which discuss these aspects in the context of the first stars and reioniztaion \cite{2005SSRv..116..625C,2018PhR...780....1D}. The fragmentation does not continue indefinitely, and is halted when the gas becomes dense and optically thick. The masses of the stars formed depend on when and how the fragmentation stops.

The distribution of the stellar masses $m_*$ is characterized by the \emph{stellar initial mass function (IMF)}. It is defined as the relative number of stars born with mass in a given range. At early times, when the cooling was not that efficient in the absence of metals, the fragmentation stops early and the stars formed tend to have much higher mass leading to a ``top-heavy'' IMF. The exact details are somewhat uncertain, however, simulation based studies put this number somewhere around $\sim 10 - 100 \Msun$ \cite{2013RPPh...76k2901B}. These heavy stars, known as \emph{Population III (PopIII) stars}, are extremely efficient producers of ionizing photons but have significantly shorter life time $\sim 10^6$ years. Their formation is also suppressed as soon as the medium gets metal polluted, hence the contribution of PopIII stars to reionization is often debated. 

Following the metal enrichment of the medium, the cooling mechanism gets modified and hence the next generation of stars, called the \emph{Population II or PopII stars} have very different properties. Their IMFs tend to be more distributed and have significant contribution at $m_* \sim \Msun$. Most of these PopII stars do not produce ionizing photons as efficiently as the PopIII stars. More details on this topic can be found in textbooks on galaxy formation \cite{2010gfe..book.....M}.

Let us assume that a fraction $f_*$ of baryons are converted into stars, without worrying about the finer details of the start formation process. This allows us to write a relation for the stellar mass
\be
M_* = f_*(M, t) \left(\f{\Omega_b}{\Omega_m}\right)~M,
\ee
where we have taken $f_*$ to depend on both $M$ and $t$ (or, equivalently $z$). Let us write the star formation rate in a halo as
\be
\dot{M}_*(M, t, t_{\mathrm{form}}) = f_*(M, t_{\mathrm{form}}) \left(\f{\Omega_b}{\Omega_m}\right)~M ~\Lambda_*(t-t_{\mathrm{form}}; t_{\mathrm{form}}),
\ee
where $t_{\mathrm{form}}$ is the formation time of the galaxy (assumed to be the same as the time of the halo collapse) and $\Lambda_*(t-t_{\mathrm{form}}; M, t_{\mathrm{form}})$ is the star formation profile. The profile is normalized as
\be
\int_0^{\infty} \de t~\Lambda_*(t; M, t_{\mathrm{form}}) = 1,
\ee
which ensures that $\int_{t_{\mathrm{form}}}^{\infty} \de t~\dot{M}_*(M, t, t_{\mathrm{form}}) = M_*$. In general, the profile depends not only on the age $t - t_{\mathrm{form}}$ of the galaxy, but also on the halo mass and the time at which the halo forms. This allows one to include any possible cosmic evolution in the nature of the star formation in the equations.

If the star formation is continuous, one can approximate $\Lambda_*(t; M, t_{\mathrm{form}}) \approx $ constant. On the other extreme, a short burst of star formation would imply $\Lambda_*(t; M, t_{\mathrm{form}}) = \delta_D(t)$. One useful form of the profile which covers a wide range of possibilities is \cite{1996ApJ...456....1G,2000ApJ...534..507C,2002MNRAS.336L..27R,2013MNRAS.435..368J,2014MNRAS.443.3341J}
\be
\Lambda_*(t; M, t_{\mathrm{form}}) = \f{t}{\kappa^2(M, t_{\mathrm{form}})~t^2_{\mathrm{dyn}}(t_{\mathrm{form}})} \exp\left(-\f{t}{\kappa(M, t_{\mathrm{form}})~t_{\mathrm{dyn}}(t_{\mathrm{form}})}\right),
\label{eq:Lambda_tdyn}
\ee
where
\be
t_{\mathrm{dyn}}(t_{\mathrm{form}}) = \sqrt{\f{3 \pi}{32 G \Delta_{\mathrm{vir}} \bar{\rho}_m (1 + z_{\mathrm{form}})^3}}
\ee
is the dynamic time-scale of a halo formed at $t_{\mathrm{form}}$ (corresponding to a redshift $z_{\mathrm{form}}$) and $\kappa(M, t_{\mathrm{form}})$ is a fudge parameter which governs the duration of the star formation. The form in \eqn{eq:Lambda_tdyn} is motivated by the understanding that the star formation in the initial stages increases with time as more and more gas cools inside the halo. This is followed by an exponential decrease because a substantial portion of the gas gets locked up in the stars and there is a scarcity of gas which can form further stars. For most calculations, one makes the simplifying assumption that $\kappa$ is independent of $M$ and $t_{\mathrm{form}}$, in that case $\Lambda_*(t; M, t_{\mathrm{form}}) = \Lambda_*(t)$.

\subsection{Radiation from stars}

The next step in modelling reionization is compute the radiation from stars inside the haloes which will eventually ionize the hydrogen. Given the star formation rate, the stellar IMF and the metallicity, it is possible to calculate the radiation from the galaxy using the \emph{population synthesis models}. Some of the popular codes, which are also publicly available are \texttt{STARBURST99}\footnote{\url{https://www.stsci.edu/science/starburst99/docs/default.htm}} \cite{1999ApJS..123....3L}, \texttt{GALAXEV}\footnote{\url{http://www.bruzual.org/bc03}} \cite{2003MNRAS.344.1000B} and \texttt{BPASS}\footnote{\url{https://bpass.auckland.ac.nz/14.html}} \cite{2017PASA...34...58E,2022MNRAS.512.5329B}. As far as reionization is concerned, the most relevant lessons from these detailed calculations are
\begin{itemize}
\item The main contributors of ionizing photons are the massive stars (because only they have high enough surface temperatures). These are known as O stars and have mass $\sim 15-90 \Msun$.
\item These stars are short-lived. Hence most of the ionizing photons are produced in $\sim 10^6$ year after the burst of star formation.
\item Also, the number of ionizing photons is more (for a given amount of stars) when the metallicity is low. This is because the stars have higher surface temperatures in the absence of cooling.
\end{itemize}

One of the most useful outputs from these codes is the luminosity $l_{\nu}(t; M, t_{\mathrm{form}})$ (the energy emitted per unit time per unit frequency range) per unit mass of stars formed for a burst of star formation at $t = 0$ inside a halo of mass $M$ and formation time $t_{\mathrm{form}}$. It is sometimes called the \emph{specific luminosity}. The total luminosity of the halo of mass $M$ is obtained by convolving the specific luminosity with the star formation rate and integrating over the lifetime of the galaxy
\bear
L_{\nu}(M, t, t_{\mathrm{form}}) &= \int_{t_{\mathrm{form}}}^{\infty} \de t'~\dot{M}_*(M, t', t_{\mathrm{form}})~l_{\nu}(t-t'; M, t_{\mathrm{form}})
\nline
&= f_*(M, t_{\mathrm{form}}) \left(\f{\Omega_b}{\Omega_m}\right) M \int_{t_{\mathrm{form}}}^{\infty} \de t'~\Lambda_*(t'-t_{\mathrm{form}})~l_{\nu}(t-t'; M, t_{\mathrm{form}}).
\label{eq:L_nu}
\ear

Our main interest, however, is the ionizing radiation from the galaxies. This is obtained by dividing the luminosity by the photon energy $h \nu$ and integrating it above $\nu = \nu_H$, the frequency corresponding to ionization potential of hydrogen. The number of ionizing photons produced per unit time in a halo is
\bear
\dot{N}_{\gamma}(M, t, t_{\mathrm{form}}) &= \int_{\nu_H}^{\infty} \de\nu~\f{L_{\nu}(M, t, t_{\mathrm{form}})}{h \nu}
\nline
&= f_*(M, t_{\mathrm{form}}) \left(\f{\Omega_b}{\Omega_m}\right) M \int_{t_{\mathrm{form}}}^{\infty} \de t'~\Lambda_*(t'-t_{\mathrm{form}}) 
\nline
&\times \int_{\nu_H}^{\infty} \de\nu~\f{l_{\nu}(t-t'; M, t_{\mathrm{form}})}{h \nu}.
\ear

Recall that the ionizing radiation is produced by massive stars which are short-lived with typical ages $\sim 10^6$ years. This is much shorter than the time-scale of expansion (or age) which is $\sim 4 \times 10^8$ years at $z = 10$ and $\sim 2 \times 10^8$ years at $z = 20$. In that case, we can approximate
\be
\int_{\nu_H}^{\infty} \de\nu~\f{l_{\nu}(t; M, t_{\mathrm{form}})}{h \nu} = \eta_{\gamma}(M, t_{\mathrm{form}})~\delta_D(t),
\ee
where
\be
\eta_{\gamma}(M, t_{\mathrm{form}}) = \int_0^{\infty} \de t~\int_{\nu_H}^{\infty} \de\nu~\f{l_{\nu}(t; M, t_{\mathrm{form}})}{h \nu}
\ee
is the total number of ionizing photons produced in the halo per unit mass of stars. Note that the quantity $\eta_{\gamma}$ can be estimated from the output of the population synthesis codes.

With this approximation, we write the expression for the production rate of ionizing photons by stars inside a halo as
\be
\dot{N}_{\gamma}(M, t, t_{\mathrm{form}}) = f_*(M, t_{\mathrm{form}})~\eta_{\gamma} (M, t_{\mathrm{form}}) \left(\f{\Omega_b}{\Omega_m}\right) M~\Lambda_*(t-t_{\mathrm{form}}).
\label{eq:Ndot}
\ee

\subsection{Escape of photons}

We now turn to another very important aspect of reionization by photons produced by stars. Until now, we have derived expressions which correspond to photons produced inside the galaxy. The point to note is that not all of these photons will escape into the surrounding IGM. A fraction will absorbed in the interstellar medium of the galaxy itself, most likely by dense neutral hydrogen clouds. Modelling this effect is daunting task, and there are almost no observational constraints at high redshifts.

So one usually quantifies this uncertainty through a parameter $f_{\mathrm{esc}}(M, t_{\mathrm{form}})$, defined as the mean fraction of photons that escape into the IGM from haloes of mass $M$ formed at $t_{\mathrm{form}}$. Clearly, it is only these photons that are important for studying reionization. In the extreme case where $f_{\mathrm{esc}} = 0$, there is no possibility of reionizing the universe through stars and one must investigate other possibilities.

Hence, we generalize \eqn{eq:Ndot} and write the number of photons produced by the galaxies and escaping to the IGM per unit time as
\be
\dot{N}_{\gamma}(M, t, t_{\mathrm{form}}) = f_{\mathrm{esc}}(M, t_{\mathrm{form}})~f_*(M, t_{\mathrm{form}})~\eta_{\gamma} (M, t_{\mathrm{form}}) \left(\f{\Omega_b}{\Omega_m}\right) M~\Lambda_*(t-t_{\mathrm{form}}).
\ee
This is an important relation which lets us calculate the rate of ionizing photons for a halo of mass $M$. The expression simplifies considerably if we assume that the duration of the star formation is significantly smaller than the age of the Universe. Typically the dynamical time-scale is a fraction of the age at high redshifts, so this may not be bad approximation if the star formation occurs at the dynamical time or shorter. In that case, we put $\Lambda_*(t-t_{\mathrm{form}}) \approx \delta_D(t-t_{\mathrm{form}})$ and obtain
\be
\dot{N}_{\gamma}(M, t, t_{\mathrm{form}}) = f_{\mathrm{esc}}(M, t_{\mathrm{form}})~f_*(M, t_{\mathrm{form}})~\eta_{\gamma} (M, t_{\mathrm{form}}) \left(\f{\Omega_b}{\Omega_m}\right) M~\delta_D(t-t_{\mathrm{form}}).
\label{eq:Ndot_general}
\ee
A related quantity is the total number of ionizing photons in the IGM from these haloes of mass $M$ at any time $t \geq t_{\mathrm{form}}$, which is obtained by integrating the rate in \eqn{eq:Ndot_general}
\be
N_{\gamma}(M, t) = f_{\mathrm{esc}}(M, t)~f_*(M, t)~\eta_{\gamma} (M, t) \left(\f{\Omega_b}{\Omega_m}\right) M.
\ee

Interestingly, the unknown quantities $f_{\mathrm{esc}}$, $f_*$ and $\eta_{\gamma}$ appear as a combination of multiplicative factors, so we do not need to know them individually for modelling reionization. The only combination which becomes relevant for reionization studies is the 
dimensionless parameter 
\be
\zeta(M, t) = \f{f_{\mathrm{esc}}(M, t) ~f_*(M, t) ~\eta_{\gamma}(M, t)~m_p}{1 - Y},
\ee
where $Y$ is helium weight fraction (usually taken to be $0.24$). The expression for $N_{\gamma}$ then simplifies to
\be
N_{\gamma}(M, t) = \zeta(M, t)~(1 - Y) \left(\f{\Omega_b}{\Omega_m}\right) \f{M}{m_p}.
\ee
The parameter $\zeta$ is called the \emph{reionization efficiency parameter}.

To understand the physical significance of $\zeta$, let us assume that the baryonic matter in the halo scales as the global fraction. In that case the mass of hydrogen inside the halo is simply $M_H = (\Omega_H / \Omega_m) M = (1 - Y) (\Omega_b / \Omega_m) M$ and hence
\be
N_{\gamma}(M, t) = \zeta(M, t) \f{M_H(M)}{m_p} = \zeta(M, t) n_H(M),
\ee
where $n_H(M) \equiv M_H(M) / m_p$ is the number density of hydrogen in the halo. Thus $\zeta(M)$ can be interpreted as the number of ionizing photons produced by the stars and escaping into the IGM per hydrogen inside the halo. 

To get a feel for what kind of values the parameter $\zeta$ can take, let us make some crude order of magnitude estimates. If we assume that the galaxy consists of stars having a Salpeter IMF, as seen in galaxies at low redshifts, the corresponding value of $\eta_{\gamma}$ turns out to be $\eta_{\gamma}~m_p \sim 7000$. Assuming typical values $f_* \sim 0.1$ and $f_{\mathrm{esc}} = 0.01$, we get $\zeta \sim 10$. On the other hand, for the metal-free PopIII stars with top-heavy IMF, one finds $\eta_{\gamma}~m_p \sim 10^5$, much higher than the normal stars. Let us assume $f_* \sim 0.01$ (very inefficient star formation because of inefficient cooling) and $f_{\mathrm{esc}} = 0.1$ (large number of ionizing photons produced leading to efficient escape), which gives $\zeta \sim 100$. The values of different quantities used for these examples are indicative only, in fact, most of these values are highly uncertain for high-redshift galaxies. The quantity $\zeta$ is usually treated as a free parameter in reionization studies and constrained through related observations, or sometimes fine-tuned to obtain a pre-decided reionization history.

The mean number of ionizing photons escaping to the IGM per unit comoving volume would then be
\bear
\bar{n}_{\gamma}(t) &= \int_{M_{\mathrm{min}}}^{\infty} \de M~N_{\gamma}(M, t)~n(M, t)
\nline
&=  \f{1 - Y}{m_p} \left(\f{\Omega_b}{\Omega_m}\right) \int_{M_{\mathrm{min}}}^{\infty} \de M~\zeta(M, t)~M~n(M, t),
\ear
where the integral is over the haloes heavier than the threshold mass that can form stars. Suppose, for the moment, we assume $\zeta$ to be independent of $M$, then
\bear
\bar{n}_{\gamma}(t) &= \f{1 - Y}{m_p} \left(\f{\Omega_b}{\Omega_m}\right) \zeta(t) \int_{M_{\mathrm{min}}}^{\infty} \de M~M~n(M, t)
\nline
&= \f{1 - Y}{m_p} \left(\f{\Omega_b}{\Omega_m}\right) \zeta(t) ~\bar{\rho}_m~f_{\mathrm{coll}}(M_{\mathrm{min}}, t)
\nline
&= \bar{n}_H~\zeta(t)~f_{\mathrm{coll}}(M_{\mathrm{min}}, t),
\ear
where the mean comoving number density of hydrogen is
\be
\bar{n}_H = \f{\bar{\rho}_H}{m_p} = \f{(1 - Y) \Omega_b}{\Omega_m} \f{\bar{\rho}_m}{m_p}. 
\ee
This shows that, under the approximations made during the discussions, ratio of the number of ionizing photons to the number of hydrogen is simply $\zeta~f_{\mathrm{coll}}$. Even when $\zeta$ has a mass dependence, we can define quantity
\be
\la \zeta~f_{\mathrm{coll}} (M_{\mathrm{min}}, t) \ra \equiv \f{1}{\bar{\rho}_m} \int_{M_{\mathrm{min}}}^{\infty} \de M~\zeta(M, t)~M~\f{\de n(M, t)}{\de M},
\ee
and write the photon number density as
\be
\bar{n}_{\gamma}(t) = \bar{n}_H~\la \zeta~f_{\mathrm{coll}}(M_{\mathrm{min}}, t) \ra
\ee
The global mean emissivity of ionizing photons, namely, the number of photons available for hydrogen reionization per unit time per unit comoving volume is the time derivative of the above quantity, i.e.,
\be
\dot{\bar{n}}_{\gamma}(t) = \bar{n}_H~\f{\de \la \zeta~f_{\mathrm{coll}}(M_{\mathrm{min}}, t) \ra}{\de t}.
\label{eq:dotn_gamma}
\ee
This emissivity is perhaps the most important quantity for any theoretical model of reionization.

We should mention here even in the detailed simulations of reionization, the galaxy formation model is assumed to be somewhat similar to what has been discussed here. Usually, one assigns the number of photons from haloes using some form of $\zeta$ and uses it to compute the reionization history. One can even tune the evolution of $\zeta$ so as to obtain the pre-decided reionization history.

\subsection{Feedback processes}
\label{sec:feedback}

Before moving on to discussing the radiative transfer in the IGM, let us discuss one important process which can affect the production of ionizing photons in the galaxies, namely, \emph{feedback}. There are different kind of feedback processes which can lead to quenching of star formation. These are
\begin{itemize}

\item \emph{Chemical feedback:} this refers to the change in metallicity of the star-forming gas. The first metal-free PopIII stars are efficient producers of ionizing radiation. However, the stars, if in the appropriate mass range, can explode into supernovae at the end of their life and pollute the medium with metals formed in their core. This will raise the metallicity of the gas which can later the nature of stars formed and thus suppress the production of ionizing photons. In semi-analytical models, the chemical feedback can be modelled using merger trees, or through simple prescriptions for transiting from PopIII stars to PopII stars.

\item \emph{Mechanical feedback:} this arises from injection of energy into the interstellar medium, e.g., through exploding supernova. This can heat up the gas thus suppressing star formation, or if the halo is small, it can eject the gas from the potential well of the halo. Modelling this effect is not straightforward. There is also controversies in the literature regarding their impact on reionization. The simplest way to include them in analytical models is to set the $M_{\mathrm{min}}$ appropriately.

\item \emph{Radiative feedback}: this occurs when the gas is heated because of reionization. This additional heating prevents gas from cooling in lighter haloes and is usually accounted for by increasing $M_{\mathrm{min}}$ in the ionized regions. The typical value, which corresponds to a heated IGM temperature $\sim 10^4$ K, is $\sim 10^9 \Msun$. Another situation where radiative feedback can play important role is in suppressing molecular cooling in minihaloes through the Lyman-Werner photons.

\end{itemize}

The above discussion on the feedback mechanisms is intentionally kept brief because of the limited scope of the article, and we refer the reader to several excellent reviews which discuss these effects in great detail \cite{2005SSRv..116..625C,2018PhR...780....1D}.

\section{Radiative transfer}
\label{sec:RT}

Once the photons escape out into the IGM, they propagate and possibly change the ionization and thermal state of the IGM. We are, if course, mostly interested in the ionization of hydrogen, however, often that is closely related to the how the thermal state of the medium evolve.

\subsection{Radiative transfer equation}

The starting point of the discussion is the radiative transfer equation which describes the propagation of photons through a medium. The equation in an expanding universe is given by 
\be
\f{\del I_{\nu}}{\del t} + \f{c}{a} \mathbf{\hat{n}} \cdot \vect{\nabla} I_{\nu} -\f{\dot{a}}{a} \nu \f{\del I_{\nu}}{\del \nu} + 3 \f{\dot{a}}{a} I_{\nu} = -c \kappa_{\nu} I_{\nu} + c j_{\nu},
\label{eq:RT_eq}
\ee
where $I_{\nu}(\vect{x}, \mathbf{\hat{n}}, t)$ is the \emph{specific intensity}, $\kappa_{\nu}$ is the absorption coefficient, $j_{\nu}$ is the emission coefficient, $\mathbf{\hat{n}}$ is the direction unit vector and the position $\vect{x}$ is measured in comoving coordinates. The equation looks almost identical to the usual radiative transfer equation except for the third and fourth terms on the left hand side. The derivation of this equation from its usual form is straightforward, one simply needs to transform from the proper coordinates to the comoving coordinates. The physical interpretation of the two new terms on the left hand side is also obvious: the third term represents the redshift of radiation as the universe expands and the fourth term corresponds to the dilution of the specific intensity with Hubble expansion.

If the emission is isotropic, or if the distribution of emitters is randomly oriented, then we can define a new quantity called the \emph{emissivity}
\be
\epsilon_{\nu} = 4 \pi j_{\nu},
\ee
and write the equation as
\be
\f{\del I_{\nu}}{\del t} + \f{c}{a} \mathbf{\hat{n}} \cdot \vect{\nabla} I_{\nu} -\f{\dot{a}}{a} \nu \f{\del I_{\nu}}{\del \nu} + 3 \f{\dot{a}}{a} I_{\nu} = -c \kappa_{\nu} I_{\nu} + \f{c}{4 \pi} \epsilon_{\nu}.
\ee
The equation can be solved using many different techniques depending on the context one is using it for. Our aim would be write an equation which captures the average properties of the medium in a cosmologically representative volume. In that case the starting point would be to average over all angles and obtain the zeroth moment
\be
\f{1}{c} \f{\del J_{\nu}}{\del t} + \vect{\nabla} \cdot \vect{F}_{\nu} -\f{\dot{a}}{a} \nu \f{\del J_{\nu}}{\del \nu} + 3 \f{\dot{a}}{a} J_{\nu}  =- \kappa_{\nu} J_{\nu} + \f{\epsilon_{\nu}}{4 \pi},
\ee
where
\be
J_{\nu}(\vect{x}, t) = \f{1}{4 \pi} \int \de \Omega~I_{\nu}(\vect{x}, \mathbf{\hat{n}}, t),~~
\vect{F}_{\nu}(\vect{x}, t) = \f{1}{4 \pi} \int \de \Omega~\mathbf{\hat{n}}~I_{\nu}(\vect{x}, \mathbf{\hat{n}}, t).
\ee
The quantity $J_{\nu}$ is the \emph{mean intensity} and $\vect{F}_{\nu}$ is the \emph{energy flux}.

At this point, let us rescale the intensity, emissivity and flux to their comoving counterparts by accounting for the volume dilution factor 
\be
\tilde{J}_{\nu} \equiv J_{\nu} a^3,
\tilde{\vect{F}}_{\nu} \equiv \vect{F}_{\nu} a^3,
\tilde{\epsilon}_{\nu} = \epsilon_{\nu} a^3.
\ee
Note that the quantities $J_{\nu}, \vect{F}_{\nu}, \epsilon_{\nu}$ are defined in terms of the physical coordinates while we use different symbols $\tilde{J}_{\nu}, \tilde{\vect{F}}_{\nu}, \tilde{\epsilon}_{\nu}$ for denoting the corresponding quantities in the comoving coordinates. However, for the number densities like $n_H, n_{\gamma}$, we \emph{always} define with respect to the comoving coordinates and hence we do not use any different symbols to distinguish quantities in physical and comoving coordinates.

In terms of the comoving quantities, the Hubble expansion term drops out and we have
\be
\f{1}{c} \f{\del \tilde{J}_{\nu}}{\del t} + \vect{\nabla} \cdot \tilde{\vect{F}}_{\nu} -\f{\dot{a}}{a} \nu \f{\del \tilde{J}_{\nu}}{\del \nu}  =- \kappa_{\nu} \tilde{J}_{\nu} + \f{\tilde{\epsilon}_{\nu}}{4 \pi}.
\label{eq:dJ_nu_dt}
\ee
Now let us consider a large representative comoving volume $V$ and let us assume that we will be interested only in quantities which are averaged over $V$. To take an example, suppose we average the mean intensity over a large cosmologically representative volume, we get
\be
\bar{\tilde{J}}_{\nu}(t) \equiv \int_V \f{\de^3 x}{V} \tilde{J}_{\nu}(t,\vect{x})
= a^3 \int_V \f{\de^3 x}{V} \int \f{\de \Omega}{4 \pi}  I_{\nu}(t,\vect{x},\hat{n}).
\ee
If we take this volume average of all the terms in \eqn{eq:dJ_nu_dt}, we find that the second term on the left hand side (the one containing a divergence of the flux) can be converted to a surface integral. This integral vanishes because the net flux flowing across the boundaries of a cosmologically representative volume will be zero (in numerical simulations, this is ensured by employing periodic boundary conditions). The angle and volume averaged radiative transfer equation is thus
\be
\f{\del \bar{\tilde{J}}_{\nu}}{\del t} = \f{\dot{a}}{a} \nu \f{\del \bar{\tilde{J}}_{\nu}}{\del \nu} - c \int_V \f{\de^3 x}{V} \kappa_{\nu}(\vect{x}) \tilde{J}_{\nu}(\vect{x})  + \f{c}{4 \pi} \bar{\tilde{\epsilon}}_{\nu},
\label{eq:delJ_nu_del_t}
\ee
where we have omitted the explicit time-dependence of all the quantities for simplicity. Because we are interested in the number of ionizing photons in the medium, we divide the terms by the photon energy $h \nu$ and integrate the equation over all frequencies $\nu \geq \nu_H \approx 3.29 \times 10^{15}$~Hz. This leads to an equation in terms of the photon number densities
\be
\dot{\bar{n}}_J = - \f{4 \pi}{c} \f{\dot{a}}{a} \f{\bar{\tilde{J}}_{\nu_H}}{h} - 4 \pi \int_V \f{\de^3 x}{V} \int_{\nu_H}^\infty \f{\de \nu}{h \nu} \kappa_{\nu}(\vect{x}) \tilde{J}_{\nu}(\vect{x}) + \dot{\bar{n}}_{\gamma},
\label{eq:dotn_J}
\ee
where
\be
\bar{n}_J = \f{4 \pi}{c} \int_{\nu_H}^\infty \de \nu \f{\bar{\tilde{J}}_{\nu}}{h \nu}
\ee
is the number of ionizing photons per unit comoving volume present in the IGM and
\be
\dot{\bar{n}}_{\gamma} = \int_{\nu_H}^{\infty} \de \nu~\f{\bar{\tilde{\epsilon}}_{\nu}}{h \nu}
\label{eq:ndot_epsilon}
\ee
is the ionizing emissivity defined as the number of photons produced by the sources and escaping into the IGM per unit comoving volume per unit time. One should realize that this is identical to the quantity used in \secn{sec:sources}, e.g., \eqn{eq:dotn_gamma}. Also note that the quantity $\bar{n}_{\gamma}$ corresponds to the ionizing photons produced by and escaping out of the galaxy-hosting haloes, while $\bar{n}_J$ corresponds to the ones existing in the IGM as an ionizing background. As can be seen from \eqn{eq:dotn_J}, not all the photons produced by the source contribute to the background. There is a decrease because of photons being redshifted below the ionization potential (the first term on the right hand side)  and also because of absorption (the second term on the right hand side). In our case, the more interesting term will be the absorption of ionizing photons as that is caused by the HI in the IGM. Let us try to work out its form in the epoch of reionization.

To make further progress, we need to understand some basic features of the reionization era. It is clear that the photons from the galaxies would start ionizing the surrounding medium and hence it will create what we can call a ionized ``bubble''. These bubbles would grow, merge and overlap and finally complete the reionization of the universe. So, even without knowing too many details of the process, we can assume that the universe during the reionization epoch would consist of ionized regions (which may contain one or multiple sources) and neutral regions (points where the ionizing radiation has not reached at that point, possibly because they are far away from the sources).

\begin{figure}
\begin{center}
\includegraphics[width=0.7\textwidth]{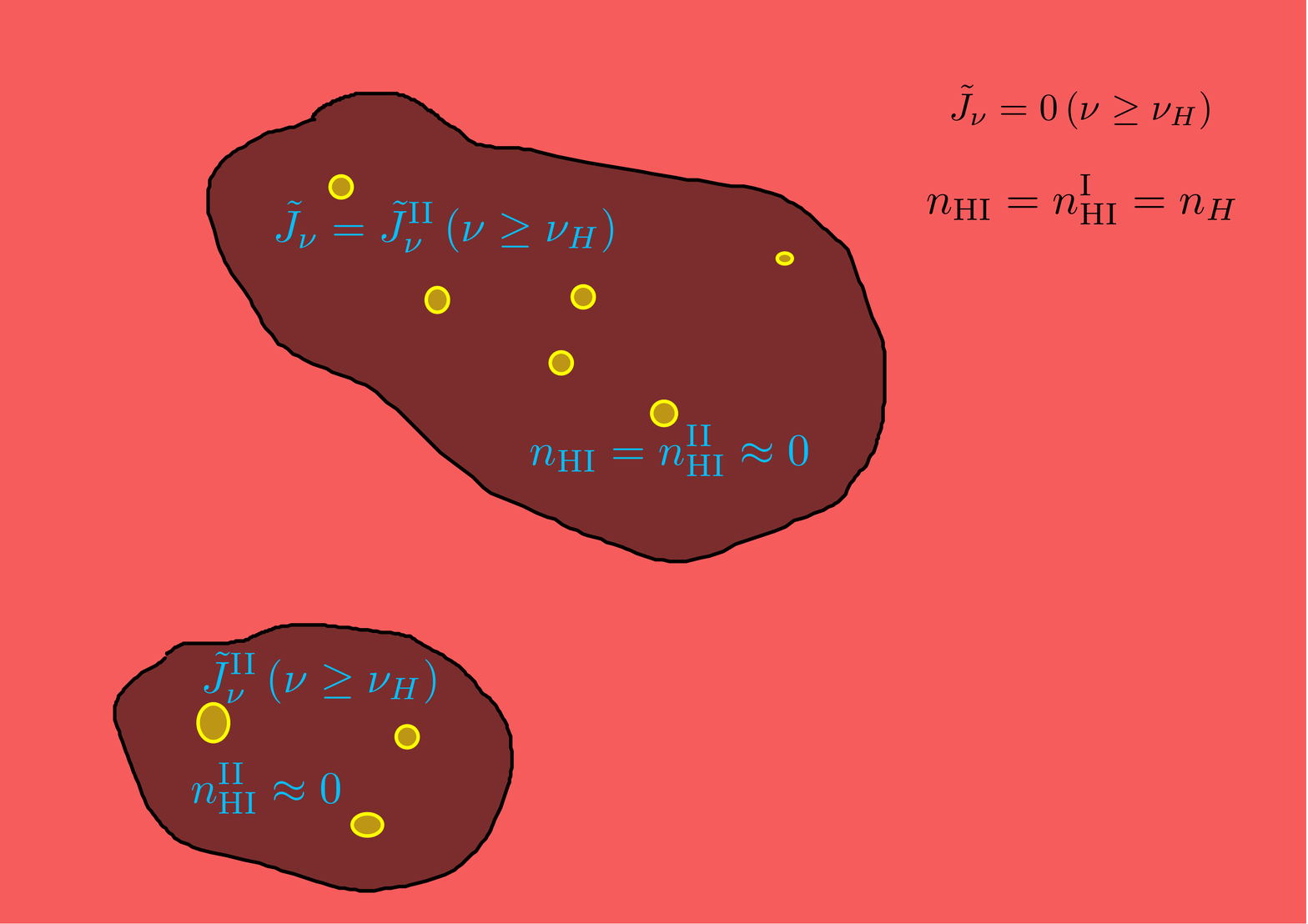}
\end{center}
\caption{A schematic diagram showing the two-phase IGM during the epoch of reionization. The light red regions represent points where hydrogen is neutral $n_{\mathrm{HI}} = n_H$ and the ionizing radiation is non-existent $\tilde{J}_{\nu} = 0$ for $\nu \geq \nu_H$. The dark regions are the ionized HII bubbles with $n_{\mathrm{HI}} \approx 0$. The yellow regions represent the sources (galaxies) of ionizing radiation, which exist only inside the HII bubbles. Any given HII region can contain multiple sources.}
\label{fig:fig_HI}
\end{figure}

Can we also have regions which are partially ionized, i.e., they are neither highly neutral nor highly ionized? To answer this, one needs to compute the mean free path of ionizing photons in a neutral medium. This mean free path is $\sim (\sigma_H(\nu) n_{H} a^3)^{-1}$, where 
\be
\sigma_H(\nu) \approx \sigma_H(\nu_H) \left(\f{\nu}{\nu_H}\right)^{-3} = 6.3 \times 10^{-18}~\text{cm}^2 \left(\f{\nu}{\nu_H}\right)^{-3}
\label{eq:sigma_H_nu}
\ee
is the absorption cross-section of ionizing photons of frequency $\nu$. A simple calculation will show that, for photons in the energy range $\sim 13.6 - 100$ eV, the mean free path in a neutral medium (having cosmological mean density) at $z \sim 5-20$ is $\sim$ kpc, much smaller than any cosmologically relevant length scale. Thus the ionizing radiation cannot penetrate the medium unless it is highly ionized. A direct consequence is that the thickness of the surface between the neutral and ionized regions is significantly smaller than the sizes of the ionized regions themselves, hence we can assume the two regions to have a sharp boundary. To a very good approximation, the IGM can be approximated as a two-phase medium during reionization if the reionization is driven by ultra-violet photons,\footnote{In the case where reionization is driven by much energetic photons, e.g., X-rays, the IGM cannot be approximated as a simple two-phase medium because these photons can have large mean free path even in a fully neutral medium. We are not exploring that possibility in this article.} one which is highly ionized, and the other is completely neutral. We present a schematic diagram in \fig{fig:fig_HI} to help visualize the two-phase structure of the hydrogen in the IGM.

Given the two-phase nature of the IGM, the absorption term in \eqn{eq:dotn_J} can pick up contribution from three regions:
\begin{itemize}
\item A volume $V_{\mathrm{HII}} = Q_{\mathrm{HII}} V$ that is ionized, $Q_{\mathrm{HII}}$ being the ionized fraction. In these regions, the absorption coefficient is essentially given by $\kappa_{\nu}(\vect{x}) = n_{\mathrm{HI}}(\vect{x})~a^{-3}~\sigma_H(\nu)$, where $n_{\mathrm{HI}}$ is the number density of residual neutral hydrogen atoms and the factor $a^{-3}$ arises because the number densities are in comoving units. These regions are represented by dark red colour in \fig{fig:fig_HI}.
\item The neutral volume $V_{\mathrm{HI}} = V - V_{\mathrm{HII}}$ that does not contain any ionizing radiation ($J_{\nu} = 0$ for $\nu \geq \nu_H$), hence it does \emph{not} contribute to the absorption term. The light red coloured areas in \fig{fig:fig_HI} correspond to these regions.
\item The boundary between the two regions, where the medium makes a sharp transition from ionized to neutral, can absorb a significant number of ionizing photons and lead to an increase in the ionized volume $V_{\mathrm{HII}}$. The black lines surrounding the dark red regions in \fig{fig:fig_HI} would represent these boundaries.
\end{itemize}

Hence, the absorption term in \eqn{eq:dotn_J} actually picks up contributions from only two of the three regions
\bear
- 4 \pi \int_V \f{\de^3 x}{V} \int_{\nu_H}^\infty \f{\de \nu}{h \nu} \kappa_{\nu}(\vect{x}) \tilde{J}_{\nu}(\vect{x}) 
&= - 4 \pi~a^{-3} \int_{V_{\mathrm{HII}}} \f{\de^3 x}{V} n_{\mathrm{HI}}(\vect{x}) \int_{\nu_H}^\infty \f{\de \nu}{h \nu} \sigma_H({\nu}) \tilde{J}_{\nu}(\vect{x}) 
\nline
&+ \left[\f{\de \bar{n}_J}{\de t} \right]_{\mathrm{boundary}},
\label{eq:abs_term}
\ear
where we use the convention that the boundary term (the second term on the right hand side) is negative if photons are lost due to absorption.

Now, if we assume that the individual HII regions contain a large number of sources (which is strictly not true at very early stages of reionization, but becomes a good approximation by the time the IGM a few per cent ionized), we can take $\tilde{J}_{\nu}(\vect{x})$ for $\nu \geq \nu_H$ to be homogeneous \emph{within the ionized regions}. Let us denote this value as $\bar{\tilde{J}}^{\mathrm{II}}_{\nu}$ (the superscript `II' represents HII or ionized regions). The relation between $\bar{\tilde{J}}^{\mathrm{II}}_{\nu}$ and $\bar{\tilde{J}}_{\nu}$ can be obtained by realising that the contribution to the ionizing intensity arises only from the ionized regions, so
\be
\bar{\tilde{J}}_{\nu} = \f{1}{V} \int_{V_{\mathrm{HII}}} \de^3 x~ \tilde{J}_{\nu}(\vect{x})
= Q_{\mathrm{HII}}~\bar{\tilde{J}}^{\mathrm{II}}_{\nu}.
\ee
Hence we can write the term representing absorption of ionizing photons in the HII regions due to residual HI, i.e., the first term on the right hand side of \eqn{eq:abs_term}, as
\bear
-\f{4 \pi}{a^{3}} \int_{V_{\mathrm{HII}}} \f{\de^3 x}{V} n_{\mathrm{HI}}(\vect{x}) \int_{\nu_H}^\infty \f{\de \nu}{h \nu} \sigma_H({\nu}) \tilde{J}_{\nu}(\vect{x})
&= -4 \pi~\bar{n}^{\mathrm{II}}_{\mathrm{HI}}~Q_{\mathrm{HII}} \int_{\nu_H}^\infty \f{\de \nu}{h \nu} \sigma_H({\nu})\left(\f{\bar{\tilde{J}}^{\mathrm{II}}_{\nu}}{a^3}\right)
\nline
&= -Q_{\mathrm{HII}}~\bar{n}^{\mathrm{II}}_{\mathrm{HI}} ~\Gamma_{\mathrm{HI}}^{\mathrm{II}},
\ear
where we have defined the average HI density within ionized regions as
\be
\bar{n}^{\mathrm{II}}_{\mathrm{HI}} \equiv \int_{V_{\mathrm{HII}}} \f{\de^3 x}{V_{\mathrm{HII}}} n_{\mathrm{HI}}(\vect{x})
= \f{1}{Q_{\mathrm{HII}}} \int_{V_{\mathrm{HII}}} \f{\de^3 x}{V} n_{\mathrm{HI}}(\vect{x}),
\ee
and the hydrogen photoionization rate in the ionized regions as
\be
\Gamma_{\mathrm{HI}}^{\mathrm{II}} \equiv 4 \pi \int_{\nu_H}^\infty \f{\de \nu}{h \nu} \sigma_H({\nu}) \bar{J}^{\mathrm{II}}_{\nu}
= 4 \pi \int_{\nu_H}^\infty \f{\de \nu}{h \nu} \sigma_H({\nu})\left(\f{\bar{\tilde{J}}^{\mathrm{II}}_{\nu}}{a^3}\right).
\label{eq:Gamma_HI}
\ee
Note that the photoionization rates are always defined in terms of the physical quantities, hence appropriate factors of $a^3$ should be accounted for when one writes the relations in terms of the comoving quantities like $\bar{\tilde{J}}^{\mathrm{II}}_{\nu}$.

For calculating the number of photons lost due to the boundary term, let as assume that the volume of ionized regions increase by $\Delta V_{\mathrm{HII}}$ in a time step $\Delta t$. During this time, one needs to ionize $\left(\bar{n}_{\mathrm{HI}}^{\mathrm{I}} - \bar{n}_{\mathrm{HI}}^{\mathrm{II}}\right)~\Delta V_{\mathrm{HII}}$ hydrogen atoms, the superscript `I' representing neutral or HI regions. Note that according to our assumption all the atoms in the neutral regions are neutral $\bar{n}_{\mathrm{HI}}^{\mathrm{I}} = \bar{n}_H$, hence the number of atoms that are ionized in time $\Delta t$ is given by $\left(\bar{n}_H - \bar{n}_{\mathrm{HI}}^{\mathrm{II}}\right) \Delta V_{\mathrm{HII}} = \bar{n}_{\mathrm{HII}}^{\mathrm{II}}~\Delta V_{\mathrm{HII}}$. So the number of ionizing photons lost in the time step per unit comoving volume is 
\be
\left[\Delta \bar{n}_J\right]_{\mathrm{boundary}} = - \bar{n}_{\mathrm{HII}}^{\mathrm{II}}~\Delta Q_{\mathrm{HII}}.
\ee
The corresponding rate is simply
\be
\left[\f{\de \bar{n}_J}{\de t} \right]_{\mathrm{boundary}} = - \bar{n}_{\mathrm{HII}}^{\mathrm{II}} \f{\de Q_{\mathrm{HII}}}{\de t}.
\ee

Including all the contributions to the absorption term, the radiative transfer \eqn{eq:dotn_J} reduces to
\be
\dot{\bar{n}}_J = - \f{4 \pi}{c} \f{\dot{a}}{a} \f{\bar{\tilde{J}}_{\nu_H}}{h} - Q_{\mathrm{HII}}~\bar{n}^{\mathrm{II}}_{\mathrm{HI}} ~\Gamma_{\mathrm{HI}}^{\mathrm{II}} - \bar{n}_{\mathrm{HII}}^{\mathrm{II}} \f{\de Q_{\mathrm{HII}}}{\de t} + \dot{\bar{n}}_{\gamma}.
\label{eq:dotn_J_GammaHI}
\ee
The interpretation of this equation is simple: the ionizing photons $\bar{n}_{\gamma}$ produced by the galaxies and escaping into the IGM are either lost due to the redshift, or absorbed while ionizing residual hydrogen in the HII regions or while creating new HII regions by ionizing atoms in the previously neutral regions; the ones which survive exist as the ionizing background $\bar{n}_J$. 

The above \eqn{eq:dotn_J_GammaHI} can be treated as a differential equation to solve for $Q_{\mathrm{HII}}(t)$, provided one has an independent way of estimating $\bar{\tilde{J}}_{\nu}$ and hence $\dot{\bar{n}}_J$ and $\Gamma_{\mathrm{HI}}^{\mathrm{II}}$. A possible way is to write down the integral solution of \eqn{eq:delJ_nu_del_t} along the line of sight (using, e.g., the method of characteristics)
\be
\bar{\tilde{J}}_{\nu}(t) = \f{c}{4 \pi} \int_0^t \de t'~ \bar{\tilde{\epsilon}}_{\nu a(t)/a(t')}(t')~\e^{-\tau(t,t';\nu)},
\label{eq:Jnu_t}
\ee
where the \emph{optical depth} along the line of sight is given by the integral
\be
\tau(t,t';\nu) = c \int_{t'}^t \de t'' ~ \kappa_{\nu(t'')} = c \int_{t'}^t \de t'' ~ \kappa_{\nu a(t)/a(t'')}.
\ee
The optical depth can also be written in terms of the mean free path $\lambda_{\mathrm{mfp}, \nu}(t) = \kappa_{\nu}^{-1}(t)$ as
\be
\tau(t,t';\nu) =  c \int_{t'}^t \f{\de t''} {\lambda_{\mathrm{mfp}, \nu a(t)/a(t'')}(t'')}.
\ee
The mean free path, defined as above, is in physical or proper units.

Given a model for the source emissivity $\bar{\tilde{\epsilon}}_{\nu}(t)$, the integral solution \eqn{eq:Jnu_t} is usually used for estimating the background radiation in the Universe over a wide range of redshifts and frequencies \cite{1996ApJ...461...20H,1999ApJ...514..648M,2009ApJ...703.1416F,2012ApJ...746..125H,2019MNRAS.484.4174K}. One of the many challenges in such calculations is to calculate the optical depth, particularly during the epoch of reionization where it is affected by the distribution of the ionized bubbles.

\subsection{Local absorption}

In many cases of interest, the photons produced by the sources are absorbed locally in the medium, i.e., the mean free path at any point of interest is much smaller than the horizon size
\be
\lambda_{\mathrm{mfp}, \nu}(t, \vect{x}) \ll \f{c}{H(t)}.
\ee
This is often called as the ``on-the-spot'' approximation. The assumption of local absorption is valid for ultra-violet photons (e.g., those produced by stars in galaxies) as long as the size of the ionized regions are not comparable to the horizon size. This, however, cannot be verified unless we work out the details of the reionization. Interestingly, even when the IGM is completely ionized, the mean free path does not rise to arbitrarily large values. In such situations, it is determined by the high density regions which are capable to absorbing all ionizing photons and still remain neutral. These regions are characterized by their high recombination rate and are self-shielded from ionizing flux. They show up prominently in quasar absorption spectra as Lyman-limit systems. The distance between these systems increase with time as the ionization fronts slowly propagate into high density regions. The local absorption assumption seems to valid till about $z \gtrsim 3$ below which the Lyman-limit systems become too sparse \cite{2014MNRAS.445.1745W}.

If the absorption is indeed local, one can show that the mean intensity is related to the emissivity through a simple relation
\be
\tilde{J}_{\nu}(t,\vect{x}) \approx \f{\lambda_{\mathrm{mfp}, \nu}(t,\vect{x})}{4 \pi} \tilde{\epsilon}_{\nu}(t,\vect{x}) = \f{\tilde{\epsilon}_{\nu}(t,\vect{x})}{4 \pi \kappa_{\nu}(t, \vect{x})},
\label{eq:local_abs}
\ee
where one needs makes the reasonable assumption that the emissivity does not evolve significantly over the small time interval $\lambda_{\mathrm{mfp}, \nu_H}/c$. The above relation is equivalent to saying that for optically thick systems the specific intensity is equal to the ``source function'', a standard result in the theory of radiative transfer \cite{1979rpa..book.....R}. One can now use this approximation in \eqn{eq:dJ_nu_dt} to show that $\de \tilde{J}_{\nu} / \de t \approx 0$, i.e., the intensity attains a quasi-steady state. The rest of the derivation follows exactly the way we have done so far and we end up with a rather simple equation
\be
\f{\de Q_{\mathrm{HII}}}{\de t} = \f{\dot{\bar{n}}_{\gamma}}{\bar{n}_{\mathrm{HII}}^{\mathrm{II}}} - Q_{\mathrm{HII}}~\bar{n}_{\mathrm{HI}}^{\mathrm{II}} \Gamma^{\mathrm{II}}_{\mathrm{HI}}.
\label{eq:dQdt}
\ee
Comparing this with the earlier, more exact, \eqn{eq:dotn_J_GammaHI}, we find that the terms corresponding to the radiation background $\bar{n}_J$ and the redshift are not present when the absorption is local. This is because when the absorption is efficient, all the photons are absorbed very close to the sources and hence there is no scope of either building up a background of ionizing radiation or the photons getting redshifted.

\subsection{Photoionization equilibrium}

In order to solve for the evolution of reionization $Q_{\mathrm{HII}}(t)$ using \eqn{eq:dQdt}, the first quantity we need is the source term $\dot{\bar{n}}_{\gamma}$. We have already discussed the sources at great length in \secn{sec:sources}. Most models use a form similar to \eqn{eq:dotn_gamma} where $\dot{\bar{n}}_{\gamma}$ is given by the time derivative of the collapse fraction. 

Besides the source, we also need to compute the second term on the right of \eqn{eq:dQdt} which in turn requires the knowledge of the residual neutral hydrogen density $n_{\mathrm{HI}}^{\mathrm{II}}$ in the ionized regions. To find this quantity, we write the evolution equation for the neutral hydrogen density $n_{\mathrm{HI}}$ at any location in the IGM
\be
\f{\de n_{\mathrm{HI}}(t,\vect{x})}{\de t} = - \Gamma_{\mathrm{HI}}(t,\vect{x}) n_{\mathrm{HI}}(t,\vect{x}) + a^{-3}~\alpha_R(T(t,\vect{x})) n_{\mathrm{HII}}(t,\vect{x}) n_e(t,\vect{x}),
\ee
where the first term on the right hand represents the depletion in HI atoms because of photoionization and the second term is the increase due to recombinations of electrons and HII, $n_e$ being the density of free electrons. The quantity $\alpha_R(T)$ is the recombination rate coefficient which depends on the temperature $T$ of the medium. Note that the densities above are in comoving units, hence a factor $a^3$ needs to be included in the recombination term. We have ignored ionizations through collisions because of the low densities of the cosmic hydrogen in the IGM. Clearly, solving the above equation becomes relevant only in HII regions.

It can be shown that if we are interested in time-scales $\gg \Gamma_{\mathrm{HI}}^{-1}$, the system relaxes to a quasi-equilibrium $\de n_{\mathrm{HI}} / \de t \approx 0$ \cite{2009RvMP...81.1405M}, and the solution to the system is given by
\be
\Gamma_{\mathrm{HI}}(t,\vect{x}) n_{\mathrm{HI}}(t,\vect{x}) \approx \alpha_R(T(t,\vect{x})) n_{\mathrm{HII}}(t,\vect{x}) n_e(t,\vect{x}).
\label{eq:PI_equi}
\ee
This condition is known as the \emph{photoionization equilibrium} and it allows one to find the ionization state of the IGM without solving the evolution equation. The assumption of photoionization equilibrium is known to be an extremely good approximation at lower redshifts $z \lesssim 5$ when the cosmologically relevant time-scales are larger than $\Gamma_{\mathrm{HI}}^{-1}$. At higher redshifts during reionization, the photionization rate can be small particularly near the boundaries of the ionized bubbles and the self-shielded regions, so one needs to check whether this photoionization equilibrium is valid \cite{2022arXiv220405362K}. Since we are concerned with the average properties of the ionized regions where the photoionization rate is usually high, the equilibrium condition can be taken to be valid.

Let us now average \eqn{eq:PI_equi} over the volume $V_{\mathrm{HII}}$ of ionized regions. We have already assumed the mean intensity to be homogeneous within the ionized regions, so we can write the photoionization equilibrium condition as
\be
\Gamma^{\mathrm{II}}_{\mathrm{HI}} \int_{V_{\mathrm{HII}}} \f{\de^3 x}{V_{\mathrm{HII}}} n_{\mathrm{HI}}(t, \vect{x}) = a^{-3}~\alpha_R(T) \int_{V_{\mathrm{HII}}} \f{\de^3 x}{V_{\mathrm{HII}}} n_{\mathrm{HII}}(t,\vect{x}) n_e(t,\vect{x}).
\ee
While writing the above equation, we have assumed that the medium is isothermal and hence taken the $T$-dependent term out of the integral. Even if this is not the case, one can ignore the temperature dependence of the recombination which is relatively weak $T^{-0.7}$. The resulting volume-averaged condition is
\be
\Gamma^{\mathrm{II}}_{\mathrm{HI}} ~\bar{n}^{\mathrm{II}}_{\mathrm{HI}} = a^{-3}~\alpha_R(T)~\mathcal{C}~\bar{n}^{\mathrm{II}}_{\mathrm{HII}}~\bar{n}^{\mathrm{II}}_e,
\ee
where the \emph{clumping factor} in the ionized regions is defined as
\be
\mathcal{C} \equiv \f{\int_{V_{\mathrm{HII}}} \left(\de^3 x / V_{\mathrm{HII}}\right) n_{\mathrm{HII}}(t,\vect{x}) n_e(t,\vect{x})}{\left[\int_{V_{\mathrm{HII}}} \left(\de^3 x / V_{\mathrm{HII}}\right) n_{\mathrm{HII}}(t,\vect{x})\right] \left[\int_{V_{\mathrm{HII}}} \left(\de^3 x / V_{\mathrm{HII}}\right) n_e(t,\vect{x})\right]}
= \f{\overline{n_{\mathrm{HII}}~n_e}^{\mathrm{II}}}{\bar{n}^{\mathrm{II}}_{\mathrm{HII}}~\bar{n}^{\mathrm{II}}_e}.
\ee
This quantity essentially characterizes the fluctuations in the ionized IGM and arises because of the $n^2$ dependence of the recombination term. For a uniform medium we have $\mathcal{C} = 1$. In general, to calculate the clumping factor, one needs to know the density structure of the IGM and the propagation of ionization fronts in the high-density regions. In reionization studies, it is not very straightforward to model it self-consistently.

Using photoionization equilibrium, we can write the evolution of $Q_{\mathrm{HII}}$ as
\be
\f{\de Q_{\mathrm{HII}}}{\de t} = \f{\dot{n}_{\gamma}}{n_{\mathrm{HII}}^{\mathrm{II}}} - Q_{\mathrm{HII}} a^{-3}~ {\cal C} \alpha_R(T) n_e^{\mathrm{II}}.
\label{eq:dQdt_main}
\ee
In the ionized regions, we have $n_{\mathrm{HII}}^{\mathrm{II}} \approx \bar{n}_H$ and $n_e^{\mathrm{II}} = \chi_{\mathrm{He}}~n_{\mathrm{HII}}^{\mathrm{II}} \approx \chi_{\mathrm{He}}~\bar{n}_H$, where $\chi_{\mathrm{He}}$ accounts for additional free electrons produced by ionized helium. Using the value of $Y = 0.24$, it is possible to show that $\chi_{\mathrm{He}} \approx 1.08$ when helium is singly ionized and $\chi_{\mathrm{He}} \approx 1.16$ when doubly ionized. Further, using \eqn{eq:dotn_gamma}, which relates the photon production rate to the dark matter collapse fraction, we can write
\be
\f{\de Q_{\mathrm{HII}}}{\de t} = \f{\de \langle\zeta f_{\mathrm{coll}}\rangle}{\de t} - Q_{\mathrm{HII}} ~\chi_{\mathrm{He}}~a^{-3} \mathcal{C} \alpha_R(T) n_H.
\ee
This equation can be solved given the two unknown parameters $\zeta$ and $\mathcal{C}$. 

\begin{figure}
\begin{center}
\includegraphics[width=0.6\textwidth]{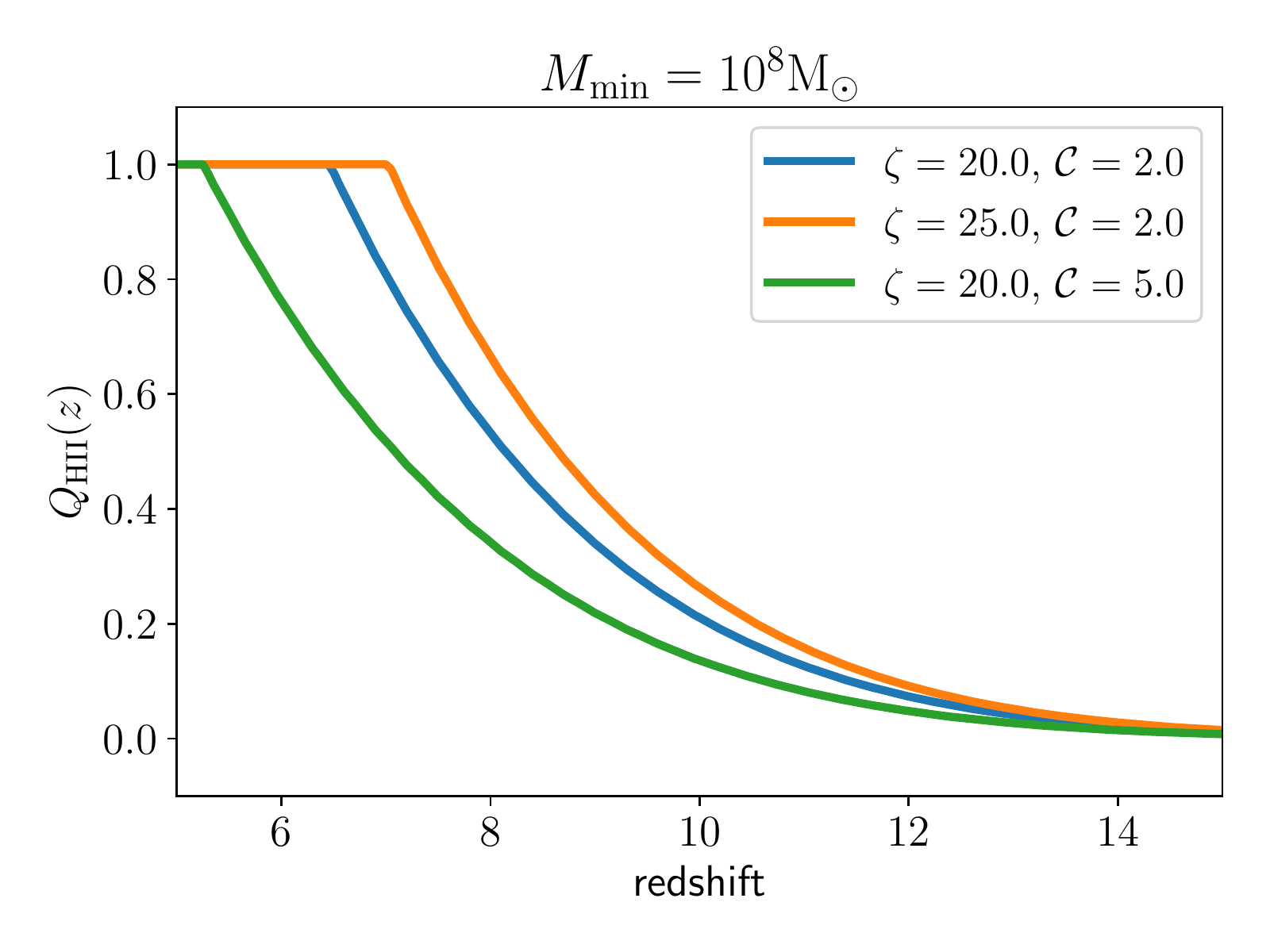}
\end{center}
\caption{Sample reionization histories for some representative values of the reionization efficiency $\zeta$ and clumping factor $\mathcal{C}$. Both have been taken to be constants. The minimum mass of haloes contributing to the ionizing photon budget is taken to be $M_{\mathrm{min}} = 10^8 \Msun$, appropriate for haloes which can cool via atomic transitions.}
\label{fig:Q_sample}
\end{figure}

To see some example reionization histories for different values of these two parameters, let us assume them to be constants (i.e., assuming no redshift evolution, and also assuming $\zeta$ to have no mass-dependence). The resulting $Q_{\mathrm{HII}}(z)$ are shown in \fig{fig:Q_sample}. As one can see from the figure, increasing the value of $\zeta$ (keeping $\mathcal{C}$ unchanged) leads to reionization completing early. This is simply because one has sources which are efficient producers of ionizing photons when $\zeta$ is large. On the other hand, increasing the value of $\mathcal{C}$ leads to slowing down of reionization. This is because the recombination rate increases with $\mathcal{C}$ and hence more ionizing photons are lost in ionizing the recombined atoms.

The evolution \eqn{eq:dQdt_main}, which forms the basis of many reionization studies, is often derived using a simplified approach starting from the growth of the cosmological ionized region around a source \cite{1986PASP...98.1014S,1987ApJ...321L.107S}. The derivation presented here is more general which does not make any specific assumption about the source distribution or overlap of the bubbles. It starts from the most basic radiative transfer equation and makes a series of approximations which are reasonable at high redshifts. It should be thus clear that \eqn{eq:dQdt_main} is nothing but the cosmological radiative transfer equation averaged over large volumes under certain well-motivated approximations.

\section{Observables related to the evolution of mean ionized fraction}
\label{sec:observables}

Now that we have some idea about how to construct an analytical model for reionization and compute globally averaged quantities like the ionized fraction $Q_{\mathrm{HII}}$, let us discuss how the outputs of these calculations can be compared with observations.

\subsection{CMB optical depth}

As the CMB photons scatter off free electrons produced during reionization, a secondary signal is produced in the CMB angular power spectrum $C_{\ell}$, not only in the temperature fluctuations, but also in the polarization. The main observational quantity one can compute is the CMB optical depth of Thomson scattering
\be
\tau_{\mathrm{el}} = c~\sigma_T~\int_{t_0}^{t_{\mathrm{LSS}}} \de t~\bar{n}_e(t)~a^{-3}(t),
\label{eq:tau_el}
\ee
where $\bar{n}_e(t)$ is the free electron number density averaged over all volume, $t_0$ is the present epoch and $t_{\mathrm{LSS}}$ is the cosmic time at the last scattering surface (i.e., during the origin of the CMB). The Thomson scattering cross section is denoted by $\sigma_T$. 

The main effects of reionization on the CMB angular power spectra are
\begin{itemize}

\item The temperature and polarization fluctuations are damped by a factor $\e^{-\tau_{\mathrm{el}}}$, hence the corresponding power spectra are damped by  $\e^{-2 \tau_{\mathrm{el}}}$. This damping, however, is completely degenerate with the amplitude $A_s$ of the primordial power spectrum of density (or scalar) perturbations and hence cannot be determined from the observations.

\item A rather direct way of measuring $\tau_{\mathrm{el}}$ is through the large-scale polarization produced by the Thomson scattering. Since the Thomson scattering is angle-dependent, it can generate polarization in the CMB photons during reionization (provided the local CMB fluctuations contain a quadropole moment to begin with). This polarization signal is completely distinct from the primary polarization anisotropies because it peaks at relatively large angular scales (i.e., low-$\ell$). The signal has been detected by successive satellites, and is constrained to a value $\tau_{\mathrm{el}} = 0.054 \pm 0.007$ according to the most recent measurements of Planck \cite{2020A&A...641A...6P}. This, in turn, indicates that the ``average'' redshift of reionization is around $z \sim 7-8$.

\end{itemize}

The CMB optical depth is straightforward to calculate in any reionization model. Since $\bar{n}_e = \chi_{\mathrm{He}} Q_{\mathrm{HII}}~\bar{n}_H$, we get
\be
\tau_{\mathrm{el}} = c~\sigma_T~\bar{n}_H \int_{t_0}^{t_{\mathrm{LSS}}} \de t~\chi_{\mathrm{He}}(t)~Q_{\mathrm{HII}}(t)~a^{-3}.
\ee
Although the integral is simple to evaluate, there are a couple of points to keep in mind. The first is that the integral is up to the present epoch, so one needs to put $Q_{\mathrm{HII}} = 1$ once the reionization is complete and continue the integral until $z = 0$. The other point is that the value of $\chi_{\mathrm{He}}$ will evolve because of reionization of singly ionized helium at $z \sim 3.5$. A good approximation is to use $\chi_{\mathrm{He}} = 1.08$ for $z > 3$ and $\chi_{\mathrm{He}} = 1.16$ for $z < 3$.

For a more detailed comparison with the CMB data, one can also use the publicly available codes for computing the CMB anisotropies, e.g., \texttt{CAMB}\footnote{\url{https://camb.info/}} and \texttt{CLASS}\footnote{\url{https://lesgourg.github.io/class_public/class.html}}, and obtain $C_{\ell}$ for a given reionization history. The only non-trivial aspect for such calculations is that the user needs to modify the relevant portions of these codes to include generic reionization histories while computing the anisotropies.

\subsection{Lyman-$\alpha$ absorption spectra of quasars}
\label{sec:lya_absorption}

Quasars are active galactic nuclei powered by gas accretion into a supermassive black hole at the centre. Because of their high luminosities, they can be observed at relatively higher redshifts. The absorption signatures produced on a quasar spectrum probe the matter distribution along the line of sight, particularly that of neutral hydrogen. With  the availability of extremely high-quality spectra of the quasars through various instruments coupled to the Very Large Telescope (VLT) and the Keck Observatory, it is now possible to probe the high-redshift universe up to $z \sim 6$ quite efficiently.

If there is no emission along the line of sight, and the specific intensity of radiation reaching us at time $t_0$ would be given by the line of sight solution of the radiative transfer \eqn{eq:RT_eq}
\be
I_{\nu(t_0)} = I_{\nu(t_i)}~a^3(t_i)~\e^{-\tau(t_0,t_i;\nu)},
\ee
where $t_i$ is the time when the light was emitted from the quasar and $I_{\nu(t_i)}$ is the corresponding emitted intensity. The factor $a^3(t_i)$ accounts for the volume dilution between the cosmic epochs $t_i$ and $t_0$. The relation between the observed frequency $\nu(t_0)$ and the emitted one $\nu(t_i)$ is simply $\nu(t_0) = \nu(t_i)~a(t_i)$.

In the intervening IGM, the electrons in the hydrogen atoms are mostly in the ground state and they can absorb the radiation of appropriate frequency to transit to the first excited state, the so-called Lyman-$\alpha$ (Ly$\alpha$) transition. Let us compute the optical depth corresponding to this transition, given by
\be
\tau_{\alpha}(t_0,t_i;\nu) = c \int_{t_i}^{t_0} \de t ~ n_{\mathrm{HI}}(t) ~ a^{-3}(t)~ \sigma_{\alpha}[\nu / a(t)],
\label{eq:tau_alpha}
\ee
where $\sigma_{\alpha}(\nu)$ is the Ly$\alpha$ absorption cross section. The integral is along a line of sight, hence $n_{\mathrm{HI}}(t)$ denotes the HI density at time $t$ at a point which is along the line of sight under consideration.

The Ly$\alpha$ is a line transition, however, the cross section has a width arising from the thermal motions and natural broadening (called the \emph{Voigt profile}). If we are interested in globally averaged quantities, we can ignore this width profile and approximate the cross-section as having a sharp profile
\be
\sigma_{\alpha}(\nu) = \sigma_{\alpha}(\nu_{\alpha}) ~ \delta_D\left(\f{\nu}{\nu_{\alpha}} - 1\right),
\ee
where $\nu_{\alpha} \approx 2.47 \times 10^{15}$~Hz is the frequency of Ly$\alpha$ transition. The main observable in the quasar spectra is the \emph{transmitted flux}
\be
F_{\nu} \equiv \f{I_{\nu}}{I^0_{\nu}} = \e^{-\tau_{\alpha}(t_0,t_i;\nu)},
\ee
where $I^0_{\nu} = I_{\nu / a(t_i)}~a^3(t_i)$ is the unabsorbed intensity or the continuum.

Converting the integration variable from $t$ to $a$ in \eqn{eq:tau_alpha} and carrying out the integral will lead to a relation
\be
\tau_{\alpha}(t_0,t_i;\nu) = \sigma_{\alpha}(\nu_{\alpha}) \f{c}{H(a=\nu / \nu_{\alpha})}~n_{\mathrm{HI}}(a=\nu/\nu_{\alpha})~\left(\f{\nu_{\alpha}}{\nu}\right)^3.
\label{eq:tau_alpha_a}
\ee
Because of the assumption of a sharp profile, we find that the optical depth observed at a frequency $\nu$ arises from neutral hydrogen atoms at a cosmic epoch $a = \nu / \nu_{\alpha}$. Hence we can interpret the optical depth purely in terms of the cosmic scale factor or redshift and write it as
\be
\tau_{\alpha}(a) = \sigma_{\alpha}(\nu_{\alpha}) \f{c}{H(a)}~n_{\mathrm{HI}}(a)~a^{-3}.
\ee

The calculation leading to \eqn{eq:tau_alpha_a} will also show that the optical depth will be non-zero only if the condition $\nu_{\alpha} a(t_i) \leq \nu \leq \nu_{\alpha}$ is satisfied. Its implication is that the Ly$\alpha$ absorption signatures will be visible blueward (i.e., at higher frequencies) of the Ly$\alpha$ emission line observed in the quasar spectra. At lower reshifts $z \lesssim 5$, these absorption features appear like a forest of lines and are called as the \emph{Lyman-$\alpha$ forest}. A simple calculation will show that $\tau_{\alpha} \sim 10^5 x_{\mathrm{HI}}$, where $x_{\mathrm{HI}}$ is the fraction of hydrogen that is neutral. The observations of Ly$\alpha$ forest imply $\tau_{\alpha} \lesssim 1$, thus implying that the universe is highly ionized $x_{\mathrm{HI}} \lesssim 10^{-5}$ at $z \lesssim 5$. This is one of the most convincing arguments for reionization. The fact that Ly$\alpha$ optical depth observations imply an ionized universe is known as the \emph{Gunn-Peterson effect} \cite{1965ApJ...142.1633G}. At higher redshifts $ \gtrsim 5$, the Ly$\alpha$ forest seems to vanish thus implying $\tau_{\alpha} \gtrsim 1$. At this point, the interpretation of the observations in terms of $x_{\mathrm{HI}}$ becomes less straightforward. More details on these aspects can be found in other review articles \cite{2009RvMP...81.1405M}.

To connect the Ly$\alpha$ optical depth to the reionization, let calculate the neutral hydrogen density in terms of the cosmological density and other physical quantities. When the IGM is highly ionized $n_{\mathrm{HI}} \ll n_H$, $n_{\mathrm{HII}} \approx n_H$, $n_e = \chi_{\mathrm{He}} n_H$, we can write a straightforward solution of the equilibrium \eqn{eq:PI_equi}
\be
n_{\mathrm{HI}} = \f{\chi_{\mathrm{He}}}{a^3} \f{\alpha_R(T)}{\Gamma_{\mathrm{HI}}} n^2_H.
\ee
The recombination rate can be well fitted by \cite{1998ARA&A..36..267R}
\be
\alpha(T) = 4.2 \times 10^{-13} \text{cm}^3 \text{s}^{-1} \left(\f{T}{10^4 \text{K}}\right)^{-0.7}.
\ee
The final piece in the calculation is to relate the temperature to the density. For the low-density IGM, relevant for the Lyman-$\alpha$ forest, one can assume a temperature-density relation of the form \cite{1997MNRAS.292...27H}
\be
T = T_0 (1 + \delta_b)^{\gamma - 1},
\ee
where $\delta_b$ is the baryonic density contrast. Collecting all the terms, we can write the optical depth in terms of the baryonic density field as
\bear
\tau_{\alpha}(\delta_b) &= 5.01~\left(\f{\chi_\mathrm{He}}{1.08}\right) \left(\f{1 - Y}{0.76}\right)^2 \left(\f{\Omega_b h^{3/2}}{0.0269}\right)^2
\left(\f{1 + z}{6.5}\right)^6 \left(\f{9.23}{H(z)/H_0}\right)
\nline 
&\times 
\left(\f{T_0}{10^4~\mathrm{K}}\right)^{-0.7}
\left(\f{10^{-12}~\mathrm{s}^{-1}}{\Gamma_{\mathrm{HI}}}\right) (1 + \delta_b)^{\beta},
\label{eq:tau_alpha_Gamma_HI}
\ear
where $\beta = 2.7 - 0.7 \gamma$. The calculation allows one to relate the Ly$\alpha$ optical depth to the cosmic matter field $\delta_b$ assuming the quantities $\Gamma_{\mathrm{HI}}$, $T_0$ and $\gamma$ are known.\footnote{Paddy was involved in developing a semi-numerical model of the Ly$\alpha$ forest which was used for constraining the photionization rate $\Gamma_{\mathrm{HI}}$ at $z \sim 3$ \cite{2001MNRAS.322..561C,2001ApJ...559...29C}.}

It is not straightforward to interpret the Ly$\alpha$ absorption observations and compare with the simple reionization model like the one we are discussing here. At $z \sim 6$, there are several ways to attempt the comparison, as discussed below:

\begin{itemize}

\item \emph{Mean transmitted flux:} The mean transmitted flux at a redshift is given by
\be
\bar{F} = \int \de \delta_b~\mathcal{P}(\delta_b)~\e^{-\tau_{\alpha}(\delta_b)},
\ee
where $\mathcal{P}(\delta_b)$ is the probability distribution of the baryonic density field. This mean flux $\bar{F}$ is a quantity that can be directly computed from the Ly$\alpha$ absorption spectra up to redshifts as high as $z \sim 6$ \cite{2018ApJ...864...53E,2020ApJ...904...26Y,2021arXiv210803699B}. Sometimes, one provides the results in terms of an \emph{effective optical depth} $\tau_{\mathrm{eff}} = - \ln \bar{F}$.

The theoretical calculation of $\bar{F}$ requires the form of the distribution $\mathcal{P}(\delta_b)$, which in principle depends on the thermal history of the medium. As a consequence, it is not straightforward to represent it by a simple analytical form. However, since most of the contribution to the integral arises from quasi-linear densities $\delta_b \lesssim 10$ (because of the $(1 + \delta_b)^{\beta}$ dependence $\tau_{\alpha}$), it is sufficient to use an approximate form that can represent the baryonic distribution at these densities. Possible approaches involve using the lognormal distribution \cite{2005MNRAS.361..577C} and a form motivated by simulations \cite{2000ApJ...530....1M}.

\item \emph{Photoionization rate:} There are studies which use high-resolution hydrodynamical simulations to compute the value of $\Gamma_{\mathrm{HI}}$ from the observational data. A possibility for the analytical models of reionization then is to compare their predictions with these inferred values of $\Gamma_{\mathrm{HI}}$. Under the assumption of local absorption, the photoionization rate is given by, see \eqns{eq:Gamma_HI}{eq:local_abs},
\be
\Gamma_{\mathrm{HI}} =  \int_{\nu_H}^{\infty} \f{\de \nu}{h \nu}~\sigma_H(\nu)~\left(\f{\bar{\tilde{\epsilon}}_{\nu}}{a^3}\right)~\lambda_{\mathrm{mfp}, \nu}.
\label{eq:Gamma_HI_epsilon}
\ee
The advantage of this method is that it does not require the knowledge of $\mathcal{P}(\delta_b)$, however, it still requires the knowledge of the mean free path. It can be determined from the observed distribution of the Lyman-limit systems, although the measurements become somewhat uncertain as we approach $z \sim 6$ \cite{2014MNRAS.445.1745W}.

\item \emph{Ionizing emissivity:}  More recently, it has been possible to put constraints on the mean free path from quasar absorption spectra combined with hydrodynamical simulations. Since both $\Gamma_{\mathrm{HI}}$ and $\lambda_{\mathrm{mfp}}$ are now available from these studies, one can compute the ionizing emissivity $\dot{\bar{n}}_{\gamma}$ \cite{2021MNRAS.508.1853B}. Let us combine \eqn{eq:Gamma_HI_epsilon} with \eqn{eq:ndot_epsilon} and write
\be
\dot{\bar{n}}_{\gamma} = \left(\f{\alpha + \beta}{\alpha}\right) \f{1}{\sigma_H(\nu_H)} \f{\Gamma_{\mathrm{HI}}}{a^3 \lambda_{\mathrm{mfp}, \nu_H}},
\ee
where we have assumed power-law $\nu$-dependence for the emissivity, the background radiation and the cross section
\bear
\epsilon(\nu) &= \epsilon(\nu_H) \left(\f{\nu}{\nu_H}\right)^{-\alpha},
\nline
\epsilon(\nu)~\lambda_{\mathrm{mfp}, \nu} &= \epsilon(\nu_H) \lambda_{\mathrm{mfp}, \nu_H}\left(\f{\nu}{\nu_H}\right)^{-\alpha_b},
\nline
\sigma_H(\nu) &= \sigma_H(\nu_H) \left(\f{\nu}{\nu_H}\right)^{-\beta}.
\ear
The photoionization cross section is well fitted by a form $\sigma_H(\nu) \propto \nu^{-3}$ at frequencies $\nu \gtrsim \nu_H$, implying $\beta = 3$, see \eqn{eq:sigma_H_nu}. The value of $\alpha$ depends on the sources of reionization; for stellar sources $\alpha \approx 2-3$. The change in the spectral slope as the photons travel through the IGM is accounted for in $\alpha_b$, which is taken to be $\alpha_b \approx 1.2-3$. In general, a value $(\alpha_b + \beta) / \alpha \approx 2$ is a good choice for calculations of this kind. 

This method provides the most direct way of comparing the reionization models with Ly$\alpha$ absorption observations. The disadvantage is that the observational constraints quoted \cite{2021MNRAS.508.1853B} are not model-independent. They require some unknown scale factors to be calibrated with high-resolution hydrodynamic simulations, which can introduce uncertainties in the quoted measurements.

\item \emph{Dark pixels:} Another probe of the neutral fraction is to use the fraction of pixels in the observed Ly$\alpha$ absorption spectra of the quasars that show no transmitted flux (i.e., they are ``dark''). This would provide an lower limit on the value of $Q_{\mathrm{HII}}$ \cite{2015MNRAS.447..499M,2011MNRAS.415.3237M}. The advantage of this method is that is relatively model-independent and hence allows direct comparison with the reionization models.

\end{itemize}

Although the comparison of the models with the Ly$\alpha$ absorption data is not that straightforward, it is still important to include them in the analysis. This is because the CMB optical depth only probes the integrated reionization history, see \eqn{eq:tau_el}, and hence is less sensitive to the details of reionization. The Ly$\alpha$ absorption, one the other hand, provides a complementary probe which is sensitive to the details of the end stages of reionization. In fact, coupling these two observations leads to rather tightened constraints on reionization models and helps in probing the unknown physics in more detail.

\subsection{Galaxy luminosity function}
\label{sec:UVLF}

During the development of the reionization source model in \secn{sec:sources}, we assigned a luminosity $L_{\nu}(M, t, t_{\mathrm{form}})$ to a halo of mass $M$ that formed at $t_{\mathrm{form}}$, see \eqn{eq:L_nu}. For a short-duration star-formation, it is given by
\be
L_{\nu}(M, t, t_{\mathrm{form}}) = f_*(M, t_{\mathrm{form}}) \left(\f{\Omega_b}{\Omega_m}\right) M~l_{\nu}(t-t_{\mathrm{form}}; M, t_{\mathrm{form}}).
\ee
The above relation which relates the luminosity of a galaxy to the mass of the host halo, when combined with the halo mass function, can be used to compute the \emph{luminosity function} of galaxies (i.e., number of galaxies per unit volume in a luminosity range). There are now direct observations of high-redshift galaxies identified using the Lyman-break technique which can be useful for constraining the reionization sources \cite{2013ApJ...768..196S,2013AJ....145....4W,2015ApJ...803...34B,2015ApJ...800...18A,2017ApJ...843..129B,2018MNRAS.479.5184A}. One point to note is that while the ionization histories are sensitive only to the efficiency $\zeta$, the UVLF requires knowledge of $f_*$ and the specific luminosity at UV wavelengths. This essentially introduces new free parameters to the model beyond what is required to model reionization \cite{2019MNRAS.484..933P,2021MNRAS.506.2390Q,2022arXiv220405268M}.

\subsection{The global 21~cm signal}

The hyperfine 21~cm radiation from HI at high redshifts is a potentially strong probe of reionization. During the reionization epoch, the signal is directly proportional to HI density (in this case, the spin temperature of HI is coupled to the gas temperature, and the gas is heated well above the CMB temperature) \cite{2006PhR...433..181F,2012RPPh...75h6901P}. In principle, the signal averaged over a large area of the sky (known as the \emph{global signal}) can be a strong probe of $Q_{\mathrm{HII}}(z)$. However, the signal is relatively weak during reionization (as opposed to cosmic dawn, where other physical processes lead to a much stronger signal), and hence is yet to be detected.

\subsection{Constraining astrophysics and cosmology using reionization}

\begin{figure}
\begin{center}
\includegraphics[width=\textwidth]{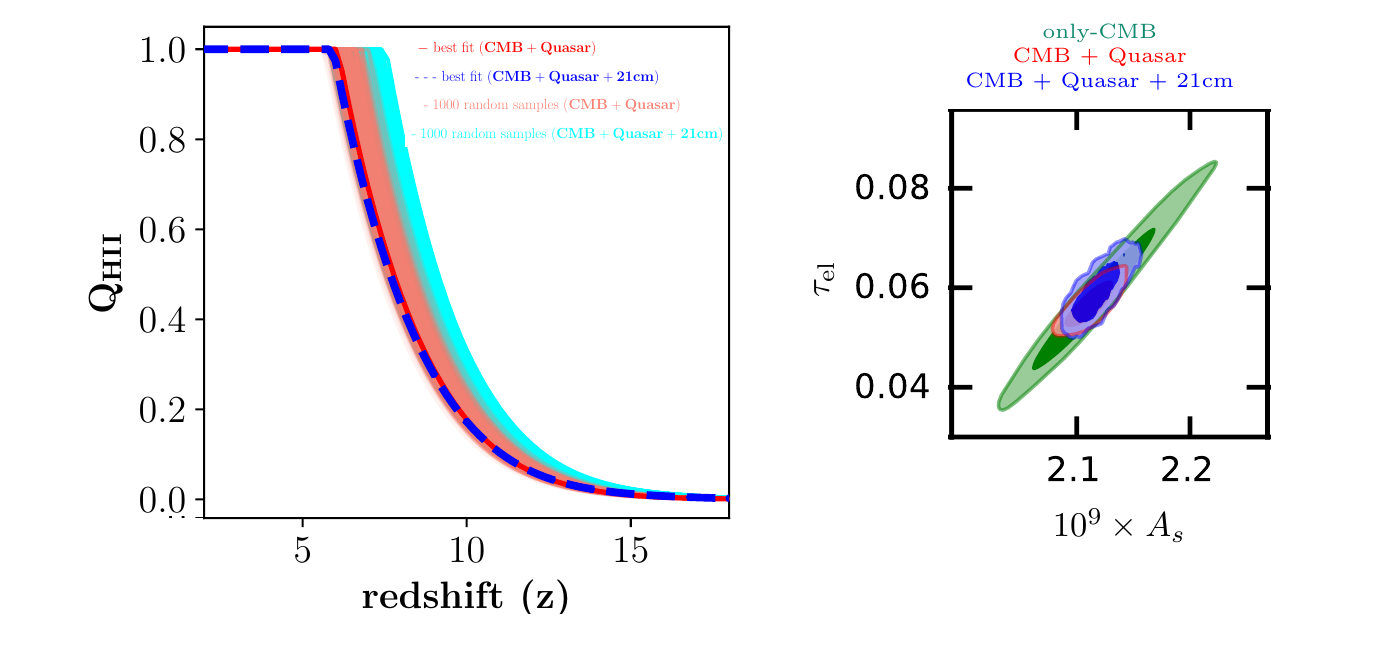}
\end{center}
\caption{Results from a parameter analysis code \texttt{CosmoReionMC}. Left panel: redshift evolution of $Q_{\mathrm{HII}}(z)$ for reionization models constrained using different data sets. Right panel: the joint posterior distribution of two of the parameters of the model, namely $A_s$ and $\tau_{\mathrm{el}}$. The plot shows strong degeneracy between these two parameters if only CMB data is used for analysis. Including the reionization data sets breaks the degeneracy and allows tighter constraints on $A_s$. The details of these results can be found in the original paper \cite{2021MNRAS.507.2405C}.}
\label{fig:cosmoreionmc}
\end{figure}

It is possible to use the simple analytical models of reionization to compare with a variety of observations and put constraints on the unknown parameters of the model \cite{2005MNRAS.361..577C}. A recent such code, named \texttt{CosmoReionMC}, attempts to use a slightly advanced version of the model discussed here and constrains both astrophysical parameters and cosmological parameter by comparing with CMB and quasar absorption observations \cite{2021MNRAS.507.2405C}. We show the constraints obtained on $Q_{\mathrm{HII}}(z)$ using a variety of observations in the left hand panel of \fig{fig:cosmoreionmc}. The right hand panel shows the joint constraints on two parameters, namely, $\tau_{\mathrm{el}}$ and $A_s$.

It is worth mentioning that reionization studies can also be used for constraining non-standard extensions to the concordance $\Lambda$CDM cosmological model. The main idea here is that many of these extensions affect the small-scale matter fluctuations, thus affecting the formation of the first dark matter haloes. This would shift the formation time of the first stars and hence the reionization history. Examples of such extensions include non-standard light dark matter particles \cite{2018ApJ...852..139V,2018MNRAS.477.2886L,2018PhRvD..98f3021S,2019JCAP...04..051N,2019PhRvD..99b3518C,2019MNRAS.487.3560C,2020PhRvD.101f3526M,2020MNRAS.497.2941S} and primordial magnetic fields \cite{2015MNRAS.451.1692P}.

\section{Fluctuations in the ionization field}
\label{sec:fluctuations}

In the previous sections, we have discussed how to construct a basic reionization model and compute the evolution of the globally averaged quantities, e.g., $Q_{\mathrm{HII}}$. An alternate way of probing reionization is to use the fluctuations in the ionized field, which can contain information related to the nature of reionization sources and the physical state of the IGM that are not possible to study using only the globally averaged quantities. 

Modelling these fluctuations would require solving all the physics we discussed in the previous sections in small cells inside a large simulation volume. Not only that, one needs to track the coupling between the different cells under consideration. One approach would be to solve the full hydrodynamic and radiative transfer equations at all locations over cosmic times, however, that is usually computationally extremely expensive. For want of space, we will limit the discussion in this article to simple and fast methods of modelling the complex physics, and avoid discussing on the more accurate and detailed simulations. This discussion should aid the reader to understand the main physical processes involved and hopefully gain some insight which they can use to build their own framework for modelling reionization.

\subsection{Dark matter haloes}

Let us begin with the sources of reionization, which in turn requires simulating the dark matter haloes. Clearly a globally averaged mass function $n(M,z)$ of haloes is not going to be sufficient to model the fluctuations in the ionization field. The most straightforward approach would be to run a $N$-body simulation (a gravity-only simulation is often sufficient for this purpose) and identify the haloes using some group-finding algorithm. However, the requirements for running such simulations are far from modest. 

Let us assume that we need to simulate a comoving volume $\gtrsim (200~\text{cMpc})^3$ to obtain a cosmologically representative sample of the Universe. This sets the largest scales to be probed. On the other extreme, as seen in \secn{sec:sources}, we need to resolve haloes $M_{\mathrm{min}} \sim 10^8 \Msun$ to track the smallest star-forming haloes in the presence of atomic cooling. This value can be much smaller if we want to track the minihaloes where the cooling is driven by molecules. Usually, one needs $N_{\mathrm{halo, min}} \sim 30-100$ particles in a group to be identify it as a collapsed halo, a group of smaller numbers could simply be transient collection of particles and may not represent a real halo in the simulation. With these requirements, a straightforward calculation would show that one would require particle numbers
\be
N_{\mathrm{part}} \approx 10^{11}~\left(\f{\Omega_m h^2}{0.15}\right)~\left(\f{L}{200 \text{cMpc}}\right)^3~\left(\f{N_{\mathrm{halo, min}}}{30}\right)~\left(\f{M_{\mathrm{min}}}{10^8 \Msun}\right)^{-1}.
\ee
This corresponds to $\sim 4640^3$ particles. Typically the $N$-body codes require $\sim 100$ bytes of RAM per particle, so this would require $\sim 10$ TB of RAM to run. Each redshift snapshot where one needs to store the output writes out the particle positions and velocities, which require $6 \times 4 = 24$ bytes. Thus each snapshot will be $\sim 2.4$ TB. These requirements are rather extreme and hence one requires access to high-performance computers to run these simulations.\footnote{It is worth noting that Paddy contributed to the development of one of the early $N$-body simulations \cite{1994MNRAS.266..227B,1996ApJ...469..470B,1997MNRAS.286.1023B,1997Prama..49..161B,1998ApJ...495...25B}.}

In recent times, the output of many of such high dynamic range simulations have been made public, e.g., \texttt{IllustrisTNG}\footnote{\url{https://www.tng-project.org/}}, \texttt{EAGLE}\footnote{\url{https://eagle.strw.leidenuniv.nl/}}, \texttt{SIMBA}\footnote{\url{http://simba.roe.ac.uk/}} and others. These simulations not only contain the collisonless dark matter particles, but also the galaxy formation modelling. These can be quite useful to identify the reionization sources in a large volume.

The alternative is to use approximation schemes as much as possible in different areas. One can, for example, use the concept of conditional mass functions to populate haloes that are not being resolved in a low-resolution simulation. This method is also known as ``subgrid prescription'' for populating haloes in a box \cite{2015MNRAS.450.1486A,2018MNRAS.481.3821C}. The main limitation of such analytical schemes is that they can be not very accurate and hence introduce errors in the calculation.

\subsection{Generation of the ionized regions}

Once the haloes have been identified in a simulation volume, either directly using group-finding algorithm or a subgrid prescription, one can use the reionization efficiency parameter $\zeta$ to assign the number of ionizing photons contributed by each halo to ionize the IGM. The next major challenge is to solve the radiative transfer equation in the IGM to generate the ionized regions.

Let us simplify the problem and concentrate on finding the number of atoms ionized by a halo. If we ignore recombinations for the time being, the number of hydrogen atoms ionized would be same as the number of ionizing photons produced by the halo over is lifetime, which is
\be
N_{\gamma}(M) = \zeta(M) (1 - Y) \left(\f{\Omega_b}{\Omega_m}\right) \f{M}{m_p}.
\ee
Note that we shall always work with haloes having $M > M_{\mathrm{min}}$. If each such halo produced a HII bubble around them, the size $R$ of the bubble would be given by the relation
\be
\f{4 \pi}{3} R^3~ \bar{n}_H \left(1 + \la \delta_b \ra_R\right) = N_{\gamma}(M),
\ee
where $\la \delta_b \ra_R$ is the baryonic density contrast (same as the density contrast in hydrogen) averaged over the sphere. At relatively large scales $\gtrsim 1$~cMpc, it is reasonable to assume $\delta_b \approx \delta_m$, the density contrast in matter fluctuations. Then a straightforward calculation will show that the total mass contained within the ionized bubble of radius $R$ will be given by
\be
M(R) = \zeta~ M.
\ee
Hence a dark matter halo of mass $M$ ionizes a region containing mass $\zeta~M$. This forms the basis for most semi-numerical models for generating ionized regions.

A simple way of generating the ionization field would be to place spheres of mass $\zeta~M$ around each halo. This however will fail if these spherical regions begin to overlap, which will indeed be the case during reionization. In fact, even during very early stages of reionization, the overlap between these individual ionized regions can be significant \cite{2004ApJ...613....1F}.

\subsubsection{Excursion set models}

An elegant way to deal with the possible overlaps of ionized regions is to use the excursion set formalism \cite{2004ApJ...613....1F}. In this case, one assumes that a ionized region can contain multiple sources (i.e., haloes). The identification of these regions can be done by the ``self-ionization condition'' where one assigns a region of size $R$ to be ionized if the number of photons produced by the sources within it exceeds the number of hydrogen atoms present. This condition can be expressed mathematically as
\be
\sum_{i \in R} N_{\gamma}(M_i) \geq  \f{4 \pi}{3} R^3~ \bar{n}_H \left(1 + \la \delta_b \ra_R\right)
\ee
where the summation on the left had is over all haloes $i$ in the sphere of radius $R$, their masses being denoted by $M_i$. A few lines of algebra will show that the relation is equivalent to
\be
\zeta~f_{\mathrm{coll}}(R) \geq  1,
\ee
where 
\be
f_{\mathrm{coll}}(R) = \f{\sum_{i \in R} M_i}{M(R)}
\ee 
is the collapsed fraction in the region under consideration. This simple condition, based on the conditional collapse fraction, can be used to identify ionized regions, either analytically or semi-numerically in a simulation volume. The algorithm can be implemented via efficient use of spherical filters in the Fourier domain to compute quantities like $f_{\mathrm{coll}}(R)$ for a range of $R$-values at every point in the box.

Because of its ease of implementation and computational efficiency, the method has been extremely popular and there exists several codes which implement the algorithm in simulations. Some of the publicly available codes based on excursion set are \texttt{21cmFAST}\footnote{\url{https://github.com/21cmfast/21cmFAST}} \cite{2011MNRAS.411..955M}, \texttt{SimFast21}\footnote{\url{https://github.com/mariogrs/Simfast21}} \cite{2010MNRAS.406.2421S,2016MNRAS.457.1550H}, \texttt{CIFOG}\footnote{\url{https://github.com/annehutter/grid-model}} \cite{2018MNRAS.477.1549H}, \texttt{ReionYuga}\footnote{\url{https://github.com/rajeshmondal18/ReionYuga}} \cite{2009MNRAS.394..960C,2014MNRAS.443.2843M,2017MNRAS.464.2992M}.

\subsubsection{Photon-conserving models}

One difficulty with the excursion set based models is the non-conservation of photons. The total number of ionizing photons produced by the haloes in a simulation volume should be equal to the number of hydrogen atoms ionized, assuming recombinations are absent. Mathematically, this corresponds to $\langle\zeta~f_{\mathrm{coll}}\rangle = 1$. However, it can be shown that the quantity deviates from unity depending on the reionization model, the amount of hydrogen ionized and also the resolution of the simulation volume \cite{2018MNRAS.481.3821C}. The reason for this non-conservation is intrinsic to excursion set algorithms which are by nature non-conserving \cite{2016MNRAS.460.1801P}. This has led people to develop algorithms which do not violate the photon-conservation.

An implementation of photon-conserving model involves generating ionized bubbles around individual sources (or around each ``bunch'' of sources, where a bunch can correspond to all sources inside a simulation cell). In the first round, these bubbles are allowed to overlap and the regions of overlap are assigned ionized fractions $ > 1$, i.e., they are ``overionized''. In the next round, the excess photons in these overionized cells are treated as sources and the photons are distributed in nearby cells. The process can be carried out iteratively until all the photons are properly distributed and no overionized cells exist \cite{2018MNRAS.481.3821C}. This algorithm, by construction, is photon-conserving. The code, named \texttt{SCRIPT}\footnote{\url{https://bitbucket.org/rctirthankar/script}}, is slower than the excursion set models, but still fast enough to explore parameter space.

There exist other photon-conserving algorithms, e.g., \texttt{ARTIST} \cite{2019MNRAS.489.5594M}, where one propagates photons accounting for the directionality and is able to match the more detailed radiative transfer results. There are also methods which modify the excursion set algorithm by shifting the ionizing field in redshift to compensate for photon non-conservation, thus retaining the computational efficiency of the excursion set models \cite{2021arXiv211205184P}.

However, there is another more important, and perhaps less appreciated, point related to the photon non-conservation which needs discussion. Most of the semi-numerical models divide the simulation box into three-dimensional grid cells and compute the quantities of interest in those cells. Clearly, the volume of these cells sets the resolution of the simulation. One can make the codes run much faster by generating the ionization fields at rather coarse resolutions, with the understanding that the scales we are interested in are much larger than the gird size. It turns out that the excursion set based models do not show numerical convergence with respect to the resolution at which the algorithm is implemented \cite{2018MNRAS.481.3821C}. As a result, the result of the calculations and the subsequent interpretation of the observations using these simulations depends on the resolution used. Interestingly, the photon-conserving algorithm in \texttt{SCRIPT} does not have this problem and provides numerically converged results with respect to the resolution.

\subsection{Observables probing the fluctuations}

The modelling of the fluctuations opens up possibilities of using more observational probes to understand reionization. Let us summarize a few important ones now.

\begin{itemize}

\item \emph{Lyman-$\alpha$ absorption:} As we have discussed in \secn{sec:lya_absorption}, the mean transmitted flux of the Ly$\alpha$ absorption spectra at $z \sim 6$ can be used to put limits on the end of reionization. However, these spectra can be used to study fluctuations in the transmitted flux and hence possibly fluctuations in the ionized field. In fact, the mean transmitted flux at $5.5 < z < 6$, when averaged over large scales ($\sim 40-50 h^{-1}$~cMpc), showed significant fluctuations \cite{2015MNRAS.447.3402B,2018MNRAS.479.1055B,2018ApJ...864...53E,2019ApJ...881...23E,2021arXiv210803699B} which could not be explained by simple models of uniform $\Gamma_{\mathrm{HI}}$. There are strong indications from radiative transfer simulations and semi-numerical studies that these fluctuations originate from neutral gas at the end stages of reionization \cite{2019MNRAS.485L..24K,2020MNRAS.497..906K,2020MNRAS.494.3080N,2021MNRAS.501.5782C}.

There is another probe of reionization based on the Ly$\alpha$ absorption spectra, namely the thermal state of the IGM at $z \gtrsim 5$ \citep{2010MNRAS.406..612B,2012MNRAS.419.2880B,2019ApJ...872...13W,2019ApJ...872..101B,2020MNRAS.494.5091G}. The temperature of the IGM at these epochs is expected to contain imprints of the reionization epoch because of the photoheating associated with hydrogen ionization and the subsequent adiabatic cooling. It has been hence proposed as a possible probe of reionization \citep{2010MNRAS.406..612B,2012MNRAS.423..558C,2012MNRAS.421.1969R,2014MNRAS.443.3761P,2022MNRAS.511.2239M,2022arXiv220405268M}.

\item \emph{Lyman-$\alpha$ emitters:} The Ly$\alpha$ emitters (LAEs) are galaxies that show strong Ly$\alpha$ emission. In principle, their luminosity function can be useful to study the galaxies during the reionization era, just like the UVLF of Lyman-break galaxies discussed in \secn{sec:UVLF}. However, there is one critical aspect of LAEs that make them rather interesting for probing the patchy reionization at $z\sim 6-7$. The essential idea is that if the LAE galaxy is situated in a surrounding medium that is primarily neutral, then the damped Lorentzian profile of the HI outside can extend to the wavelengths corresponding the Ly$\alpha$ emission of the galaxy. This can lead to significant decrease in the luminosity of the LAE. Thus any significant drop in the number of LAEs around the end of reionization could point towards an independent probe of patchy reionization. The interpretation of the observations, however, are complicated by the exact topology of the ionized regions, the velocity profile of the LAEs, presence of high-density absorbers along the line of sight to the LAE and other effects. A large amount of work has already been done in using the LAEs as a probe of patchy reionization, we refer the reader to some recent review articles \cite{2017arXiv170403416D,2020ARA&A..58..617O}.

\item \emph{kSZ signal from patchy reionization:} With the inclusion of Planck and several ground-based experiments like the SPT, it is possible to probe the CMB anisotropies at relatively smaller angular scales where the temperature and polarization power spectra are not dominated by primary fluctuations from the last scattering surface at $z \sim 1100$. An important source of these secondary anisotropies is the kinetic Sunyaev-Zeldovich (kSZ) effect arising from the Doppler effect due to bulk motion of free electrons in HII bubbles during reionization \cite{1980ARA&A..18..537S}. The kSZ anisotropy depends not only the ionized field but also the velocity field of the free electrons and hence can be a probe of the details of the reionization topology. The signal has already been detected by SPT \cite{2021ApJ...908..199R} and has been used for understanding the reionization epoch \cite{2021MNRAS.501L...7C,2022arXiv220208698G,2022arXiv220304337C}.

\item \emph{Fast radio bursts:} An interesting possibility of studying reionization is to use the fast radio bursts (FRBs) at $z \gtrsim 6$ \citep{2021MNRAS.502.5134B,2021MNRAS.505.2195P,2022ApJ...933...57H}. The dispersion measure of the FRBs provides an estimate of the ionized matter along the line of sight, and hence can potentially provide a complementary probe of the integrated electron density.

\item \emph{21~cm fluctuations:} In the near future, the details of reionization can be studied using the fluctuations in the 21~cm emission from HI \cite{2006PhR...433..181F,2012RPPh...75h6901P}. These are best studied using low-frequency intererometers. There have already been data from telescopes like the GMRT, PAPER, LOFAR and MWA, however, their sensitivities are still not at the level where one can rule out interesting reionzation models. In the future, telescopes like HERA and SKA should allow one to do detailed studies of reionization using 21~cm fluctuations. We refer the reader to a different review for the status of these experiments \cite{2019arXiv190912491T}.

\end{itemize}

\section{Future prospects}
\label{sec:future}

The study of reionization is going to receive continuous boost thanks to a number of upcoming facilities. In fact, it is difficult to think of an astrophysical probe which does not have some relevance to the epoch of reionization. The recently launched JWST is likely to probe the first galaxies responsible for reionization. These will be followed up by the extremely large optical telescopes, namely, EELT, GMT, TMT,\footnote{Interestingly, Paddy played an important role in India joining one of these mega-projects, TMT. He was the convener of one of the advisory groups to facilitate India's entry to the international collaboration.} which are planning to detect the galaxies as well the Ly$\alpha$ absorption along sight lines towards the galaxies. The upcoming radio interferometers like SKA and HERA will detect the HI fluctuations with great sensitivity which should track the topology of the ionized regions. The CMB probes like LiteBIRD, CMB-S4 and PICO would also be extremely useful for studying reionization through the secondary temperature and polarization anisotropies at different angular scales.

The probes will need to be appropriately complemented by theoretical understanding and calculations. It is still a challenge to develop calculations that are accurate and efficient at the same time, so a lot of work still needs to be done. With rather sophisticated techniques, based on Machine Learning, one hopes to break the barrier and come up with accurate and efficient methods of studying the early Universe.

I end this article by remembering Paddy once more who is of course very well-known for his work on gravity. What is often less known is that he used to be quite interested in issues related to reionization and intergalactic medium, and I have tried to highlight his contributions at various places in this article. The hope is this article in the special issue of the journal provides readers some interesting aspects of this growing field of research.

\section*{Data availability}

Data sharing not applicable to this article as no datasets were generated or analysed during the current study.



\begin{thebibliography}{116}
\ifx \bisbn   \undefined \def \bisbn  #1{ISBN #1}\fi
\ifx \binits  \undefined \def \binits#1{#1}\fi
\ifx \bauthor  \undefined \def \bauthor#1{#1}\fi
\ifx \batitle  \undefined \def \batitle#1{#1}\fi
\ifx \bjtitle  \undefined \def \bjtitle#1{#1}\fi
\ifx \bvolume  \undefined \def \bvolume#1{\textbf{#1}}\fi
\ifx \byear  \undefined \def \byear#1{#1}\fi
\ifx \bissue  \undefined \def \bissue#1{#1}\fi
\ifx \bfpage  \undefined \def \bfpage#1{#1}\fi
\ifx \blpage  \undefined \def \blpage #1{#1}\fi
\ifx \burl  \undefined \def \burl#1{\textsf{#1}}\fi
\ifx \doiurl  \undefined \def \doiurl#1{\url{https://doi.org/#1}}\fi
\ifx \betal  \undefined \def \betal{\textit{et al.}}\fi
\ifx \binstitute  \undefined \def \binstitute#1{#1}\fi
\ifx \binstitutionaled  \undefined \def \binstitutionaled#1{#1}\fi
\ifx \bctitle  \undefined \def \bctitle#1{#1}\fi
\ifx \beditor  \undefined \def \beditor#1{#1}\fi
\ifx \bpublisher  \undefined \def \bpublisher#1{#1}\fi
\ifx \bbtitle  \undefined \def \bbtitle#1{#1}\fi
\ifx \bedition  \undefined \def \bedition#1{#1}\fi
\ifx \bseriesno  \undefined \def \bseriesno#1{#1}\fi
\ifx \blocation  \undefined \def \blocation#1{#1}\fi
\ifx \bsertitle  \undefined \def \bsertitle#1{#1}\fi
\ifx \bsnm \undefined \def \bsnm#1{#1}\fi
\ifx \bsuffix \undefined \def \bsuffix#1{#1}\fi
\ifx \bparticle \undefined \def \bparticle#1{#1}\fi
\ifx \barticle \undefined \def \barticle#1{#1}\fi
\bibcommenthead
\ifx \bconfdate \undefined \def \bconfdate #1{#1}\fi
\ifx \botherref \undefined \def \botherref #1{#1}\fi
\ifx \url \undefined \def \url#1{\textsf{#1}}\fi
\ifx \bchapter \undefined \def \bchapter#1{#1}\fi
\ifx \bbook \undefined \def \bbook#1{#1}\fi
\ifx \bcomment \undefined \def \bcomment#1{#1}\fi
\ifx \oauthor \undefined \def \oauthor#1{#1}\fi
\ifx \citeauthoryear \undefined \def \citeauthoryear#1{#1}\fi
\ifx \endbibitem  \undefined \def \endbibitem {}\fi
\ifx \bconflocation  \undefined \def \bconflocation#1{#1}\fi
\ifx \arxivurl  \undefined \def \arxivurl#1{\textsf{#1}}\fi
\csname PreBibitemsHook\endcsname

{\footnotesize{

\bibitem{2018MNRAS.481.3821C}
\begin{barticle}
\bauthor{\bsnm{{Choudhury}}, \binits{T.R.}},
\bauthor{\bsnm{{Paranjape}}, \binits{A.}}:
\batitle{{Photon number conservation and the large-scale 21 cm power spectrum
  in seminumerical models of reionization}}.
\bjtitle{\mnras}
\bvolume{481}(\bissue{3}),
\bfpage{3821}--\blpage{3837}
(\byear{2018})
{\href{https://arxiv.org/abs/1807.00836}{{arXiv:1807.00836}}}
{[astro-ph.CO]}.
\doiurl{10.1093/mnras/sty2551}
\end{barticle}
\endbibitem

\bibitem{2001PhR...349..125B}
\begin{barticle}
\bauthor{\bsnm{{Barkana}}, \binits{R.}},
\bauthor{\bsnm{{Loeb}}, \binits{A.}}:
\batitle{{In the beginning: the first sources of light and the reionization of
  the universe}}.
\bjtitle{\physrep}
\bvolume{349}(\bissue{2}),
\bfpage{125}--\blpage{238}
(\byear{2001})
{\href{https://arxiv.org/abs/astro-ph/0010468}{{arXiv:astro-ph/0010468}}}
{[astro-ph]}.
\doiurl{10.1016/S0370-1573(01)00019-9}
\end{barticle}
\endbibitem

\bibitem{2005SSRv..116..625C}
\begin{barticle}
\bauthor{\bsnm{{Ciardi}}, \binits{B.}},
\bauthor{\bsnm{{Ferrara}}, \binits{A.}}:
\batitle{{The First Cosmic Structures and Their Effects}}.
\bjtitle{\ssr}
\bvolume{116}(\bissue{3-4}),
\bfpage{625}--\blpage{705}
(\byear{2005})
{\href{https://arxiv.org/abs/astro-ph/0409018}{{arXiv:astro-ph/0409018}}}
{[astro-ph]}.
\doiurl{10.1007/s11214-005-3592-0}
\end{barticle}
\endbibitem

\bibitem{2018PhR...780....1D}
\begin{barticle}
\bauthor{\bsnm{{Dayal}}, \binits{P.}},
\bauthor{\bsnm{{Ferrara}}, \binits{A.}}:
\batitle{{Early galaxy formation and its large-scale effects}}.
\bjtitle{\physrep}
\bvolume{780},
\bfpage{1}--\blpage{64}
(\byear{2018})
{\href{https://arxiv.org/abs/1809.09136}{{arXiv:1809.09136}}}
{[astro-ph.GA]}.
\doiurl{10.1016/j.physrep.2018.10.002}
\end{barticle}
\endbibitem

\bibitem{2022arXiv220802260G}
\begin{botherref}
\oauthor{\bsnm{{Gnedin}}, \binits{N.Y.}},
\oauthor{\bsnm{{Madau}}, \binits{P.}}:
{Modeling Cosmic Reionization}.
arXiv e-prints,
2208--02260
(2022)
{\href{https://arxiv.org/abs/2208.02260}{{arXiv:2208.02260}}}
{[astro-ph.CO]}
\end{botherref}
\endbibitem

\bibitem{2002thas.book.....P}
\begin{bbook}
\bauthor{\bsnm{{Padmanabhan}}, \binits{T.}}:
\bbtitle{{Theoretical Astrophysics - Volume 3, Galaxies and Cosmology}}
vol. \bseriesno{3},
(\byear{2002}).
\doiurl{10.2277/0521562422}
\end{bbook}
\endbibitem

\bibitem{1999coph.book.....P}
\begin{bbook}
\bauthor{\bsnm{{Peacock}}, \binits{J.A.}}:
\bbtitle{{Cosmological Physics}},
(\byear{1999})
\end{bbook}
\endbibitem

\bibitem{2010gfe..book.....M}
\begin{bbook}
\bauthor{\bsnm{{Mo}}, \binits{H.}},
\bauthor{\bsnm{{van den Bosch}}, \binits{F.C.}},
\bauthor{\bsnm{{White}}, \binits{S.}}:
\bbtitle{{Galaxy Formation and Evolution}},
(\byear{2010})
\end{bbook}
\endbibitem

\bibitem{1993sfu..book.....P}
\begin{bbook}
\bauthor{\bsnm{{Padmanabhan}}, \binits{T.}}:
\bbtitle{{Structure Formation in the Universe}},
(\byear{1993})
\end{bbook}
\endbibitem

\bibitem{1991ApJ...379..440B}
\begin{barticle}
\bauthor{\bsnm{{Bond}}, \binits{J.R.}},
\bauthor{\bsnm{{Cole}}, \binits{S.}},
\bauthor{\bsnm{{Efstathiou}}, \binits{G.}},
\bauthor{\bsnm{{Kaiser}}, \binits{N.}}:
\batitle{{Excursion Set Mass Functions for Hierarchical Gaussian
  Fluctuations}}.
\bjtitle{\apj}
\bvolume{379},
\bfpage{440}
(\byear{1991}).
\doiurl{10.1086/170520}
\end{barticle}
\endbibitem

\bibitem{1974ApJ...187..425P}
\begin{barticle}
\bauthor{\bsnm{{Press}}, \binits{W.H.}},
\bauthor{\bsnm{{Schechter}}, \binits{P.}}:
\batitle{{Formation of Galaxies and Clusters of Galaxies by Self-Similar
  Gravitational Condensation}}.
\bjtitle{\apj}
\bvolume{187},
\bfpage{425}--\blpage{438}
(\byear{1974}).
\doiurl{10.1086/152650}
\end{barticle}
\endbibitem

\bibitem{2001MNRAS.323....1S}
\begin{barticle}
\bauthor{\bsnm{{Sheth}}, \binits{R.K.}},
\bauthor{\bsnm{{Mo}}, \binits{H.J.}},
\bauthor{\bsnm{{Tormen}}, \binits{G.}}:
\batitle{{Ellipsoidal collapse and an improved model for the number and spatial
  distribution of dark matter haloes}}.
\bjtitle{\mnras}
\bvolume{323}(\bissue{1}),
\bfpage{1}--\blpage{12}
(\byear{2001})
{\href{https://arxiv.org/abs/astro-ph/9907024}{{arXiv:astro-ph/9907024}}}
{[astro-ph]}.
\doiurl{10.1046/j.1365-8711.2001.04006.x}
\end{barticle}
\endbibitem

\bibitem{2002MNRAS.329...61S}
\begin{barticle}
\bauthor{\bsnm{{Sheth}}, \binits{R.K.}},
\bauthor{\bsnm{{Tormen}}, \binits{G.}}:
\batitle{{An excursion set model of hierarchical clustering: ellipsoidal
  collapse and the moving barrier}}.
\bjtitle{\mnras}
\bvolume{329}(\bissue{1}),
\bfpage{61}--\blpage{75}
(\byear{2002})
{\href{https://arxiv.org/abs/astro-ph/0105113}{{arXiv:astro-ph/0105113}}}
{[astro-ph]}.
\doiurl{10.1046/j.1365-8711.2002.04950.x}
\end{barticle}
\endbibitem

\bibitem{2001MNRAS.321..372J}
\begin{barticle}
\bauthor{\bsnm{{Jenkins}}, \binits{A.}},
\bauthor{\bsnm{{Frenk}}, \binits{C.S.}},
\bauthor{\bsnm{{White}}, \binits{S.D.M.}},
\bauthor{\bsnm{{Colberg}}, \binits{J.M.}},
\bauthor{\bsnm{{Cole}}, \binits{S.}},
\bauthor{\bsnm{{Evrard}}, \binits{A.E.}},
\bauthor{\bsnm{{Couchman}}, \binits{H.M.P.}},
\bauthor{\bsnm{{Yoshida}}, \binits{N.}}:
\batitle{{The mass function of dark matter haloes}}.
\bjtitle{\mnras}
\bvolume{321}(\bissue{2}),
\bfpage{372}--\blpage{384}
(\byear{2001})
{\href{https://arxiv.org/abs/astro-ph/0005260}{{arXiv:astro-ph/0005260}}}
{[astro-ph]}.
\doiurl{10.1046/j.1365-8711.2001.04029.x}
\end{barticle}
\endbibitem

\bibitem{2013MNRAS.433.1230W}
\begin{barticle}
\bauthor{\bsnm{{Watson}}, \binits{W.A.}},
\bauthor{\bsnm{{Iliev}}, \binits{I.T.}},
\bauthor{\bsnm{{D'Aloisio}}, \binits{A.}},
\bauthor{\bsnm{{Knebe}}, \binits{A.}},
\bauthor{\bsnm{{Shapiro}}, \binits{P.R.}},
\bauthor{\bsnm{{Yepes}}, \binits{G.}}:
\batitle{{The halo mass function through the cosmic ages}}.
\bjtitle{\mnras}
\bvolume{433}(\bissue{2}),
\bfpage{1230}--\blpage{1245}
(\byear{2013})
{\href{https://arxiv.org/abs/1212.0095}{{arXiv:1212.0095}}}
{[astro-ph.CO]}.
\doiurl{10.1093/mnras/stt791}
\end{barticle}
\endbibitem

\bibitem{2005MNRAS.364.1105S}
\begin{barticle}
\bauthor{\bsnm{{Springel}}, \binits{V.}}:
\batitle{{The cosmological simulation code GADGET-2}}.
\bjtitle{\mnras}
\bvolume{364}(\bissue{4}),
\bfpage{1105}--\blpage{1134}
(\byear{2005})
{\href{https://arxiv.org/abs/astro-ph/0505010}{{arXiv:astro-ph/0505010}}}
{[astro-ph]}.
\doiurl{10.1111/j.1365-2966.2005.09655.x}
\end{barticle}
\endbibitem

\bibitem{2013RPPh...76k2901B}
\begin{barticle}
\bauthor{\bsnm{{Bromm}}, \binits{V.}}:
\batitle{{Formation of the first stars}}.
\bjtitle{Reports on Progress in Physics}
\bvolume{76}(\bissue{11}),
\bfpage{112901}
(\byear{2013})
{\href{https://arxiv.org/abs/1305.5178}{{arXiv:1305.5178}}}
{[astro-ph.CO]}.
\doiurl{10.1088/0034-4885/76/11/112901}
\end{barticle}
\endbibitem

\bibitem{1996ApJ...456....1G}
\begin{barticle}
\bauthor{\bsnm{{Gnedin}}, \binits{N.Y.}}:
\batitle{{Galaxy Formation in a CDM + Lambda Universe. I. Properties of Gas and
  Galaxies}}.
\bjtitle{\apj}
\bvolume{456},
\bfpage{1}
(\byear{1996}).
\doiurl{10.1086/176623}
\end{barticle}
\endbibitem

\bibitem{2000ApJ...534..507C}
\begin{barticle}
\bauthor{\bsnm{{Chiu}}, \binits{W.A.}},
\bauthor{\bsnm{{Ostriker}}, \binits{J.P.}}:
\batitle{{A Semianalytic Model for Cosmological Reheating and Reionization Due
  to the Gravitational Collapse of Structure}}.
\bjtitle{\apj}
\bvolume{534}(\bissue{2}),
\bfpage{507}--\blpage{532}
(\byear{2000})
{\href{https://arxiv.org/abs/astro-ph/9907220}{{arXiv:astro-ph/9907220}}}
{[astro-ph]}.
\doiurl{10.1086/308780}
\end{barticle}
\endbibitem

\bibitem{2002MNRAS.336L..27R}
\begin{barticle}
\bauthor{\bsnm{{Choudhury}}, \binits{T.R.}},
\bauthor{\bsnm{{Srianand}}, \binits{R.}}:
\batitle{{Probing the dark ages with redshift distribution of gamma-ray
  bursts}}.
\bjtitle{\mnras}
\bvolume{336}(\bissue{2}),
\bfpage{27}--\blpage{31}
(\byear{2002})
{\href{https://arxiv.org/abs/astro-ph/0205446}{{arXiv:astro-ph/0205446}}}
{[astro-ph]}.
\doiurl{10.1046/j.1365-8711.2002.05984.x}
\end{barticle}
\endbibitem

\bibitem{2013MNRAS.435..368J}
\begin{barticle}
\bauthor{\bsnm{{Jose}}, \binits{C.}},
\bauthor{\bsnm{{Srianand}}, \binits{R.}},
\bauthor{\bsnm{{Subramanian}}, \binits{K.}}:
\batitle{{Clustering at high redshift: the connection between Lyman
  {\ensuremath{\alpha}} emitters and Lyman break galaxies}}.
\bjtitle{\mnras}
\bvolume{435}(\bissue{1}),
\bfpage{368}--\blpage{377}
(\byear{2013})
{\href{https://arxiv.org/abs/1304.7458}{{arXiv:1304.7458}}}
{[astro-ph.CO]}.
\doiurl{10.1093/mnras/stt1299}
\end{barticle}
\endbibitem

\bibitem{2014MNRAS.443.3341J}
\begin{barticle}
\bauthor{\bsnm{{Jose}}, \binits{C.}},
\bauthor{\bsnm{{Srianand}}, \binits{R.}},
\bauthor{\bsnm{{Subramanian}}, \binits{K.}}:
\batitle{{A physical model for the redshift evolution of high-z Lyman-break
  galaxies}}.
\bjtitle{\mnras}
\bvolume{443}(\bissue{4}),
\bfpage{3341}--\blpage{3350}
(\byear{2014})
{\href{https://arxiv.org/abs/1310.3917}{{arXiv:1310.3917}}}
{[astro-ph.CO]}.
\doiurl{10.1093/mnras/stu1339}
\end{barticle}
\endbibitem

\bibitem{1999ApJS..123....3L}
\begin{barticle}
\bauthor{\bsnm{{Leitherer}}, \binits{C.}},
\bauthor{\bsnm{{Schaerer}}, \binits{D.}},
\bauthor{\bsnm{{Goldader}}, \binits{J.D.}},
\bauthor{\bsnm{{Delgado}}, \binits{R.M.G.}},
\bauthor{\bsnm{{Robert}}, \binits{C.}},
\bauthor{\bsnm{{Kune}}, \binits{D.F.}},
\bauthor{\bsnm{{de Mello}}, \binits{D.F.}},
\bauthor{\bsnm{{Devost}}, \binits{D.}},
\bauthor{\bsnm{{Heckman}}, \binits{T.M.}}:
\batitle{{Starburst99: Synthesis Models for Galaxies with Active Star
  Formation}}.
\bjtitle{\apjs}
\bvolume{123}(\bissue{1}),
\bfpage{3}--\blpage{40}
(\byear{1999})
{\href{https://arxiv.org/abs/astro-ph/9902334}{{arXiv:astro-ph/9902334}}}
{[astro-ph]}.
\doiurl{10.1086/313233}
\end{barticle}
\endbibitem

\bibitem{2003MNRAS.344.1000B}
\begin{barticle}
\bauthor{\bsnm{{Bruzual}}, \binits{G.}},
\bauthor{\bsnm{{Charlot}}, \binits{S.}}:
\batitle{{Stellar population synthesis at the resolution of 2003}}.
\bjtitle{\mnras}
\bvolume{344}(\bissue{4}),
\bfpage{1000}--\blpage{1028}
(\byear{2003})
{\href{https://arxiv.org/abs/astro-ph/0309134}{{arXiv:astro-ph/0309134}}}
{[astro-ph]}.
\doiurl{10.1046/j.1365-8711.2003.06897.x}
\end{barticle}
\endbibitem

\bibitem{2017PASA...34...58E}
\begin{barticle}
\bauthor{\bsnm{{Eldridge}}, \binits{J.J.}},
\bauthor{\bsnm{{Stanway}}, \binits{E.R.}},
\bauthor{\bsnm{{Xiao}}, \binits{L.}},
\bauthor{\bsnm{{McClelland}}, \binits{L.A.S.}},
\bauthor{\bsnm{{Taylor}}, \binits{G.}},
\bauthor{\bsnm{{Ng}}, \binits{M.}},
\bauthor{\bsnm{{Greis}}, \binits{S.M.L.}},
\bauthor{\bsnm{{Bray}}, \binits{J.C.}}:
\batitle{{Binary Population and Spectral Synthesis Version 2.1: Construction,
  Observational Verification, and New Results}}.
\bjtitle{\pasa}
\bvolume{34},
\bfpage{058}
(\byear{2017})
{\href{https://arxiv.org/abs/1710.02154}{{arXiv:1710.02154}}}
{[astro-ph.SR]}.
\doiurl{10.1017/pasa.2017.51}
\end{barticle}
\endbibitem

\bibitem{2022MNRAS.512.5329B}
\begin{barticle}
\bauthor{\bsnm{{Byrne}}, \binits{C.M.}},
\bauthor{\bsnm{{Stanway}}, \binits{E.R.}},
\bauthor{\bsnm{{Eldridge}}, \binits{J.J.}},
\bauthor{\bsnm{{McSwiney}}, \binits{L.}},
\bauthor{\bsnm{{Townsend}}, \binits{O.T.}}:
\batitle{{The dependence of theoretical synthetic spectra on
  {\ensuremath{\alpha}}-enhancement in young, binary stellar populations}}.
\bjtitle{\mnras}
\bvolume{512}(\bissue{4}),
\bfpage{5329}--\blpage{5338}
(\byear{2022})
{\href{https://arxiv.org/abs/2203.13275}{{arXiv:2203.13275}}}
{[astro-ph.SR]}.
\doiurl{10.1093/mnras/stac807}
\end{barticle}
\endbibitem

\bibitem{1996ApJ...461...20H}
\begin{barticle}
\bauthor{\bsnm{{Haardt}}, \binits{F.}},
\bauthor{\bsnm{{Madau}}, \binits{P.}}:
\batitle{{Radiative Transfer in a Clumpy Universe. II. The Ultraviolet
  Extragalactic Background}}.
\bjtitle{\apj}
\bvolume{461},
\bfpage{20}
(\byear{1996})
{\href{https://arxiv.org/abs/astro-ph/9509093}{{arXiv:astro-ph/9509093}}}
{[astro-ph]}.
\doiurl{10.1086/177035}
\end{barticle}
\endbibitem

\bibitem{1999ApJ...514..648M}
\begin{barticle}
\bauthor{\bsnm{{Madau}}, \binits{P.}},
\bauthor{\bsnm{{Haardt}}, \binits{F.}},
\bauthor{\bsnm{{Rees}}, \binits{M.J.}}:
\batitle{{Radiative Transfer in a Clumpy Universe. III. The Nature of
  Cosmological Ionizing Sources}}.
\bjtitle{\apj}
\bvolume{514}(\bissue{2}),
\bfpage{648}--\blpage{659}
(\byear{1999})
{\href{https://arxiv.org/abs/astro-ph/9809058}{{arXiv:astro-ph/9809058}}}
{[astro-ph]}.
\doiurl{10.1086/306975}
\end{barticle}
\endbibitem

\bibitem{2009ApJ...703.1416F}
\begin{barticle}
\bauthor{\bsnm{{Faucher-Gigu{\`e}re}}, \binits{C.-A.}},
\bauthor{\bsnm{{Lidz}}, \binits{A.}},
\bauthor{\bsnm{{Zaldarriaga}}, \binits{M.}},
\bauthor{\bsnm{{Hernquist}}, \binits{L.}}:
\batitle{{A New Calculation of the Ionizing Background Spectrum and the Effects
  of He II Reionization}}.
\bjtitle{\apj}
\bvolume{703}(\bissue{2}),
\bfpage{1416}--\blpage{1443}
(\byear{2009})
{\href{https://arxiv.org/abs/0901.4554}{{arXiv:0901.4554}}}
{[astro-ph.CO]}.
\doiurl{10.1088/0004-637X/703/2/1416}
\end{barticle}
\endbibitem

\bibitem{2012ApJ...746..125H}
\begin{barticle}
\bauthor{\bsnm{{Haardt}}, \binits{F.}},
\bauthor{\bsnm{{Madau}}, \binits{P.}}:
\batitle{{Radiative Transfer in a Clumpy Universe. IV. New Synthesis Models of
  the Cosmic UV/X-Ray Background}}.
\bjtitle{\apj}
\bvolume{746}(\bissue{2}),
\bfpage{125}
(\byear{2012})
{\href{https://arxiv.org/abs/1105.2039}{{arXiv:1105.2039}}}
{[astro-ph.CO]}.
\doiurl{10.1088/0004-637X/746/2/125}
\end{barticle}
\endbibitem

\bibitem{2019MNRAS.484.4174K}
\begin{barticle}
\bauthor{\bsnm{{Khaire}}, \binits{V.}},
\bauthor{\bsnm{{Srianand}}, \binits{R.}}:
\batitle{{New synthesis models of consistent extragalactic background light
  over cosmic time}}.
\bjtitle{\mnras}
\bvolume{484}(\bissue{3}),
\bfpage{4174}--\blpage{4199}
(\byear{2019})
{\href{https://arxiv.org/abs/1801.09693}{{arXiv:1801.09693}}}
{[astro-ph.GA]}.
\doiurl{10.1093/mnras/stz174}
\end{barticle}
\endbibitem

\bibitem{2014MNRAS.445.1745W}
\begin{barticle}
\bauthor{\bsnm{{Worseck}}, \binits{G.}},
\bauthor{\bsnm{{Prochaska}}, \binits{J.X.}},
\bauthor{\bsnm{{O'Meara}}, \binits{J.M.}},
\bauthor{\bsnm{{Becker}}, \binits{G.D.}},
\bauthor{\bsnm{{Ellison}}, \binits{S.L.}},
\bauthor{\bsnm{{Lopez}}, \binits{S.}},
\bauthor{\bsnm{{Meiksin}}, \binits{A.}},
\bauthor{\bsnm{{M{\'e}nard}}, \binits{B.}},
\bauthor{\bsnm{{Murphy}}, \binits{M.T.}},
\bauthor{\bsnm{{Fumagalli}}, \binits{M.}}:
\batitle{{The Giant Gemini GMOS survey of z$_{em}$ > 4.4 quasars - I. Measuring
  the mean free path across cosmic time}}.
\bjtitle{\mnras}
\bvolume{445}(\bissue{2}),
\bfpage{1745}--\blpage{1760}
(\byear{2014})
{\href{https://arxiv.org/abs/1402.4154}{{arXiv:1402.4154}}}
{[astro-ph.CO]}.
\doiurl{10.1093/mnras/stu1827}
\end{barticle}
\endbibitem

\bibitem{1979rpa..book.....R}
\begin{bbook}
\bauthor{\bsnm{{Rybicki}}, \binits{G.B.}},
\bauthor{\bsnm{{Lightman}}, \binits{A.P.}}:
\bbtitle{{Radiative Processes in Astrophysics}},
(\byear{1979})
\end{bbook}
\endbibitem

\bibitem{2009RvMP...81.1405M}
\begin{barticle}
\bauthor{\bsnm{{Meiksin}}, \binits{A.A.}}:
\batitle{{The physics of the intergalactic medium}}.
\bjtitle{Reviews of Modern Physics}
\bvolume{81}(\bissue{4}),
\bfpage{1405}--\blpage{1469}
(\byear{2009})
{\href{https://arxiv.org/abs/0711.3358}{{arXiv:0711.3358}}}
{[astro-ph]}.
\doiurl{10.1103/RevModPhys.81.1405}
\end{barticle}
\endbibitem

\bibitem{2022arXiv220405362K}
\begin{barticle}
\bauthor{\bsnm{{Ku{\v{s}}mi{\'c}}}, \binits{S.}},
\bauthor{\bsnm{{Finlator}}, \binits{K.}},
\bauthor{\bsnm{{Keating}}, \binits{L.}},
\bauthor{\bsnm{{Huscher}}, \binits{E.}}:
\batitle{{Assuming Ionization Equilibrium and the Impact on the
  Ly{\ensuremath{\alpha}} Forest Power Spectrum during the End of Reionization
  at 8 {\ensuremath{\geq}} z {\ensuremath{\geq}} 5}}.
\bjtitle{\apj}
\bvolume{931}(\bissue{1}),
\bfpage{46}
(\byear{2022})
{\href{https://arxiv.org/abs/2204.05362}{{arXiv:2204.05362}}}
{[astro-ph.CO]}.
\doiurl{10.3847/1538-4357/ac66e3}
\end{barticle}
\endbibitem

\bibitem{1986PASP...98.1014S}
\begin{barticle}
\bauthor{\bsnm{{Shapiro}}, \binits{P.R.}}:
\batitle{{Cosmological H II regions and the photoionization of the
  intergalactic medium.}}
\bjtitle{\pasp}
\bvolume{98},
\bfpage{1014}--\blpage{1017}
(\byear{1986}).
\doiurl{10.1086/131865}
\end{barticle}
\endbibitem

\bibitem{1987ApJ...321L.107S}
\begin{barticle}
\bauthor{\bsnm{{Shapiro}}, \binits{P.R.}},
\bauthor{\bsnm{{Giroux}}, \binits{M.L.}}:
\batitle{{Cosmological H II Regions and the Photoionization of the
  Intergalactic Medium}}.
\bjtitle{\apjl}
\bvolume{321},
\bfpage{107}
(\byear{1987}).
\doiurl{10.1086/185015}
\end{barticle}
\endbibitem

\bibitem{2020A&A...641A...6P}
\begin{barticle}
\bauthor{\bsnm{{Planck Collaboration}}},
\bauthor{\bsnm{{Aghanim}}, \binits{N.}},
\bauthor{\bsnm{{Akrami}}, \binits{Y.}},
\bauthor{\bsnm{{Ashdown}}, \binits{M.}},
\bauthor{\bsnm{{Aumont}}, \binits{J.}},
\bauthor{\bsnm{{Baccigalupi}}, \binits{C.}},
\bauthor{\bsnm{{Ballardini}}, \binits{M.}},
\bauthor{\bsnm{{Banday}}, \binits{A.J.}},
\bauthor{\bsnm{{Barreiro}}, \binits{R.B.}},
\bauthor{\bsnm{{Bartolo}}, \binits{N.}},
\bauthor{\bsnm{{Basak}}, \binits{S.}},
\bauthor{\bsnm{{Battye}}, \binits{R.}},
\bauthor{\bsnm{{Benabed}}, \binits{K.}},
\bauthor{\bsnm{{Bernard}}, \binits{J.-P.}},
\bauthor{\bsnm{{Bersanelli}}, \binits{M.}},
\bauthor{\bsnm{{Bielewicz}}, \binits{P.}},
\bauthor{\bsnm{{Bock}}, \binits{J.J.}},
\bauthor{\bsnm{{Bond}}, \binits{J.R.}},
\bauthor{\bsnm{{Borrill}}, \binits{J.}},
\bauthor{\bsnm{{Bouchet}}, \binits{F.R.}},
\bauthor{\bsnm{{Boulanger}}, \binits{F.}},
\bauthor{\bsnm{{Bucher}}, \binits{M.}},
\bauthor{\bsnm{{Burigana}}, \binits{C.}},
\bauthor{\bsnm{{Butler}}, \binits{R.C.}},
\bauthor{\bsnm{{Calabrese}}, \binits{E.}},
\bauthor{\bsnm{{Cardoso}}, \binits{J.-F.}},
\bauthor{\bsnm{{Carron}}, \binits{J.}},
\bauthor{\bsnm{{Challinor}}, \binits{A.}},
\bauthor{\bsnm{{Chiang}}, \binits{H.C.}},
\bauthor{\bsnm{{Chluba}}, \binits{J.}},
\bauthor{\bsnm{{Colombo}}, \binits{L.P.L.}},
\bauthor{\bsnm{{Combet}}, \binits{C.}},
\bauthor{\bsnm{{Contreras}}, \binits{D.}},
\bauthor{\bsnm{{Crill}}, \binits{B.P.}},
\bauthor{\bsnm{{Cuttaia}}, \binits{F.}},
\bauthor{\bsnm{{de Bernardis}}, \binits{P.}},
\bauthor{\bsnm{{de Zotti}}, \binits{G.}},
\bauthor{\bsnm{{Delabrouille}}, \binits{J.}},
\bauthor{\bsnm{{Delouis}}, \binits{J.-M.}},
\bauthor{\bsnm{{Di Valentino}}, \binits{E.}},
\bauthor{\bsnm{{Diego}}, \binits{J.M.}},
\bauthor{\bsnm{{Dor{\'e}}}, \binits{O.}},
\bauthor{\bsnm{{Douspis}}, \binits{M.}},
\bauthor{\bsnm{{Ducout}}, \binits{A.}},
\bauthor{\bsnm{{Dupac}}, \binits{X.}},
\bauthor{\bsnm{{Dusini}}, \binits{S.}},
\bauthor{\bsnm{{Efstathiou}}, \binits{G.}},
\bauthor{\bsnm{{Elsner}}, \binits{F.}},
\bauthor{\bsnm{{En{\ss}lin}}, \binits{T.A.}},
\bauthor{\bsnm{{Eriksen}}, \binits{H.K.}},
\bauthor{\bsnm{{Fantaye}}, \binits{Y.}},
\bauthor{\bsnm{{Farhang}}, \binits{M.}},
\bauthor{\bsnm{{Fergusson}}, \binits{J.}},
\bauthor{\bsnm{{Fernandez-Cobos}}, \binits{R.}},
\bauthor{\bsnm{{Finelli}}, \binits{F.}},
\bauthor{\bsnm{{Forastieri}}, \binits{F.}},
\bauthor{\bsnm{{Frailis}}, \binits{M.}},
\bauthor{\bsnm{{Fraisse}}, \binits{A.A.}},
\bauthor{\bsnm{{Franceschi}}, \binits{E.}},
\bauthor{\bsnm{{Frolov}}, \binits{A.}},
\bauthor{\bsnm{{Galeotta}}, \binits{S.}},
\bauthor{\bsnm{{Galli}}, \binits{S.}},
\bauthor{\bsnm{{Ganga}}, \binits{K.}},
\bauthor{\bsnm{{G{\'e}nova-Santos}}, \binits{R.T.}},
\bauthor{\bsnm{{Gerbino}}, \binits{M.}},
\bauthor{\bsnm{{Ghosh}}, \binits{T.}},
\bauthor{\bsnm{{Gonz{\'a}lez-Nuevo}}, \binits{J.}},
\bauthor{\bsnm{{G{\'o}rski}}, \binits{K.M.}},
\bauthor{\bsnm{{Gratton}}, \binits{S.}},
\bauthor{\bsnm{{Gruppuso}}, \binits{A.}},
\bauthor{\bsnm{{Gudmundsson}}, \binits{J.E.}},
\bauthor{\bsnm{{Hamann}}, \binits{J.}},
\bauthor{\bsnm{{Handley}}, \binits{W.}},
\bauthor{\bsnm{{Hansen}}, \binits{F.K.}},
\bauthor{\bsnm{{Herranz}}, \binits{D.}},
\bauthor{\bsnm{{Hildebrandt}}, \binits{S.R.}},
\bauthor{\bsnm{{Hivon}}, \binits{E.}},
\bauthor{\bsnm{{Huang}}, \binits{Z.}},
\bauthor{\bsnm{{Jaffe}}, \binits{A.H.}},
\bauthor{\bsnm{{Jones}}, \binits{W.C.}},
\bauthor{\bsnm{{Karakci}}, \binits{A.}},
\bauthor{\bsnm{{Keih{\"a}nen}}, \binits{E.}},
\bauthor{\bsnm{{Keskitalo}}, \binits{R.}},
\bauthor{\bsnm{{Kiiveri}}, \binits{K.}},
\bauthor{\bsnm{{Kim}}, \binits{J.}},
\bauthor{\bsnm{{Kisner}}, \binits{T.S.}},
\bauthor{\bsnm{{Knox}}, \binits{L.}},
\bauthor{\bsnm{{Krachmalnicoff}}, \binits{N.}},
\bauthor{\bsnm{{Kunz}}, \binits{M.}},
\bauthor{\bsnm{{Kurki-Suonio}}, \binits{H.}},
\bauthor{\bsnm{{Lagache}}, \binits{G.}},
\bauthor{\bsnm{{Lamarre}}, \binits{J.-M.}},
\bauthor{\bsnm{{Lasenby}}, \binits{A.}},
\bauthor{\bsnm{{Lattanzi}}, \binits{M.}},
\bauthor{\bsnm{{Lawrence}}, \binits{C.R.}},
\bauthor{\bsnm{{Le Jeune}}, \binits{M.}},
\bauthor{\bsnm{{Lemos}}, \binits{P.}},
\bauthor{\bsnm{{Lesgourgues}}, \binits{J.}},
\bauthor{\bsnm{{Levrier}}, \binits{F.}},
\bauthor{\bsnm{{Lewis}}, \binits{A.}},
\bauthor{\bsnm{{Liguori}}, \binits{M.}},
\bauthor{\bsnm{{Lilje}}, \binits{P.B.}},
\bauthor{\bsnm{{Lilley}}, \binits{M.}},
\bauthor{\bsnm{{Lindholm}}, \binits{V.}},
\bauthor{\bsnm{{L{\'o}pez-Caniego}}, \binits{M.}},
\bauthor{\bsnm{{Lubin}}, \binits{P.M.}},
\bauthor{\bsnm{{Ma}}, \binits{Y.-Z.}},
\bauthor{\bsnm{{Mac{\'\i}as-P{\'e}rez}}, \binits{J.F.}},
\bauthor{\bsnm{{Maggio}}, \binits{G.}},
\bauthor{\bsnm{{Maino}}, \binits{D.}},
\bauthor{\bsnm{{Mandolesi}}, \binits{N.}},
\bauthor{\bsnm{{Mangilli}}, \binits{A.}},
\bauthor{\bsnm{{Marcos-Caballero}}, \binits{A.}},
\bauthor{\bsnm{{Maris}}, \binits{M.}},
\bauthor{\bsnm{{Martin}}, \binits{P.G.}},
\bauthor{\bsnm{{Martinelli}}, \binits{M.}},
\bauthor{\bsnm{{Mart{\'\i}nez-Gonz{\'a}lez}}, \binits{E.}},
\bauthor{\bsnm{{Matarrese}}, \binits{S.}},
\bauthor{\bsnm{{Mauri}}, \binits{N.}},
\bauthor{\bsnm{{McEwen}}, \binits{J.D.}},
\bauthor{\bsnm{{Meinhold}}, \binits{P.R.}},
\bauthor{\bsnm{{Melchiorri}}, \binits{A.}},
\bauthor{\bsnm{{Mennella}}, \binits{A.}},
\bauthor{\bsnm{{Migliaccio}}, \binits{M.}},
\bauthor{\bsnm{{Millea}}, \binits{M.}},
\bauthor{\bsnm{{Mitra}}, \binits{S.}},
\bauthor{\bsnm{{Miville-Desch{\^e}nes}}, \binits{M.-A.}},
\bauthor{\bsnm{{Molinari}}, \binits{D.}},
\bauthor{\bsnm{{Montier}}, \binits{L.}},
\bauthor{\bsnm{{Morgante}}, \binits{G.}},
\bauthor{\bsnm{{Moss}}, \binits{A.}},
\bauthor{\bsnm{{Natoli}}, \binits{P.}},
\bauthor{\bsnm{{N{\o}rgaard-Nielsen}}, \binits{H.U.}},
\bauthor{\bsnm{{Pagano}}, \binits{L.}},
\bauthor{\bsnm{{Paoletti}}, \binits{D.}},
\bauthor{\bsnm{{Partridge}}, \binits{B.}},
\bauthor{\bsnm{{Patanchon}}, \binits{G.}},
\bauthor{\bsnm{{Peiris}}, \binits{H.V.}},
\bauthor{\bsnm{{Perrotta}}, \binits{F.}},
\bauthor{\bsnm{{Pettorino}}, \binits{V.}},
\bauthor{\bsnm{{Piacentini}}, \binits{F.}},
\bauthor{\bsnm{{Polastri}}, \binits{L.}},
\bauthor{\bsnm{{Polenta}}, \binits{G.}},
\bauthor{\bsnm{{Puget}}, \binits{J.-L.}},
\bauthor{\bsnm{{Rachen}}, \binits{J.P.}},
\bauthor{\bsnm{{Reinecke}}, \binits{M.}},
\bauthor{\bsnm{{Remazeilles}}, \binits{M.}},
\bauthor{\bsnm{{Renzi}}, \binits{A.}},
\bauthor{\bsnm{{Rocha}}, \binits{G.}},
\bauthor{\bsnm{{Rosset}}, \binits{C.}},
\bauthor{\bsnm{{Roudier}}, \binits{G.}},
\bauthor{\bsnm{{Rubi{\~n}o-Mart{\'\i}n}}, \binits{J.A.}},
\bauthor{\bsnm{{Ruiz-Granados}}, \binits{B.}},
\bauthor{\bsnm{{Salvati}}, \binits{L.}},
\bauthor{\bsnm{{Sandri}}, \binits{M.}},
\bauthor{\bsnm{{Savelainen}}, \binits{M.}},
\bauthor{\bsnm{{Scott}}, \binits{D.}},
\bauthor{\bsnm{{Shellard}}, \binits{E.P.S.}},
\bauthor{\bsnm{{Sirignano}}, \binits{C.}},
\bauthor{\bsnm{{Sirri}}, \binits{G.}},
\bauthor{\bsnm{{Spencer}}, \binits{L.D.}},
\bauthor{\bsnm{{Sunyaev}}, \binits{R.}},
\bauthor{\bsnm{{Suur-Uski}}, \binits{A.-S.}},
\bauthor{\bsnm{{Tauber}}, \binits{J.A.}},
\bauthor{\bsnm{{Tavagnacco}}, \binits{D.}},
\bauthor{\bsnm{{Tenti}}, \binits{M.}},
\bauthor{\bsnm{{Toffolatti}}, \binits{L.}},
\bauthor{\bsnm{{Tomasi}}, \binits{M.}},
\bauthor{\bsnm{{Trombetti}}, \binits{T.}},
\bauthor{\bsnm{{Valenziano}}, \binits{L.}},
\bauthor{\bsnm{{Valiviita}}, \binits{J.}},
\bauthor{\bsnm{{Van Tent}}, \binits{B.}},
\bauthor{\bsnm{{Vibert}}, \binits{L.}},
\bauthor{\bsnm{{Vielva}}, \binits{P.}},
\bauthor{\bsnm{{Villa}}, \binits{F.}},
\bauthor{\bsnm{{Vittorio}}, \binits{N.}},
\bauthor{\bsnm{{Wandelt}}, \binits{B.D.}},
\bauthor{\bsnm{{Wehus}}, \binits{I.K.}},
\bauthor{\bsnm{{White}}, \binits{M.}},
\bauthor{\bsnm{{White}}, \binits{S.D.M.}},
\bauthor{\bsnm{{Zacchei}}, \binits{A.}},
\bauthor{\bsnm{{Zonca}}, \binits{A.}}:
\batitle{{Planck 2018 results. VI. Cosmological parameters}}.
\bjtitle{\aap}
\bvolume{641},
\bfpage{6}
(\byear{2020})
{\href{https://arxiv.org/abs/1807.06209}{{arXiv:1807.06209}}}
{[astro-ph.CO]}.
\doiurl{10.1051/0004-6361/201833910}
\end{barticle}
\endbibitem

\bibitem{1965ApJ...142.1633G}
\begin{barticle}
\bauthor{\bsnm{{Gunn}}, \binits{J.E.}},
\bauthor{\bsnm{{Peterson}}, \binits{B.A.}}:
\batitle{{On the Density of Neutral Hydrogen in Intergalactic Space.}}
\bjtitle{\apj}
\bvolume{142},
\bfpage{1633}--\blpage{1636}
(\byear{1965}).
\doiurl{10.1086/148444}
\end{barticle}
\endbibitem

\bibitem{1998ARA&A..36..267R}
\begin{barticle}
\bauthor{\bsnm{{Rauch}}, \binits{M.}}:
\batitle{{The Lyman Alpha Forest in the Spectra of QSOs}}.
\bjtitle{\araa}
\bvolume{36},
\bfpage{267}--\blpage{316}
(\byear{1998})
{\href{https://arxiv.org/abs/astro-ph/9806286}{{arXiv:astro-ph/9806286}}}
{[astro-ph]}.
\doiurl{10.1146/annurev.astro.36.1.267}
\end{barticle}
\endbibitem

\bibitem{1997MNRAS.292...27H}
\begin{barticle}
\bauthor{\bsnm{{Hui}}, \binits{L.}},
\bauthor{\bsnm{{Gnedin}}, \binits{N.Y.}}:
\batitle{{Equation of state of the photoionized intergalactic medium}}.
\bjtitle{\mnras}
\bvolume{292}(\bissue{1}),
\bfpage{27}--\blpage{42}
(\byear{1997})
{\href{https://arxiv.org/abs/astro-ph/9612232}{{arXiv:astro-ph/9612232}}}
{[astro-ph]}.
\doiurl{10.1093/mnras/292.1.27}
\end{barticle}
\endbibitem

\bibitem{2001MNRAS.322..561C}
\begin{barticle}
\bauthor{\bsnm{{Choudhury}}, \binits{T.R.}},
\bauthor{\bsnm{{Padmanabhan}}, \binits{T.}},
\bauthor{\bsnm{{Srianand}}, \binits{R.}}:
\batitle{{Semi-analytic approach to understanding the distribution of neutral
  hydrogen in the Universe}}.
\bjtitle{\mnras}
\bvolume{322}(\bissue{3}),
\bfpage{561}--\blpage{575}
(\byear{2001})
{\href{https://arxiv.org/abs/astro-ph/0005252}{{arXiv:astro-ph/0005252}}}
{[astro-ph]}.
\doiurl{10.1046/j.1365-8711.2001.04108.x}
\end{barticle}
\endbibitem

\bibitem{2001ApJ...559...29C}
\begin{barticle}
\bauthor{\bsnm{{Choudhury}}, \binits{T.R.}},
\bauthor{\bsnm{{Srianand}}, \binits{R.}},
\bauthor{\bsnm{{Padmanabhan}}, \binits{T.}}:
\batitle{{Semianalytic Approach to Understanding the Distribution of Neutral
  Hydrogen in the Universe: Comparison of Simulations with Observations}}.
\bjtitle{\apj}
\bvolume{559}(\bissue{1}),
\bfpage{29}--\blpage{40}
(\byear{2001})
{\href{https://arxiv.org/abs/astro-ph/0012498}{{arXiv:astro-ph/0012498}}}
{[astro-ph]}.
\doiurl{10.1086/322327}
\end{barticle}
\endbibitem

\bibitem{2018ApJ...864...53E}
\begin{barticle}
\bauthor{\bsnm{{Eilers}}, \binits{A.-C.}},
\bauthor{\bsnm{{Davies}}, \binits{F.B.}},
\bauthor{\bsnm{{Hennawi}}, \binits{J.F.}}:
\batitle{{The Opacity of the Intergalactic Medium Measured along Quasar
  Sightlines at z {\ensuremath{\sim}} 6}}.
\bjtitle{\apj}
\bvolume{864}(\bissue{1}),
\bfpage{53}
(\byear{2018})
{\href{https://arxiv.org/abs/1807.04229}{{arXiv:1807.04229}}}
{[astro-ph.GA]}.
\doiurl{10.3847/1538-4357/aad4fd}
\end{barticle}
\endbibitem

\bibitem{2020ApJ...904...26Y}
\begin{barticle}
\bauthor{\bsnm{{Yang}}, \binits{J.}},
\bauthor{\bsnm{{Wang}}, \binits{F.}},
\bauthor{\bsnm{{Fan}}, \binits{X.}},
\bauthor{\bsnm{{Hennawi}}, \binits{J.F.}},
\bauthor{\bsnm{{Davies}}, \binits{F.B.}},
\bauthor{\bsnm{{Yue}}, \binits{M.}},
\bauthor{\bsnm{{Eilers}}, \binits{A.-C.}},
\bauthor{\bsnm{{Farina}}, \binits{E.P.}},
\bauthor{\bsnm{{Wu}}, \binits{X.-B.}},
\bauthor{\bsnm{{Bian}}, \binits{F.}},
\bauthor{\bsnm{{Pacucci}}, \binits{F.}},
\bauthor{\bsnm{{Lee}}, \binits{K.-G.}}:
\batitle{{Measurements of the z {\ensuremath{\sim}} 6 Intergalactic Medium
  Optical Depth and Transmission Spikes Using a New z > 6.3 Quasar Sample}}.
\bjtitle{\apj}
\bvolume{904}(\bissue{1}),
\bfpage{26}
(\byear{2020})
{\href{https://arxiv.org/abs/2009.13544}{{arXiv:2009.13544}}}
{[astro-ph.GA]}.
\doiurl{10.3847/1538-4357/abbc1b}
\end{barticle}
\endbibitem

\bibitem{2021arXiv210803699B}
\begin{barticle}
\bauthor{\bsnm{{Bosman}}, \binits{S.E.I.}},
\bauthor{\bsnm{{Davies}}, \binits{F.B.}},
\bauthor{\bsnm{{Becker}}, \binits{G.D.}},
\bauthor{\bsnm{{Keating}}, \binits{L.C.}},
\bauthor{\bsnm{{Davies}}, \binits{R.L.}},
\bauthor{\bsnm{{Zhu}}, \binits{Y.}},
\bauthor{\bsnm{{Eilers}}, \binits{A.-C.}},
\bauthor{\bsnm{{D'Odorico}}, \binits{V.}},
\bauthor{\bsnm{{Bian}}, \binits{F.}},
\bauthor{\bsnm{{Bischetti}}, \binits{M.}},
\bauthor{\bsnm{{Cristiani}}, \binits{S.V.}},
\bauthor{\bsnm{{Fan}}, \binits{X.}},
\bauthor{\bsnm{{Farina}}, \binits{E.P.}},
\bauthor{\bsnm{{Haehnelt}}, \binits{M.G.}},
\bauthor{\bsnm{{Hennawi}}, \binits{J.F.}},
\bauthor{\bsnm{{Kulkarni}}, \binits{G.}},
\bauthor{\bsnm{{Mesinger}}, \binits{A.}},
\bauthor{\bsnm{{Meyer}}, \binits{R.A.}},
\bauthor{\bsnm{{Onoue}}, \binits{M.}},
\bauthor{\bsnm{{Pallottini}}, \binits{A.}},
\bauthor{\bsnm{{Qin}}, \binits{Y.}},
\bauthor{\bsnm{{Ryan-Weber}}, \binits{E.}},
\bauthor{\bsnm{{Schindler}}, \binits{J.-T.}},
\bauthor{\bsnm{{Walter}}, \binits{F.}},
\bauthor{\bsnm{{Wang}}, \binits{F.}},
\bauthor{\bsnm{{Yang}}, \binits{J.}}:
\batitle{{Hydrogen reionization ends by z = 5.3: Lyman-{\ensuremath{\alpha}}
  optical depth measured by the XQR-30 sample}}.
\bjtitle{\mnras}
\bvolume{514}(\bissue{1}),
\bfpage{55}--\blpage{76}
(\byear{2022})
{\href{https://arxiv.org/abs/2108.03699}{{arXiv:2108.03699}}}
{[astro-ph.CO]}.
\doiurl{10.1093/mnras/stac1046}
\end{barticle}
\endbibitem

\bibitem{2005MNRAS.361..577C}
\begin{barticle}
\bauthor{\bsnm{{Choudhury}}, \binits{T.R.}},
\bauthor{\bsnm{{Ferrara}}, \binits{A.}}:
\batitle{{Experimental constraints on self-consistent reionization models}}.
\bjtitle{\mnras}
\bvolume{361}(\bissue{2}),
\bfpage{577}--\blpage{594}
(\byear{2005})
{\href{https://arxiv.org/abs/astro-ph/0411027}{{arXiv:astro-ph/0411027}}}
{[astro-ph]}.
\doiurl{10.1111/j.1365-2966.2005.09196.x}
\end{barticle}
\endbibitem

\bibitem{2000ApJ...530....1M}
\begin{barticle}
\bauthor{\bsnm{{Miralda-Escud{\'e}}}, \binits{J.}},
\bauthor{\bsnm{{Haehnelt}}, \binits{M.}},
\bauthor{\bsnm{{Rees}}, \binits{M.J.}}:
\batitle{{Reionization of the Inhomogeneous Universe}}.
\bjtitle{\apj}
\bvolume{530}(\bissue{1}),
\bfpage{1}--\blpage{16}
(\byear{2000})
{\href{https://arxiv.org/abs/astro-ph/9812306}{{arXiv:astro-ph/9812306}}}
{[astro-ph]}.
\doiurl{10.1086/308330}
\end{barticle}
\endbibitem

\bibitem{2021MNRAS.508.1853B}
\begin{barticle}
\bauthor{\bsnm{{Becker}}, \binits{G.D.}},
\bauthor{\bsnm{{D'Aloisio}}, \binits{A.}},
\bauthor{\bsnm{{Christenson}}, \binits{H.M.}},
\bauthor{\bsnm{{Zhu}}, \binits{Y.}},
\bauthor{\bsnm{{Worseck}}, \binits{G.}},
\bauthor{\bsnm{{Bolton}}, \binits{J.S.}}:
\batitle{{The mean free path of ionizing photons at 5 < z < 6: evidence for
  rapid evolution near reionization}}.
\bjtitle{\mnras}
\bvolume{508}(\bissue{2}),
\bfpage{1853}--\blpage{1869}
(\byear{2021})
{\href{https://arxiv.org/abs/2103.16610}{{arXiv:2103.16610}}}
{[astro-ph.CO]}.
\doiurl{10.1093/mnras/stab2696}
\end{barticle}
\endbibitem

\bibitem{2015MNRAS.447..499M}
\begin{barticle}
\bauthor{\bsnm{{McGreer}}, \binits{I.D.}},
\bauthor{\bsnm{{Mesinger}}, \binits{A.}},
\bauthor{\bsnm{{D'Odorico}}, \binits{V.}}:
\batitle{{Model-independent evidence in favour of an end to reionization by z
  {\ensuremath{\approx}} 6}}.
\bjtitle{\mnras}
\bvolume{447}(\bissue{1}),
\bfpage{499}--\blpage{505}
(\byear{2015})
{\href{https://arxiv.org/abs/1411.5375}{{arXiv:1411.5375}}}
{[astro-ph.CO]}.
\doiurl{10.1093/mnras/stu2449}
\end{barticle}
\endbibitem

\bibitem{2011MNRAS.415.3237M}
\begin{barticle}
\bauthor{\bsnm{{McGreer}}, \binits{I.D.}},
\bauthor{\bsnm{{Mesinger}}, \binits{A.}},
\bauthor{\bsnm{{Fan}}, \binits{X.}}:
\batitle{{The first (nearly) model-independent constraint on the neutral
  hydrogen fraction at z {\ensuremath{\sim}} 6}}.
\bjtitle{\mnras}
\bvolume{415}(\bissue{4}),
\bfpage{3237}--\blpage{3246}
(\byear{2011})
{\href{https://arxiv.org/abs/1101.3314}{{arXiv:1101.3314}}}
{[astro-ph.CO]}.
\doiurl{10.1111/j.1365-2966.2011.18935.x}
\end{barticle}
\endbibitem

\bibitem{2013ApJ...768..196S}
\begin{barticle}
\bauthor{\bsnm{{Schenker}}, \binits{M.A.}},
\bauthor{\bsnm{{Robertson}}, \binits{B.E.}},
\bauthor{\bsnm{{Ellis}}, \binits{R.S.}},
\bauthor{\bsnm{{Ono}}, \binits{Y.}},
\bauthor{\bsnm{{McLure}}, \binits{R.J.}},
\bauthor{\bsnm{{Dunlop}}, \binits{J.S.}},
\bauthor{\bsnm{{Koekemoer}}, \binits{A.}},
\bauthor{\bsnm{{Bowler}}, \binits{R.A.A.}},
\bauthor{\bsnm{{Ouchi}}, \binits{M.}},
\bauthor{\bsnm{{Curtis-Lake}}, \binits{E.}},
\bauthor{\bsnm{{Rogers}}, \binits{A.B.}},
\bauthor{\bsnm{{Schneider}}, \binits{E.}},
\bauthor{\bsnm{{Charlot}}, \binits{S.}},
\bauthor{\bsnm{{Stark}}, \binits{D.P.}},
\bauthor{\bsnm{{Furlanetto}}, \binits{S.R.}},
\bauthor{\bsnm{{Cirasuolo}}, \binits{M.}}:
\batitle{{The UV Luminosity Function of Star-forming Galaxies via Dropout
  Selection at Redshifts z \raisebox{-0.5ex}\textasciitilde 7 and 8 from the
  2012 Ultra Deep Field Campaign}}.
\bjtitle{\apj}
\bvolume{768}(\bissue{2}),
\bfpage{196}
(\byear{2013})
{\href{https://arxiv.org/abs/1212.4819}{{arXiv:1212.4819}}}
{[astro-ph.CO]}.
\doiurl{10.1088/0004-637X/768/2/196}
\end{barticle}
\endbibitem

\bibitem{2013AJ....145....4W}
\begin{barticle}
\bauthor{\bsnm{{Willott}}, \binits{C.J.}},
\bauthor{\bsnm{{McLure}}, \binits{R.J.}},
\bauthor{\bsnm{{Hibon}}, \binits{P.}},
\bauthor{\bsnm{{Bielby}}, \binits{R.}},
\bauthor{\bsnm{{McCracken}}, \binits{H.J.}},
\bauthor{\bsnm{{Kneib}}, \binits{J.-P.}},
\bauthor{\bsnm{{Ilbert}}, \binits{O.}},
\bauthor{\bsnm{{Bonfield}}, \binits{D.G.}},
\bauthor{\bsnm{{Bruce}}, \binits{V.A.}},
\bauthor{\bsnm{{Jarvis}}, \binits{M.J.}}:
\batitle{{An Exponential Decline at the Bright End of the z = 6 Galaxy
  Luminosity Function}}.
\bjtitle{\aj}
\bvolume{145}(\bissue{1}),
\bfpage{4}
(\byear{2013})
{\href{https://arxiv.org/abs/1202.5330}{{arXiv:1202.5330}}}
{[astro-ph.CO]}.
\doiurl{10.1088/0004-6256/145/1/4}
\end{barticle}
\endbibitem

\bibitem{2015ApJ...803...34B}
\begin{barticle}
\bauthor{\bsnm{{Bouwens}}, \binits{R.J.}},
\bauthor{\bsnm{{Illingworth}}, \binits{G.D.}},
\bauthor{\bsnm{{Oesch}}, \binits{P.A.}},
\bauthor{\bsnm{{Trenti}}, \binits{M.}},
\bauthor{\bsnm{{Labb{\'e}}}, \binits{I.}},
\bauthor{\bsnm{{Bradley}}, \binits{L.}},
\bauthor{\bsnm{{Carollo}}, \binits{M.}},
\bauthor{\bsnm{{van Dokkum}}, \binits{P.G.}},
\bauthor{\bsnm{{Gonzalez}}, \binits{V.}},
\bauthor{\bsnm{{Holwerda}}, \binits{B.}},
\bauthor{\bsnm{{Franx}}, \binits{M.}},
\bauthor{\bsnm{{Spitler}}, \binits{L.}},
\bauthor{\bsnm{{Smit}}, \binits{R.}},
\bauthor{\bsnm{{Magee}}, \binits{D.}}:
\batitle{{UV Luminosity Functions at Redshifts z {\ensuremath{\sim}} 4 to z
  {\ensuremath{\sim}} 10: 10,000 Galaxies from HST Legacy Fields}}.
\bjtitle{\apj}
\bvolume{803}(\bissue{1}),
\bfpage{34}
(\byear{2015})
{\href{https://arxiv.org/abs/1403.4295}{{arXiv:1403.4295}}}
{[astro-ph.CO]}.
\doiurl{10.1088/0004-637X/803/1/34}
\end{barticle}
\endbibitem

\bibitem{2015ApJ...800...18A}
\begin{barticle}
\bauthor{\bsnm{{Atek}}, \binits{H.}},
\bauthor{\bsnm{{Richard}}, \binits{J.}},
\bauthor{\bsnm{{Kneib}}, \binits{J.-P.}},
\bauthor{\bsnm{{Jauzac}}, \binits{M.}},
\bauthor{\bsnm{{Schaerer}}, \binits{D.}},
\bauthor{\bsnm{{Clement}}, \binits{B.}},
\bauthor{\bsnm{{Limousin}}, \binits{M.}},
\bauthor{\bsnm{{Jullo}}, \binits{E.}},
\bauthor{\bsnm{{Natarajan}}, \binits{P.}},
\bauthor{\bsnm{{Egami}}, \binits{E.}},
\bauthor{\bsnm{{Ebeling}}, \binits{H.}}:
\batitle{{New Constraints on the Faint End of the UV Luminosity Function at z
  \raisebox{-0.5ex}\textasciitilde 7-8 Using the Gravitational Lensing of the
  Hubble Frontier Fields Cluster A2744}}.
\bjtitle{\apj}
\bvolume{800}(\bissue{1}),
\bfpage{18}
(\byear{2015})
{\href{https://arxiv.org/abs/1409.0512}{{arXiv:1409.0512}}}
{[astro-ph.GA]}.
\doiurl{10.1088/0004-637X/800/1/18}
\end{barticle}
\endbibitem

\bibitem{2017ApJ...843..129B}
\begin{barticle}
\bauthor{\bsnm{{Bouwens}}, \binits{R.J.}},
\bauthor{\bsnm{{Oesch}}, \binits{P.A.}},
\bauthor{\bsnm{{Illingworth}}, \binits{G.D.}},
\bauthor{\bsnm{{Ellis}}, \binits{R.S.}},
\bauthor{\bsnm{{Stefanon}}, \binits{M.}}:
\batitle{{The z {\ensuremath{\sim}} 6 Luminosity Function Fainter than -15 mag
  from the Hubble Frontier Fields: The Impact of Magnification Uncertainties}}.
\bjtitle{\apj}
\bvolume{843}(\bissue{2}),
\bfpage{129}
(\byear{2017})
{\href{https://arxiv.org/abs/1610.00283}{{arXiv:1610.00283}}}
{[astro-ph.GA]}.
\doiurl{10.3847/1538-4357/aa70a4}
\end{barticle}
\endbibitem

\bibitem{2018MNRAS.479.5184A}
\begin{barticle}
\bauthor{\bsnm{{Atek}}, \binits{H.}},
\bauthor{\bsnm{{Richard}}, \binits{J.}},
\bauthor{\bsnm{{Kneib}}, \binits{J.-P.}},
\bauthor{\bsnm{{Schaerer}}, \binits{D.}}:
\batitle{{The extreme faint end of the UV luminosity function at z
  {\ensuremath{\sim}} 6 through gravitational telescopes: a comprehensive
  assessment of strong lensing uncertainties}}.
\bjtitle{\mnras}
\bvolume{479}(\bissue{4}),
\bfpage{5184}--\blpage{5195}
(\byear{2018})
{\href{https://arxiv.org/abs/1803.09747}{{arXiv:1803.09747}}}
{[astro-ph.GA]}.
\doiurl{10.1093/mnras/sty1820}
\end{barticle}
\endbibitem

\bibitem{2019MNRAS.484..933P}
\begin{barticle}
\bauthor{\bsnm{{Park}}, \binits{J.}},
\bauthor{\bsnm{{Mesinger}}, \binits{A.}},
\bauthor{\bsnm{{Greig}}, \binits{B.}},
\bauthor{\bsnm{{Gillet}}, \binits{N.}}:
\batitle{{Inferring the astrophysics of reionization and cosmic dawn from
  galaxy luminosity functions and the 21-cm signal}}.
\bjtitle{\mnras}
\bvolume{484}(\bissue{1}),
\bfpage{933}--\blpage{949}
(\byear{2019})
{\href{https://arxiv.org/abs/1809.08995}{{arXiv:1809.08995}}}
{[astro-ph.GA]}.
\doiurl{10.1093/mnras/stz032}
\end{barticle}
\endbibitem

\bibitem{2021MNRAS.506.2390Q}
\begin{barticle}
\bauthor{\bsnm{{Qin}}, \binits{Y.}},
\bauthor{\bsnm{{Mesinger}}, \binits{A.}},
\bauthor{\bsnm{{Bosman}}, \binits{S.E.I.}},
\bauthor{\bsnm{{Viel}}, \binits{M.}}:
\batitle{{Reionization and galaxy inference from the high-redshift Ly
  {\ensuremath{\alpha}} forest}}.
\bjtitle{\mnras}
\bvolume{506}(\bissue{2}),
\bfpage{2390}--\blpage{2407}
(\byear{2021})
{\href{https://arxiv.org/abs/2101.09033}{{arXiv:2101.09033}}}
{[astro-ph.CO]}.
\doiurl{10.1093/mnras/stab1833}
\end{barticle}
\endbibitem

\bibitem{2022arXiv220405268M}
\begin{barticle}
\bauthor{\bsnm{{Maity}}, \binits{B.}},
\bauthor{\bsnm{{Choudhury}}, \binits{T.R.}}:
\batitle{{Constraining the reionization and thermal history of the Universe
  using a seminumerical photon-conserving code SCRIPT}}.
\bjtitle{\mnras}
\bvolume{515}(\bissue{1}),
\bfpage{617}--\blpage{630}
(\byear{2022})
{\href{https://arxiv.org/abs/2204.05268}{{arXiv:2204.05268}}}
{[astro-ph.CO]}.
\doiurl{10.1093/mnras/stac1847}
\end{barticle}
\endbibitem

\bibitem{2006PhR...433..181F}
\begin{barticle}
\bauthor{\bsnm{{Furlanetto}}, \binits{S.R.}},
\bauthor{\bsnm{{Oh}}, \binits{S.P.}},
\bauthor{\bsnm{{Briggs}}, \binits{F.H.}}:
\batitle{{Cosmology at low frequencies: The 21 cm transition and the
  high-redshift Universe}}.
\bjtitle{\physrep}
\bvolume{433}(\bissue{4-6}),
\bfpage{181}--\blpage{301}
(\byear{2006})
{\href{https://arxiv.org/abs/astro-ph/0608032}{{arXiv:astro-ph/0608032}}}
{[astro-ph]}.
\doiurl{10.1016/j.physrep.2006.08.002}
\end{barticle}
\endbibitem

\bibitem{2012RPPh...75h6901P}
\begin{barticle}
\bauthor{\bsnm{{Pritchard}}, \binits{J.R.}},
\bauthor{\bsnm{{Loeb}}, \binits{A.}}:
\batitle{{21 cm cosmology in the 21st century}}.
\bjtitle{Reports on Progress in Physics}
\bvolume{75}(\bissue{8}),
\bfpage{086901}
(\byear{2012})
{\href{https://arxiv.org/abs/1109.6012}{{arXiv:1109.6012}}}
{[astro-ph.CO]}.
\doiurl{10.1088/0034-4885/75/8/086901}
\end{barticle}
\endbibitem

\bibitem{2021MNRAS.507.2405C}
\begin{barticle}
\bauthor{\bsnm{{Chatterjee}}, \binits{A.}},
\bauthor{\bsnm{{Choudhury}}, \binits{T.R.}},
\bauthor{\bsnm{{Mitra}}, \binits{S.}}:
\batitle{{CosmoReionMC: a package for estimating cosmological and astrophysical
  parameters using CMB, Lyman-{\ensuremath{\alpha}} absorption, and global 21
  cm data}}.
\bjtitle{\mnras}
\bvolume{507}(\bissue{2}),
\bfpage{2405}--\blpage{2422}
(\byear{2021})
{\href{https://arxiv.org/abs/2101.11088}{{arXiv:2101.11088}}}
{[astro-ph.CO]}.
\doiurl{10.1093/mnras/stab2316}
\end{barticle}
\endbibitem

\bibitem{2018ApJ...852..139V}
\begin{barticle}
\bauthor{\bsnm{{Villanueva-Domingo}}, \binits{P.}},
\bauthor{\bsnm{{Gnedin}}, \binits{N.Y.}},
\bauthor{\bsnm{{Mena}}, \binits{O.}}:
\batitle{{Warm Dark Matter and Cosmic Reionization}}.
\bjtitle{\apj}
\bvolume{852}(\bissue{2}),
\bfpage{139}
(\byear{2018})
{\href{https://arxiv.org/abs/1708.08277}{{arXiv:1708.08277}}}
{[astro-ph.CO]}.
\doiurl{10.3847/1538-4357/aa9ff5}
\end{barticle}
\endbibitem

\bibitem{2018MNRAS.477.2886L}
\begin{barticle}
\bauthor{\bsnm{{Lovell}}, \binits{M.R.}},
\bauthor{\bsnm{{Zavala}}, \binits{J.}},
\bauthor{\bsnm{{Vogelsberger}}, \binits{M.}},
\bauthor{\bsnm{{Shen}}, \binits{X.}},
\bauthor{\bsnm{{Cyr-Racine}}, \binits{F.-Y.}},
\bauthor{\bsnm{{Pfrommer}}, \binits{C.}},
\bauthor{\bsnm{{Sigurdson}}, \binits{K.}},
\bauthor{\bsnm{{Boylan-Kolchin}}, \binits{M.}},
\bauthor{\bsnm{{Pillepich}}, \binits{A.}}:
\batitle{{ETHOS - an effective theory of structure formation: predictions for
  the high-redshift Universe - abundance of galaxies and reionization}}.
\bjtitle{\mnras}
\bvolume{477}(\bissue{3}),
\bfpage{2886}--\blpage{2899}
(\byear{2018})
{\href{https://arxiv.org/abs/1711.10497}{{arXiv:1711.10497}}}
{[astro-ph.CO]}.
\doiurl{10.1093/mnras/sty818}
\end{barticle}
\endbibitem

\bibitem{2018PhRvD..98f3021S}
\begin{barticle}
\bauthor{\bsnm{{Schneider}}, \binits{A.}}:
\batitle{{Constraining noncold dark matter models with the global 21-cm
  signal}}.
\bjtitle{\prd}
\bvolume{98}(\bissue{6}),
\bfpage{063021}
(\byear{2018})
{\href{https://arxiv.org/abs/1805.00021}{{arXiv:1805.00021}}}
{[astro-ph.CO]}.
\doiurl{10.1103/PhysRevD.98.063021}
\end{barticle}
\endbibitem

\bibitem{2019JCAP...04..051N}
\begin{barticle}
\bauthor{\bsnm{{Nebrin}}, \binits{O.}},
\bauthor{\bsnm{{Ghara}}, \binits{R.}},
\bauthor{\bsnm{{Mellema}}, \binits{G.}}:
\batitle{{Fuzzy dark matter at cosmic dawn: new 21-cm constraints}}.
\bjtitle{\jcap}
\bvolume{2019}(\bissue{4}),
\bfpage{051}
(\byear{2019})
{\href{https://arxiv.org/abs/1812.09760}{{arXiv:1812.09760}}}
{[astro-ph.CO]}.
\doiurl{10.1088/1475-7516/2019/04/051}
\end{barticle}
\endbibitem

\bibitem{2019PhRvD..99b3518C}
\begin{barticle}
\bauthor{\bsnm{{Carucci}}, \binits{I.P.}},
\bauthor{\bsnm{{Corasaniti}}, \binits{P.-S.}}:
\batitle{{Cosmic reionization history and dark matter scenarios}}.
\bjtitle{\prd}
\bvolume{99}(\bissue{2}),
\bfpage{023518}
(\byear{2019})
{\href{https://arxiv.org/abs/1811.07904}{{arXiv:1811.07904}}}
{[astro-ph.CO]}.
\doiurl{10.1103/PhysRevD.99.023518}
\end{barticle}
\endbibitem

\bibitem{2019MNRAS.487.3560C}
\begin{barticle}
\bauthor{\bsnm{{Chatterjee}}, \binits{A.}},
\bauthor{\bsnm{{Dayal}}, \binits{P.}},
\bauthor{\bsnm{{Choudhury}}, \binits{T.R.}},
\bauthor{\bsnm{{Hutter}}, \binits{A.}}:
\batitle{{Ruling out 3 keV warm dark matter using 21 cm EDGES data}}.
\bjtitle{\mnras}
\bvolume{487}(\bissue{3}),
\bfpage{3560}--\blpage{3567}
(\byear{2019})
{\href{https://arxiv.org/abs/1902.09562}{{arXiv:1902.09562}}}
{[astro-ph.CO]}.
\doiurl{10.1093/mnras/stz1444}
\end{barticle}
\endbibitem

\bibitem{2020PhRvD.101f3526M}
\begin{barticle}
\bauthor{\bsnm{{Mu{\~n}oz}}, \binits{J.B.}},
\bauthor{\bsnm{{Dvorkin}}, \binits{C.}},
\bauthor{\bsnm{{Cyr-Racine}}, \binits{F.-Y.}}:
\batitle{{Probing the small-scale matter power spectrum with large-scale 21-cm
  data}}.
\bjtitle{\prd}
\bvolume{101}(\bissue{6}),
\bfpage{063526}
(\byear{2020})
{\href{https://arxiv.org/abs/1911.11144}{{arXiv:1911.11144}}}
{[astro-ph.CO]}.
\doiurl{10.1103/PhysRevD.101.063526}
\end{barticle}
\endbibitem

\bibitem{2020MNRAS.497.2941S}
\begin{barticle}
\bauthor{\bsnm{{Saxena}}, \binits{A.}},
\bauthor{\bsnm{{Majumdar}}, \binits{S.}},
\bauthor{\bsnm{{Kamran}}, \binits{M.}},
\bauthor{\bsnm{{Viel}}, \binits{M.}}:
\batitle{{Impact of dark matter models on the EoR 21-cm signal bispectrum}}.
\bjtitle{\mnras}
\bvolume{497}(\bissue{3}),
\bfpage{2941}--\blpage{2953}
(\byear{2020})
{\href{https://arxiv.org/abs/2004.04808}{{arXiv:2004.04808}}}
{[astro-ph.CO]}.
\doiurl{10.1093/mnras/staa1768}
\end{barticle}
\endbibitem

\bibitem{2015MNRAS.451.1692P}
\begin{barticle}
\bauthor{\bsnm{{Pandey}}, \binits{K.L.}},
\bauthor{\bsnm{{Choudhury}}, \binits{T.R.}},
\bauthor{\bsnm{{Sethi}}, \binits{S.K.}},
\bauthor{\bsnm{{Ferrara}}, \binits{A.}}:
\batitle{{Reionization constraints on primordial magnetic fields}}.
\bjtitle{\mnras}
\bvolume{451}(\bissue{2}),
\bfpage{1692}--\blpage{1700}
(\byear{2015})
{\href{https://arxiv.org/abs/1410.0368}{{arXiv:1410.0368}}}
{[astro-ph.CO]}.
\doiurl{10.1093/mnras/stv1055}
\end{barticle}
\endbibitem

\bibitem{1994MNRAS.266..227B}
\begin{barticle}
\bauthor{\bsnm{{Bagla}}, \binits{J.S.}},
\bauthor{\bsnm{{Padmanabhan}}, \binits{T.}}:
\batitle{{Nonlinear Evolution of Density Perturbations Using the Approximate
  Constancy of the Gravitational Potential}}.
\bjtitle{\mnras}
\bvolume{266},
\bfpage{227}
(\byear{1994})
{\href{https://arxiv.org/abs/gr-qc/9304021}{{arXiv:gr-qc/9304021}}}
{[gr-qc]}.
\doiurl{10.1093/mnras/266.1.227}
\end{barticle}
\endbibitem

\bibitem{1996ApJ...469..470B}
\begin{barticle}
\bauthor{\bsnm{{Bagla}}, \binits{J.S.}},
\bauthor{\bsnm{{Padmanabhan}}, \binits{T.}}:
\batitle{{A New Statistical Indicator to Study Nonlinear Gravitational
  Clustering and Structure Formation}}.
\bjtitle{\apj}
\bvolume{469},
\bfpage{470}
(\byear{1996})
{\href{https://arxiv.org/abs/astro-ph/9503121}{{arXiv:astro-ph/9503121}}}
{[astro-ph]}.
\doiurl{10.1086/177796}
\end{barticle}
\endbibitem

\bibitem{1997MNRAS.286.1023B}
\begin{barticle}
\bauthor{\bsnm{{Bagla}}, \binits{J.S.}},
\bauthor{\bsnm{{Padmanabhan}}, \binits{T.}}:
\batitle{{Transfer of power in non-linear gravitational clustering}}.
\bjtitle{\mnras}
\bvolume{286}(\bissue{4}),
\bfpage{1023}--\blpage{1031}
(\byear{1997})
{\href{https://arxiv.org/abs/astro-ph/9605202}{{arXiv:astro-ph/9605202}}}
{[astro-ph]}.
\doiurl{10.1093/mnras/286.4.1023}
\end{barticle}
\endbibitem

\bibitem{1997Prama..49..161B}
\begin{barticle}
\bauthor{\bsnm{{Bagla}}, \binits{J.S.}},
\bauthor{\bsnm{{Padmanabhan}}, \binits{T.}}:
\batitle{{Cosmological N-body simulations.}}
\bjtitle{Pramana}
\bvolume{49},
\bfpage{161}
(\byear{1997})
{\href{https://arxiv.org/abs/astro-ph/0411730}{{arXiv:astro-ph/0411730}}}
{[astro-ph]}.
\doiurl{10.1007/BF02845853}
\end{barticle}
\endbibitem

\bibitem{1998ApJ...495...25B}
\begin{barticle}
\bauthor{\bsnm{{Bagla}}, \binits{J.S.}},
\bauthor{\bsnm{{Engineer}}, \binits{S.}},
\bauthor{\bsnm{{Padmanabhan}}, \binits{T.}}:
\batitle{{Scaling Relations for Gravitational Clustering in Two Dimensions}}.
\bjtitle{\apj}
\bvolume{495}(\bissue{1}),
\bfpage{25}--\blpage{28}
(\byear{1998})
{\href{https://arxiv.org/abs/astro-ph/9707330}{{arXiv:astro-ph/9707330}}}
{[astro-ph]}.
\doiurl{10.1086/305292}
\end{barticle}
\endbibitem

\bibitem{2015MNRAS.450.1486A}
\begin{barticle}
\bauthor{\bsnm{{Ahn}}, \binits{K.}},
\bauthor{\bsnm{{Iliev}}, \binits{I.T.}},
\bauthor{\bsnm{{Shapiro}}, \binits{P.R.}},
\bauthor{\bsnm{{Srisawat}}, \binits{C.}}:
\batitle{{Non-linear bias of cosmological halo formation in the early
  universe}}.
\bjtitle{\mnras}
\bvolume{450}(\bissue{2}),
\bfpage{1486}--\blpage{1502}
(\byear{2015})
{\href{https://arxiv.org/abs/1407.2637}{{arXiv:1407.2637}}}
{[astro-ph.CO]}.
\doiurl{10.1093/mnras/stv704}
\end{barticle}
\endbibitem

\bibitem{2004ApJ...613....1F}
\begin{barticle}
\bauthor{\bsnm{{Furlanetto}}, \binits{S.R.}},
\bauthor{\bsnm{{Zaldarriaga}}, \binits{M.}},
\bauthor{\bsnm{{Hernquist}}, \binits{L.}}:
\batitle{{The Growth of H II Regions During Reionization}}.
\bjtitle{\apj}
\bvolume{613}(\bissue{1}),
\bfpage{1}--\blpage{15}
(\byear{2004})
{\href{https://arxiv.org/abs/astro-ph/0403697}{{arXiv:astro-ph/0403697}}}
{[astro-ph]}.
\doiurl{10.1086/423025}
\end{barticle}
\endbibitem

\bibitem{2011MNRAS.411..955M}
\begin{barticle}
\bauthor{\bsnm{{Mesinger}}, \binits{A.}},
\bauthor{\bsnm{{Furlanetto}}, \binits{S.}},
\bauthor{\bsnm{{Cen}}, \binits{R.}}:
\batitle{{21CMFAST: a fast, seminumerical simulation of the high-redshift 21-cm
  signal}}.
\bjtitle{\mnras}
\bvolume{411}(\bissue{2}),
\bfpage{955}--\blpage{972}
(\byear{2011})
{\href{https://arxiv.org/abs/1003.3878}{{arXiv:1003.3878}}}
{[astro-ph.CO]}.
\doiurl{10.1111/j.1365-2966.2010.17731.x}
\end{barticle}
\endbibitem

\bibitem{2010MNRAS.406.2421S}
\begin{barticle}
\bauthor{\bsnm{{Santos}}, \binits{M.G.}},
\bauthor{\bsnm{{Ferramacho}}, \binits{L.}},
\bauthor{\bsnm{{Silva}}, \binits{M.B.}},
\bauthor{\bsnm{{Amblard}}, \binits{A.}},
\bauthor{\bsnm{{Cooray}}, \binits{A.}}:
\batitle{{Fast large volume simulations of the 21-cm signal from the
  reionization and pre-reionization epochs}}.
\bjtitle{\mnras}
\bvolume{406}(\bissue{4}),
\bfpage{2421}--\blpage{2432}
(\byear{2010})
{\href{https://arxiv.org/abs/0911.2219}{{arXiv:0911.2219}}}
{[astro-ph.CO]}.
\doiurl{10.1111/j.1365-2966.2010.16898.x}
\end{barticle}
\endbibitem

\bibitem{2016MNRAS.457.1550H}
\begin{barticle}
\bauthor{\bsnm{{Hassan}}, \binits{S.}},
\bauthor{\bsnm{{Dav{\'e}}}, \binits{R.}},
\bauthor{\bsnm{{Finlator}}, \binits{K.}},
\bauthor{\bsnm{{Santos}}, \binits{M.G.}}:
\batitle{{Simulating the 21 cm signal from reionization including non-linear
  ionizations and inhomogeneous recombinations}}.
\bjtitle{\mnras}
\bvolume{457}(\bissue{2}),
\bfpage{1550}--\blpage{1567}
(\byear{2016})
{\href{https://arxiv.org/abs/1510.04280}{{arXiv:1510.04280}}}
{[astro-ph.CO]}.
\doiurl{10.1093/mnras/stv3001}
\end{barticle}
\endbibitem

\bibitem{2018MNRAS.477.1549H}
\begin{barticle}
\bauthor{\bsnm{{Hutter}}, \binits{A.}}:
\batitle{{The accuracy of seminumerical reionization models in comparison with
  radiative transfer simulations}}.
\bjtitle{\mnras}
\bvolume{477}(\bissue{2}),
\bfpage{1549}--\blpage{1566}
(\byear{2018})
{\href{https://arxiv.org/abs/1803.00088}{{arXiv:1803.00088}}}
{[astro-ph.CO]}.
\doiurl{10.1093/mnras/sty683}
\end{barticle}
\endbibitem

\bibitem{2009MNRAS.394..960C}
\begin{barticle}
\bauthor{\bsnm{{Choudhury}}, \binits{T.R.}},
\bauthor{\bsnm{{Haehnelt}}, \binits{M.G.}},
\bauthor{\bsnm{{Regan}}, \binits{J.}}:
\batitle{{Inside-out or outside-in: the topology of reionization in the
  photon-starved regime suggested by Ly{\ensuremath{\alpha}} forest data}}.
\bjtitle{\mnras}
\bvolume{394}(\bissue{2}),
\bfpage{960}--\blpage{977}
(\byear{2009})
{\href{https://arxiv.org/abs/0806.1524}{{arXiv:0806.1524}}}
{[astro-ph]}.
\doiurl{10.1111/j.1365-2966.2008.14383.x}
\end{barticle}
\endbibitem

\bibitem{2014MNRAS.443.2843M}
\begin{barticle}
\bauthor{\bsnm{{Majumdar}}, \binits{S.}},
\bauthor{\bsnm{{Mellema}}, \binits{G.}},
\bauthor{\bsnm{{Datta}}, \binits{K.K.}},
\bauthor{\bsnm{{Jensen}}, \binits{H.}},
\bauthor{\bsnm{{Choudhury}}, \binits{T.R.}},
\bauthor{\bsnm{{Bharadwaj}}, \binits{S.}},
\bauthor{\bsnm{{Friedrich}}, \binits{M.M.}}:
\batitle{{On the use of seminumerical simulations in predicting the 21-cm
  signal from the epoch of reionization}}.
\bjtitle{\mnras}
\bvolume{443}(\bissue{4}),
\bfpage{2843}--\blpage{2861}
(\byear{2014})
{\href{https://arxiv.org/abs/1403.0941}{{arXiv:1403.0941}}}
{[astro-ph.CO]}.
\doiurl{10.1093/mnras/stu1342}
\end{barticle}
\endbibitem

\bibitem{2017MNRAS.464.2992M}
\begin{barticle}
\bauthor{\bsnm{{Mondal}}, \binits{R.}},
\bauthor{\bsnm{{Bharadwaj}}, \binits{S.}},
\bauthor{\bsnm{{Majumdar}}, \binits{S.}}:
\batitle{{Statistics of the epoch of reionization (EoR) 21-cm signal - II. The
  evolution of the power-spectrum error-covariance}}.
\bjtitle{\mnras}
\bvolume{464}(\bissue{3}),
\bfpage{2992}--\blpage{3004}
(\byear{2017})
{\href{https://arxiv.org/abs/1606.03874}{{arXiv:1606.03874}}}
{[astro-ph.CO]}.
\doiurl{10.1093/mnras/stw2599}
\end{barticle}
\endbibitem

\bibitem{2016MNRAS.460.1801P}
\begin{barticle}
\bauthor{\bsnm{{Paranjape}}, \binits{A.}},
\bauthor{\bsnm{{Choudhury}}, \binits{T.R.}},
\bauthor{\bsnm{{Padmanabhan}}, \binits{H.}}:
\batitle{{Photon number conserving models of H II bubbles during
  reionization}}.
\bjtitle{\mnras}
\bvolume{460}(\bissue{2}),
\bfpage{1801}--\blpage{1810}
(\byear{2016})
{\href{https://arxiv.org/abs/1512.01345}{{arXiv:1512.01345}}}
{[astro-ph.CO]}.
\doiurl{10.1093/mnras/stw1060}
\end{barticle}
\endbibitem

\bibitem{2019MNRAS.489.5594M}
\begin{barticle}
\bauthor{\bsnm{{Molaro}}, \binits{M.}},
\bauthor{\bsnm{{Dav{\'e}}}, \binits{R.}},
\bauthor{\bsnm{{Hassan}}, \binits{S.}},
\bauthor{\bsnm{{Santos}}, \binits{M.G.}},
\bauthor{\bsnm{{Finlator}}, \binits{K.}}:
\batitle{{Artist: fast radiative transfer for large-scale simulations of the
  epoch of reionization}}.
\bjtitle{\mnras}
\bvolume{489}(\bissue{4}),
\bfpage{5594}--\blpage{5611}
(\byear{2019})
{\href{https://arxiv.org/abs/1901.03340}{{arXiv:1901.03340}}}
{[astro-ph.CO]}.
\doiurl{10.1093/mnras/stz2171}
\end{barticle}
\endbibitem

\bibitem{2021arXiv211205184P}
\begin{botherref}
\oauthor{\bsnm{{Park}}, \binits{J.}},
\oauthor{\bsnm{{Greig}}, \binits{B.}},
\oauthor{\bsnm{{Mesinger}}, \binits{A.}}:
{Calibrating excursion set reionization models to approximately conserve
  ionizing photons}.
arXiv e-prints,
2112--05184
(2021)
{\href{https://arxiv.org/abs/2112.05184}{{arXiv:2112.05184}}}
{[astro-ph.CO]}
\end{botherref}
\endbibitem

\bibitem{2015MNRAS.447.3402B}
\begin{barticle}
\bauthor{\bsnm{{Becker}}, \binits{G.D.}},
\bauthor{\bsnm{{Bolton}}, \binits{J.S.}},
\bauthor{\bsnm{{Madau}}, \binits{P.}},
\bauthor{\bsnm{{Pettini}}, \binits{M.}},
\bauthor{\bsnm{{Ryan-Weber}}, \binits{E.V.}},
\bauthor{\bsnm{{Venemans}}, \binits{B.P.}}:
\batitle{{Evidence of patchy hydrogen reionization from an extreme
  Ly{\ensuremath{\alpha}} trough below redshift six}}.
\bjtitle{\mnras}
\bvolume{447}(\bissue{4}),
\bfpage{3402}--\blpage{3419}
(\byear{2015})
{\href{https://arxiv.org/abs/1407.4850}{{arXiv:1407.4850}}}
{[astro-ph.CO]}.
\doiurl{10.1093/mnras/stu2646}
\end{barticle}
\endbibitem

\bibitem{2018MNRAS.479.1055B}
\begin{barticle}
\bauthor{\bsnm{{Bosman}}, \binits{S.E.I.}},
\bauthor{\bsnm{{Fan}}, \binits{X.}},
\bauthor{\bsnm{{Jiang}}, \binits{L.}},
\bauthor{\bsnm{{Reed}}, \binits{S.}},
\bauthor{\bsnm{{Matsuoka}}, \binits{Y.}},
\bauthor{\bsnm{{Becker}}, \binits{G.}},
\bauthor{\bsnm{{Haehnelt}}, \binits{M.}}:
\batitle{{New constraints on Lyman-{\ensuremath{\alpha}} opacity with a sample
  of 62 quasarsat z > 5.7}}.
\bjtitle{\mnras}
\bvolume{479}(\bissue{1}),
\bfpage{1055}--\blpage{1076}
(\byear{2018})
{\href{https://arxiv.org/abs/1802.08177}{{arXiv:1802.08177}}}
{[astro-ph.GA]}.
\doiurl{10.1093/mnras/sty1344}
\end{barticle}
\endbibitem

\bibitem{2019ApJ...881...23E}
\begin{barticle}
\bauthor{\bsnm{{Eilers}}, \binits{A.-C.}},
\bauthor{\bsnm{{Hennawi}}, \binits{J.F.}},
\bauthor{\bsnm{{Davies}}, \binits{F.B.}},
\bauthor{\bsnm{{O{\~n}orbe}}, \binits{J.}}:
\batitle{{Anomaly in the Opacity of the Post-reionization Intergalactic Medium
  in the Ly{\ensuremath{\alpha}} and Ly{\ensuremath{\beta}} Forest}}.
\bjtitle{\apj}
\bvolume{881}(\bissue{1}),
\bfpage{23}
(\byear{2019})
{\href{https://arxiv.org/abs/1906.05874}{{arXiv:1906.05874}}}
{[astro-ph.GA]}.
\doiurl{10.3847/1538-4357/ab2b3f}
\end{barticle}
\endbibitem

\bibitem{2019MNRAS.485L..24K}
\begin{barticle}
\bauthor{\bsnm{{Kulkarni}}, \binits{G.}},
\bauthor{\bsnm{{Keating}}, \binits{L.C.}},
\bauthor{\bsnm{{Haehnelt}}, \binits{M.G.}},
\bauthor{\bsnm{{Bosman}}, \binits{S.E.I.}},
\bauthor{\bsnm{{Puchwein}}, \binits{E.}},
\bauthor{\bsnm{{Chardin}}, \binits{J.}},
\bauthor{\bsnm{{Aubert}}, \binits{D.}}:
\batitle{{Large Ly {\ensuremath{\alpha}} opacity fluctuations and low CMB
  {\ensuremath{\tau}} in models of late reionization with large islands of
  neutral hydrogen extending to z < 5.5}}.
\bjtitle{\mnras}
\bvolume{485}(\bissue{1}),
\bfpage{24}--\blpage{28}
(\byear{2019})
{\href{https://arxiv.org/abs/1809.06374}{{arXiv:1809.06374}}}
{[astro-ph.CO]}.
\doiurl{10.1093/mnrasl/slz025}
\end{barticle}
\endbibitem

\bibitem{2020MNRAS.497..906K}
\begin{barticle}
\bauthor{\bsnm{{Keating}}, \binits{L.C.}},
\bauthor{\bsnm{{Kulkarni}}, \binits{G.}},
\bauthor{\bsnm{{Haehnelt}}, \binits{M.G.}},
\bauthor{\bsnm{{Chardin}}, \binits{J.}},
\bauthor{\bsnm{{Aubert}}, \binits{D.}}:
\batitle{{Constraining the second half of reionization with the Ly
  {\ensuremath{\beta}} forest}}.
\bjtitle{\mnras}
\bvolume{497}(\bissue{1}),
\bfpage{906}--\blpage{915}
(\byear{2020})
{\href{https://arxiv.org/abs/1912.05582}{{arXiv:1912.05582}}}
{[astro-ph.CO]}.
\doiurl{10.1093/mnras/staa1909}
\end{barticle}
\endbibitem

\bibitem{2020MNRAS.494.3080N}
\begin{barticle}
\bauthor{\bsnm{{Nasir}}, \binits{F.}},
\bauthor{\bsnm{{D'Aloisio}}, \binits{A.}}:
\batitle{{Observing the tail of reionization: neutral islands in the z = 5.5
  Lyman-{\ensuremath{\alpha}} forest}}.
\bjtitle{\mnras}
\bvolume{494}(\bissue{3}),
\bfpage{3080}--\blpage{3094}
(\byear{2020})
{\href{https://arxiv.org/abs/1910.03570}{{arXiv:1910.03570}}}
{[astro-ph.CO]}.
\doiurl{10.1093/mnras/staa894}
\end{barticle}
\endbibitem

\bibitem{2021MNRAS.501.5782C}
\begin{barticle}
\bauthor{\bsnm{{Choudhury}}, \binits{T.R.}},
\bauthor{\bsnm{{Paranjape}}, \binits{A.}},
\bauthor{\bsnm{{Bosman}}, \binits{S.E.I.}}:
\batitle{{Studying the Lyman {\ensuremath{\alpha}} optical depth fluctuations
  at z {\ensuremath{\sim}} 5.5 using fast semi-numerical methods}}.
\bjtitle{\mnras}
\bvolume{501}(\bissue{4}),
\bfpage{5782}--\blpage{5796}
(\byear{2021})
{\href{https://arxiv.org/abs/2003.08958}{{arXiv:2003.08958}}}
{[astro-ph.CO]}.
\doiurl{10.1093/mnras/stab045}
\end{barticle}
\endbibitem

\bibitem{2010MNRAS.406..612B}
\begin{barticle}
\bauthor{\bsnm{{Bolton}}, \binits{J.S.}},
\bauthor{\bsnm{{Becker}}, \binits{G.D.}},
\bauthor{\bsnm{{Wyithe}}, \binits{J.S.B.}},
\bauthor{\bsnm{{Haehnelt}}, \binits{M.G.}},
\bauthor{\bsnm{{Sargent}}, \binits{W.L.W.}}:
\batitle{{A first direct measurement of the intergalactic medium temperature
  around a quasar at z = 6}}.
\bjtitle{\mnras}
\bvolume{406}(\bissue{1}),
\bfpage{612}--\blpage{625}
(\byear{2010})
{\href{https://arxiv.org/abs/1001.3415}{{arXiv:1001.3415}}}
{[astro-ph.CO]}.
\doiurl{10.1111/j.1365-2966.2010.16701.x}
\end{barticle}
\endbibitem

\bibitem{2012MNRAS.419.2880B}
\begin{barticle}
\bauthor{\bsnm{{Bolton}}, \binits{J.S.}},
\bauthor{\bsnm{{Becker}}, \binits{G.D.}},
\bauthor{\bsnm{{Raskutti}}, \binits{S.}},
\bauthor{\bsnm{{Wyithe}}, \binits{J.S.B.}},
\bauthor{\bsnm{{Haehnelt}}, \binits{M.G.}},
\bauthor{\bsnm{{Sargent}}, \binits{W.L.W.}}:
\batitle{{Improved measurements of the intergalactic medium temperature around
  quasars: possible evidence for the initial stages of He II reionization at z
  $\approx$ 6}}.
\bjtitle{\mnras}
\bvolume{419}(\bissue{4}),
\bfpage{2880}--\blpage{2892}
(\byear{2012})
{\href{https://arxiv.org/abs/1110.0539}{{arXiv:1110.0539}}}
{[astro-ph.CO]}.
\doiurl{10.1111/j.1365-2966.2011.19929.x}
\end{barticle}
\endbibitem

\bibitem{2019ApJ...872...13W}
\begin{barticle}
\bauthor{\bsnm{{Walther}}, \binits{M.}},
\bauthor{\bsnm{{O{\~n}orbe}}, \binits{J.}},
\bauthor{\bsnm{{Hennawi}}, \binits{J.F.}},
\bauthor{\bsnm{{Luki{\'c}}}, \binits{Z.}}:
\batitle{{New Constraints on IGM Thermal Evolution from the
  Ly{\ensuremath{\alpha}} Forest Power Spectrum}}.
\bjtitle{\apj}
\bvolume{872}(\bissue{1}),
\bfpage{13}
(\byear{2019})
{\href{https://arxiv.org/abs/1808.04367}{{arXiv:1808.04367}}}
{[astro-ph.CO]}.
\doiurl{10.3847/1538-4357/aafad1}
\end{barticle}
\endbibitem

\bibitem{2019ApJ...872..101B}
\begin{barticle}
\bauthor{\bsnm{{Boera}}, \binits{E.}},
\bauthor{\bsnm{{Becker}}, \binits{G.D.}},
\bauthor{\bsnm{{Bolton}}, \binits{J.S.}},
\bauthor{\bsnm{{Nasir}}, \binits{F.}}:
\batitle{{Revealing Reionization with the Thermal History of the Intergalactic
  Medium: New Constraints from the Ly{\ensuremath{\alpha}} Flux Power
  Spectrum}}.
\bjtitle{\apj}
\bvolume{872}(\bissue{1}),
\bfpage{101}
(\byear{2019})
{\href{https://arxiv.org/abs/1809.06980}{{arXiv:1809.06980}}}
{[astro-ph.CO]}.
\doiurl{10.3847/1538-4357/aafee4}
\end{barticle}
\endbibitem

\bibitem{2020MNRAS.494.5091G}
\begin{barticle}
\bauthor{\bsnm{{Gaikwad}}, \binits{P.}},
\bauthor{\bsnm{{Rauch}}, \binits{M.}},
\bauthor{\bsnm{{Haehnelt}}, \binits{M.G.}},
\bauthor{\bsnm{{Puchwein}}, \binits{E.}},
\bauthor{\bsnm{{Bolton}}, \binits{J.S.}},
\bauthor{\bsnm{{Keating}}, \binits{L.C.}},
\bauthor{\bsnm{{Kulkarni}}, \binits{G.}},
\bauthor{\bsnm{{Ir{\v{s}}i{\v{c}}}}, \binits{V.}},
\bauthor{\bsnm{{Ba{\~n}ados}}, \binits{E.}},
\bauthor{\bsnm{{Becker}}, \binits{G.D.}},
\bauthor{\bsnm{{Boera}}, \binits{E.}},
\bauthor{\bsnm{{Zahedy}}, \binits{F.S.}},
\bauthor{\bsnm{{Chen}}, \binits{H.-W.}},
\bauthor{\bsnm{{Carswell}}, \binits{R.F.}},
\bauthor{\bsnm{{Chardin}}, \binits{J.}},
\bauthor{\bsnm{{Rorai}}, \binits{A.}}:
\batitle{{Probing the thermal state of the intergalactic medium at z > 5 with
  the transmission spikes in high-resolution Ly {\ensuremath{\alpha}} forest
  spectra}}.
\bjtitle{\mnras}
\bvolume{494}(\bissue{4}),
\bfpage{5091}--\blpage{5109}
(\byear{2020})
{\href{https://arxiv.org/abs/2001.10018}{{arXiv:2001.10018}}}
{[astro-ph.CO]}.
\doiurl{10.1093/mnras/staa907}
\end{barticle}
\endbibitem

\bibitem{2012MNRAS.423..558C}
\begin{barticle}
\bauthor{\bsnm{{Ciardi}}, \binits{B.}},
\bauthor{\bsnm{{Bolton}}, \binits{J.S.}},
\bauthor{\bsnm{{Maselli}}, \binits{A.}},
\bauthor{\bsnm{{Graziani}}, \binits{L.}}:
\batitle{{The effect of intergalactic helium on hydrogen reionization:
  implications for the sources of ionizing photons at z>6}}.
\bjtitle{\mnras}
\bvolume{423}(\bissue{1}),
\bfpage{558}--\blpage{574}
(\byear{2012})
{\href{https://arxiv.org/abs/1112.4646}{{arXiv:1112.4646}}}
{[astro-ph.CO]}.
\doiurl{10.1111/j.1365-2966.2012.20902.x}
\end{barticle}
\endbibitem

\bibitem{2012MNRAS.421.1969R}
\begin{barticle}
\bauthor{\bsnm{{Raskutti}}, \binits{S.}},
\bauthor{\bsnm{{Bolton}}, \binits{J.S.}},
\bauthor{\bsnm{{Wyithe}}, \binits{J.S.B.}},
\bauthor{\bsnm{{Becker}}, \binits{G.D.}}:
\batitle{{Thermal constraints on the reionization of hydrogen by Population II
  stellar sources}}.
\bjtitle{\mnras}
\bvolume{421}(\bissue{3}),
\bfpage{1969}--\blpage{1981}
(\byear{2012})
{\href{https://arxiv.org/abs/1201.5138}{{arXiv:1201.5138}}}
{[astro-ph.CO]}.
\doiurl{10.1111/j.1365-2966.2011.20401.x}
\end{barticle}
\endbibitem

\bibitem{2014MNRAS.443.3761P}
\begin{barticle}
\bauthor{\bsnm{{Padmanabhan}}, \binits{H.}},
\bauthor{\bsnm{{Choudhury}}, \binits{T.R.}},
\bauthor{\bsnm{{Srianand}}, \binits{R.}}:
\batitle{{Probing reionization using quasar near-zones at redshift z
  {\ensuremath{\sim}} 6}}.
\bjtitle{\mnras}
\bvolume{443}(\bissue{4}),
\bfpage{3761}--\blpage{3779}
(\byear{2014})
{\href{https://arxiv.org/abs/1403.0221}{{arXiv:1403.0221}}}
{[astro-ph.CO]}.
\doiurl{10.1093/mnras/stu1433}
\end{barticle}
\endbibitem

\bibitem{2022MNRAS.511.2239M}
\begin{barticle}
\bauthor{\bsnm{{Maity}}, \binits{B.}},
\bauthor{\bsnm{{Choudhury}}, \binits{T.R.}}:
\batitle{{Probing the thermal history during reionization using a seminumerical
  photon-conserving code SCRIPT}}.
\bjtitle{\mnras}
\bvolume{511}(\bissue{2}),
\bfpage{2239}--\blpage{2258}
(\byear{2022})
{\href{https://arxiv.org/abs/2110.14231}{{arXiv:2110.14231}}}
{[astro-ph.CO]}.
\doiurl{10.1093/mnras/stac182}
\end{barticle}
\endbibitem

\bibitem{2017arXiv170403416D}
\begin{botherref}
\oauthor{\bsnm{{Dijkstra}}, \binits{M.}}:
{Saas-Fee Lecture Notes: Physics of Lyman Alpha Radiative Transfer}.
arXiv e-prints,
1704--03416
(2017)
{\href{https://arxiv.org/abs/1704.03416}{{arXiv:1704.03416}}}
{[astro-ph.GA]}
\end{botherref}
\endbibitem

\bibitem{2020ARA&A..58..617O}
\begin{barticle}
\bauthor{\bsnm{{Ouchi}}, \binits{M.}},
\bauthor{\bsnm{{Ono}}, \binits{Y.}},
\bauthor{\bsnm{{Shibuya}}, \binits{T.}}:
\batitle{{Observations of the Lyman-{\ensuremath{\alpha}} Universe}}.
\bjtitle{\araa}
\bvolume{58},
\bfpage{617}--\blpage{659}
(\byear{2020})
{\href{https://arxiv.org/abs/2012.07960}{{arXiv:2012.07960}}}
{[astro-ph.GA]}.
\doiurl{10.1146/annurev-astro-032620-021859}
\end{barticle}
\endbibitem

\bibitem{1980ARA&A..18..537S}
\begin{barticle}
\bauthor{\bsnm{{Sunyaev}}, \binits{R.A.}},
\bauthor{\bsnm{{Zeldovich}}, \binits{I.B.}}:
\batitle{{Microwave background radiation as a probe of the contemporary
  structure and history of the universe}}.
\bjtitle{\araa}
\bvolume{18},
\bfpage{537}--\blpage{560}
(\byear{1980}).
\doiurl{10.1146/annurev.aa.18.090180.002541}
\end{barticle}
\endbibitem

\bibitem{2021ApJ...908..199R}
\begin{barticle}
\bauthor{\bsnm{{Reichardt}}, \binits{C.L.}},
\bauthor{\bsnm{{Patil}}, \binits{S.}},
\bauthor{\bsnm{{Ade}}, \binits{P.A.R.}},
\bauthor{\bsnm{{Anderson}}, \binits{A.J.}},
\bauthor{\bsnm{{Austermann}}, \binits{J.E.}},
\bauthor{\bsnm{{Avva}}, \binits{J.S.}},
\bauthor{\bsnm{{Baxter}}, \binits{E.}},
\bauthor{\bsnm{{Beall}}, \binits{J.A.}},
\bauthor{\bsnm{{Bender}}, \binits{A.N.}},
\bauthor{\bsnm{{Benson}}, \binits{B.A.}},
\bauthor{\bsnm{{Bianchini}}, \binits{F.}},
\bauthor{\bsnm{{Bleem}}, \binits{L.E.}},
\bauthor{\bsnm{{Carlstrom}}, \binits{J.E.}},
\bauthor{\bsnm{{Chang}}, \binits{C.L.}},
\bauthor{\bsnm{{Chaubal}}, \binits{P.}},
\bauthor{\bsnm{{Chiang}}, \binits{H.C.}},
\bauthor{\bsnm{{Chou}}, \binits{T.L.}},
\bauthor{\bsnm{{Citron}}, \binits{R.}},
\bauthor{\bsnm{{Moran}}, \binits{C.C.}},
\bauthor{\bsnm{{Crawford}}, \binits{T.M.}},
\bauthor{\bsnm{{Crites}}, \binits{A.T.}},
\bauthor{\bsnm{{de Haan}}, \binits{T.}},
\bauthor{\bsnm{{Dobbs}}, \binits{M.A.}},
\bauthor{\bsnm{{Everett}}, \binits{W.}},
\bauthor{\bsnm{{Gallicchio}}, \binits{J.}},
\bauthor{\bsnm{{George}}, \binits{E.M.}},
\bauthor{\bsnm{{Gilbert}}, \binits{A.}},
\bauthor{\bsnm{{Gupta}}, \binits{N.}},
\bauthor{\bsnm{{Halverson}}, \binits{N.W.}},
\bauthor{\bsnm{{Harrington}}, \binits{N.}},
\bauthor{\bsnm{{Henning}}, \binits{J.W.}},
\bauthor{\bsnm{{Hilton}}, \binits{G.C.}},
\bauthor{\bsnm{{Holder}}, \binits{G.P.}},
\bauthor{\bsnm{{Holzapfel}}, \binits{W.L.}},
\bauthor{\bsnm{{Hrubes}}, \binits{J.D.}},
\bauthor{\bsnm{{Huang}}, \binits{N.}},
\bauthor{\bsnm{{Hubmayr}}, \binits{J.}},
\bauthor{\bsnm{{Irwin}}, \binits{K.D.}},
\bauthor{\bsnm{{Knox}}, \binits{L.}},
\bauthor{\bsnm{{Lee}}, \binits{A.T.}},
\bauthor{\bsnm{{Li}}, \binits{D.}},
\bauthor{\bsnm{{Lowitz}}, \binits{A.}},
\bauthor{\bsnm{{Luong-Van}}, \binits{D.}},
\bauthor{\bsnm{{McMahon}}, \binits{J.J.}},
\bauthor{\bsnm{{Mehl}}, \binits{J.}},
\bauthor{\bsnm{{Meyer}}, \binits{S.S.}},
\bauthor{\bsnm{{Millea}}, \binits{M.}},
\bauthor{\bsnm{{Mocanu}}, \binits{L.M.}},
\bauthor{\bsnm{{Mohr}}, \binits{J.J.}},
\bauthor{\bsnm{{Montgomery}}, \binits{J.}},
\bauthor{\bsnm{{Nadolski}}, \binits{A.}},
\bauthor{\bsnm{{Natoli}}, \binits{T.}},
\bauthor{\bsnm{{Nibarger}}, \binits{J.P.}},
\bauthor{\bsnm{{Noble}}, \binits{G.}},
\bauthor{\bsnm{{Novosad}}, \binits{V.}},
\bauthor{\bsnm{{Omori}}, \binits{Y.}},
\bauthor{\bsnm{{Padin}}, \binits{S.}},
\bauthor{\bsnm{{Pryke}}, \binits{C.}},
\bauthor{\bsnm{{Ruhl}}, \binits{J.E.}},
\bauthor{\bsnm{{Saliwanchik}}, \binits{B.R.}},
\bauthor{\bsnm{{Sayre}}, \binits{J.T.}},
\bauthor{\bsnm{{Schaffer}}, \binits{K.K.}},
\bauthor{\bsnm{{Shirokoff}}, \binits{E.}},
\bauthor{\bsnm{{Sievers}}, \binits{C.}},
\bauthor{\bsnm{{Smecher}}, \binits{G.}},
\bauthor{\bsnm{{Spieler}}, \binits{H.G.}},
\bauthor{\bsnm{{Staniszewski}}, \binits{Z.}},
\bauthor{\bsnm{{Stark}}, \binits{A.A.}},
\bauthor{\bsnm{{Tucker}}, \binits{C.}},
\bauthor{\bsnm{{Vanderlinde}}, \binits{K.}},
\bauthor{\bsnm{{Veach}}, \binits{T.}},
\bauthor{\bsnm{{Vieira}}, \binits{J.D.}},
\bauthor{\bsnm{{Wang}}, \binits{G.}},
\bauthor{\bsnm{{Whitehorn}}, \binits{N.}},
\bauthor{\bsnm{{Williamson}}, \binits{R.}},
\bauthor{\bsnm{{Wu}}, \binits{W.L.K.}},
\bauthor{\bsnm{{Yefremenko}}, \binits{V.}}:
\batitle{{An Improved Measurement of the Secondary Cosmic Microwave Background
  Anisotropies from the SPT-SZ + SPTpol Surveys}}.
\bjtitle{\apj}
\bvolume{908}(\bissue{2}),
\bfpage{199}
(\byear{2021})
{\href{https://arxiv.org/abs/2002.06197}{{arXiv:2002.06197}}}
{[astro-ph.CO]}.
\doiurl{10.3847/1538-4357/abd407}
\end{barticle}
\endbibitem

\bibitem{2021MNRAS.501L...7C}
\begin{barticle}
\bauthor{\bsnm{{Choudhury}}, \binits{T.R.}},
\bauthor{\bsnm{{Mukherjee}}, \binits{S.}},
\bauthor{\bsnm{{Paul}}, \binits{S.}}:
\batitle{{Cosmic microwave background constraints on a physical model of
  reionization}}.
\bjtitle{\mnras}
\bvolume{501}(\bissue{1}),
\bfpage{7}--\blpage{11}
(\byear{2021})
{\href{https://arxiv.org/abs/2007.03705}{{arXiv:2007.03705}}}
{[astro-ph.CO]}.
\doiurl{10.1093/mnrasl/slaa185}
\end{barticle}
\endbibitem

\bibitem{2022arXiv220208698G}
\begin{barticle}
\bauthor{\bsnm{{Gorce}}, \binits{A.}},
\bauthor{\bsnm{{Douspis}}, \binits{M.}},
\bauthor{\bsnm{{Salvati}}, \binits{L.}}:
\batitle{{Retrieving cosmological information from small-scale CMB foregrounds.
  II. The kinetic Sunyaev Zel'dovich effect}}.
\bjtitle{\aap}
\bvolume{662},
\bfpage{122}
(\byear{2022})
{\href{https://arxiv.org/abs/2202.08698}{{arXiv:2202.08698}}}
{[astro-ph.CO]}.
\doiurl{10.1051/0004-6361/202243351}
\end{barticle}
\endbibitem

\bibitem{2022arXiv220304337C}
\begin{botherref}
\oauthor{\bsnm{{Chen}}, \binits{N.}},
\oauthor{\bsnm{{Trac}}, \binits{H.}},
\oauthor{\bsnm{{Mukherjee}}, \binits{S.}},
\oauthor{\bsnm{{Cen}}, \binits{R.}}:
{Patchy Kinetic Sunyaev-Zel'dovich Effect with Controlled Reionization History
  and Morphology}.
arXiv e-prints,
2203--04337
(2022)
{\href{https://arxiv.org/abs/2203.04337}{{arXiv:2203.04337}}}
{[astro-ph.CO]}
\end{botherref}
\endbibitem

\bibitem{2021MNRAS.502.5134B}
\begin{barticle}
\bauthor{\bsnm{{Beniamini}}, \binits{P.}},
\bauthor{\bsnm{{Kumar}}, \binits{P.}},
\bauthor{\bsnm{{Ma}}, \binits{X.}},
\bauthor{\bsnm{{Quataert}}, \binits{E.}}:
\batitle{{Exploring the epoch of hydrogen reionization using FRBs}}.
\bjtitle{\mnras}
\bvolume{502}(\bissue{4}),
\bfpage{5134}--\blpage{5146}
(\byear{2021})
{\href{https://arxiv.org/abs/2011.11643}{{arXiv:2011.11643}}}
{[astro-ph.CO]}.
\doiurl{10.1093/mnras/stab309}
\end{barticle}
\endbibitem

\bibitem{2021MNRAS.505.2195P}
\begin{barticle}
\bauthor{\bsnm{{Pagano}}, \binits{M.}},
\bauthor{\bsnm{{Fronenberg}}, \binits{H.}}:
\batitle{{Constraining the epoch of reionization with highly dispersed fast
  radio bursts}}.
\bjtitle{\mnras}
\bvolume{505}(\bissue{2}),
\bfpage{2195}--\blpage{2206}
(\byear{2021})
{\href{https://arxiv.org/abs/2103.03252}{{arXiv:2103.03252}}}
{[astro-ph.CO]}.
\doiurl{10.1093/mnras/stab1438}
\end{barticle}
\endbibitem

\bibitem{2022ApJ...933...57H}
\begin{barticle}
\bauthor{\bsnm{{Heimersheim}}, \binits{S.}},
\bauthor{\bsnm{{Sartorio}}, \binits{N.S.}},
\bauthor{\bsnm{{Fialkov}}, \binits{A.}},
\bauthor{\bsnm{{Lorimer}}, \binits{D.R.}}:
\batitle{{What It Takes to Measure Reionization with Fast Radio Bursts}}.
\bjtitle{\apj}
\bvolume{933}(\bissue{1}),
\bfpage{57}
(\byear{2022})
{\href{https://arxiv.org/abs/2203.12645}{{arXiv:2203.12645}}}
{[astro-ph.CO]}.
\doiurl{10.3847/1538-4357/ac70c9}
\end{barticle}
\endbibitem

\bibitem{2019arXiv190912491T}
\begin{botherref}
\oauthor{\bsnm{{Trott}}, \binits{C.M.}},
\oauthor{\bsnm{{Pober}}, \binits{J.C.}}:
{The status of 21cm interferometric experiments}.
arXiv e-prints,
1909--12491
(2019)
{\href{https://arxiv.org/abs/1909.12491}{{arXiv:1909.12491}}}
{[astro-ph.IM]}
\end{botherref}
\endbibitem

}
}

\end{thebibliography}

\end{document}